\documentclass[twocolumn]{aastex63}

\usepackage{lineno}
\received{00 00, 0000}
\revised{00 00, 0000}
\accepted{\today}
\submitjournal{ApJ}

\shorttitle{Diagnosing Circumburst Environment with Multiband GRB Radio Afterglows}
\shortauthors{Zhang et al.}
\begin{document}

\title{Diagnosing Circumburst Environment with Multiband Gamma-Ray Burst Radio Afterglows}

\correspondingauthor{Bo Zhang}
\email{zhangbo@nao.cas.cn}
\correspondingauthor{Xue-Feng Wu}
\email{xfwu@pmo.ac.cn}

\author[0000-0002-4781-7056]{Bo Zhang}
\affiliation{National Astronomical Observatories, Chinese Academy of Sciences, 20A Datun Road, Beijing, 100012, People’s Republic of China}
\affiliation{CAS Key Laboratory of FAST, National Astronomical Observatories, Chinese Academy of Sciences, People’s Republic of China}
\affiliation{Purple Mountain Observatory, Chinese Academy of Sciences, 
Nanjing, Jiangsu, 210034, People’s Republic of China}

\author{Liang-Duan Liu}
\affiliation{Institute of Astrophysics, Central China Normal University, Wuhan 430079, People’s Republic of China}
 \affiliation{Department of Astronomy, Beijing Normal University, 
 Beijing, 100875, People’s Republic of China}
% \affiliation{School of Astronomy and Space Science, Nanjing University\\
% Nanjing, Jiangsu, 210093, China}

%\collaboration{1}{(AAS Journals Data Scientists collaboration)}

\author{Tian-Rui Sun}
\affiliation{Purple Mountain Observatory, Chinese Academy of Sciences, Nanjing, Jiangsu, 210034, People’s Republic of China}
\affiliation{School of Astronomy and Space Science, University of Science and Technology of China, Hefei, Anhui, 230026, People’s Republic of China}
%\nocollaboration{1}

\author{Fen Lyu}
\affiliation{Astronomical Research Center, Shanghai Science and Technology Museum, Shanghai, 201306, People’s Republic of China}
\affiliation{Purple Mountain Observatory, Chinese Academy of Sciences, Nanjing, Jiangsu, 210034, People’s Republic of China}
%\affiliation{University of Chinese Academy of Sciences\\
%Beijing, 100049, China}

%\collaboration{1}{(LaTeX collaboration)}

\author{Xue-Feng Wu}
\affiliation{Purple Mountain Observatory, Chinese Academy of Sciences,
Nanjing, Jiangsu, 210034, People’s Republic of China}

%\author{Scott Chernoff}
%\affiliation{IOP Publishing, Washington, DC 20005}




\begin{abstract}

It has been widely recognized that gamma-ray burst (GRB) afterglows arise from interactions between the GRB outflow and circumburst medium, while their evolution follows the behaviors of relativistic shock waves. Assuming the distribution of circumburst medium follows a general power-law form, that is, $n = A_{\ast} R^{-k}$, where $R$ denotes the distance from the burst, it is obvious that the value of the density-distribution index $k$ can affect the behaviors of the afterglow. In this paper, we analyze the temporal and spectral behaviors of GRB radio afterglows with arbitrary $k$ values. In the radio band, a standard GRB afterglow produced by a forward shock exhibits a late-time flux peak, and the relative peak fluxes as well as peak times at different frequencies show dependencies on $k$. Thus, with multiband radio-peak observations, one can determine the density profile of the circumburst medium by comparing the relations between peak flux/time and frequency at each observing band. Also, the effects of transrelativistic shock waves, as well as jets in afterglows are discussed. By analyzing 31 long and 1 short GRB with multiband data of radio afterglows, we find that nearly half of them can be explained with uniform interstellar medium ($k=0$), $\sim 1/5$ can be constrained to exhibiting a stellar-wind environment  ($k=2$), while less than $\sim 1/3$ of the samples show $0< k< 2$.\\
\textit{Unified Astronomy Thesaurus concepts: Burst astrophysics (187), Gamma-ray bursts (629)}\\

\end{abstract}

%\keywords{editorials, notices --- 
%miscellaneous --- catalogs --- surveys}


\section{Introduction} \label{sec:intro}

Gamma-ray bursts (GRBs) are stellar explosions arising from the collapses of massive stars (for long bursts, e.g., see \citealt{Woosley1993}; \citealt{Paczynski1998}; \citealt{MacFadyen1999}) or mergers of binary compact stars (for short bursts, e.g., see \citealt{Paczynski1986}; \citealt{Narayan1992}; \citealt{Gehrels2009}; \citealt{Abbott2017}). The afterglows of GRBs are the results of interactions between relativistic ejecta from central engines and the circumburst medium. The shock waves produced during such interactions can accelerate the swept-up electrons, thus giving rise to nonthermal radiation, such as synchrotron emission (e.g., see \citealt{Piran1999}; \citealt{vanParadijs2000}; \citealt{Meszaros2002}). Our understanding of GRB afterglows has been greatly improved with the discoveries of GRB multiband afterglows (\citealt{Costa1997}; \citealt{vanParadijs1997}; \citealt{Frail1997}; \citealt{Zhang2007}).

Assuming the profile of the circumburst density $n$ follows a power-law form, that is, $n = A_{\ast} R^{-k}$, where $R$ denotes the distance from GRB central engine, it is obvious that different values of the power-law index $k$ lead to different light-curve behaviors (e.g., see \citealt{Wu2005}). The most widely discussed circumburst density model is the homogeneous interstellar medium (ISM) case with $k=0$ (e.g., see \citealt{Sari1998}; \citealt{Kobayashi2000}; \citealt{Panaitescu2004}), while the stellar-wind environment with $k=2$ has also been proposed (e.g., see \citealt{Dai1998}; \citealt{Meszaros1998}; \citealt{Chevalier2000}; \citealt{Panaitescu2000, Panaitescu2004}; \citealt{Kobayashi2003}; \citealt{Wu2003, Wu2004}; \citealt{Zou2005}). Previously, it was thought that because the long GRBs originate from massive stars with significant mass losses via stellar winds before their demise, such events should preferably occur in stellar-wind environments. For example, \cite{Starling2008} performed a joint fitting for X-ray to IR afterglows of 10 GRBs, and put constraints on $k$ for half of the sample, with 4 windlike cases and 1 ISM one. However, \cite{Panaitescu2002} found that half of their 10 GRBs favor a $k=0$ medium; while \cite{Curran2009} made $k$ constraints for 6 out of 10 GRBs, with 2 consistent with the ISM environment, 2 with the wind case, and another 2 samples compatible with both $k=0$ and $2$ cases. \cite{Yi2013} and \cite{vanEerten2014} presented various relationships between observables in GRB X-ray and optical afterglows for arbitrary $k$. However, \cite{Yi2013} have drawn a typical value of $k \sim 1$ for 19 \textit{Swift} long bursts by analyzing their early afterglows that originated from forward-reverse-shock interactions, implying a mass-loss scheme of the GRB progenitors other than stellar wind.  

In order to give a better understanding of circumburst medium distributions, it is necessary to investigate this issue with methods other than early X-ray/Optical observations. In this work, we utilize a new way to constrain the density distribution of the circumburst medium, that is, to put a constraint on $k$ by comparing multiband GRB radio afterglow peak times/fluxes with the corresponding frequencies. Observationally, the light curves of GRB radio afterglows usually show a double-peaked structure. In the rest frame, the typical peak times are $0.1-0.2$ days and 2 days after triggers of prompt emissions, respectively \citep{Chandra2012}. Generally speaking, the first peak can be explained with forward--reverse-shock interactions at early times, while the second peak is due to late-time forward-shock evolution and is unique for the radio band. In fact, because a strong dependency exists between GRB afterglow behaviors in the radio band and the underlying synchrotron radiation, \cite{BarniolDuran2014} and \cite{Beniamini2017} have already put constraints on the GRB microphysical parameters with GRB radio peaks, including the fraction of electron and magnetic energy among shock energy, that is, $\varepsilon_e$ and $\varepsilon_b$, respectively.

In our circumburst density analysis, we ignore the effects of early radio peaks due to reverse shocks and only take the second peak into consideration. Since the peak flux $F_{{\rm peak}, \nu}$ and the peak time $t_{{\rm peak},\nu}$ of this late peak depend on $k$, the value of $k$ can be deduced with multiband radio-peak observations. Although in theory, the $k$ value can also change the evolution of radio afterglows, it should be noted that radio observations are largely constrained by instrumental sensitivities, refractive interstellar scintillation \citep{Chandra2012}, or the intrinsic properties of GRBs \citep{Hancock2013}. Thus, the precise shapes of the complete light curves are usually hard to obtain. Hence, the main advantage of our radio-peak method is that the peak times/fluxes can be determined more easily and reliably due to the high fluxes of such peaks, as long as the afterglow light curves are sampled frequently enough. And because radio afterglows occur at a larger radius than early forward-reverse shocks, combining density profiles for earlier afterglows using the method proposed by \cite{Yi2013}, and those for later radio peaks as shown in this work, a more complete circumburst medium profile at various $R$ can be obtained.

This paper is organized as follows. In Section \ref{sec:theory}, we search for the peak time $t_{\rm peak, \nu}$ in the analytical light curve of the GRB radio afterglow with an arbitrary $k$, and calculate the relations between $t_{\rm peak, \nu}$ and $F_{\rm peak, \nu}$ with observing frequencies, with the role played by the $k$ value clarified. Meanwhile, the effects of transrelativistic shocks, as well as jets, are discussed. In Section \ref{sec:obs}, we apply our theoretical predictions to 31 long GRBs, as well as 1 short burst, with multiband radio afterglow observations, and draw a conclusion that most of them can be explained with transrelativistic shock waves under ISM environments, which is different from \cite{Yi2013}. This implies that the circumburst density profile may change with radius. In Section \ref{sec:summary} our results are discussed and summarized. In our calculations, the $\Lambda$CDM cosmology is adopted, with cosmological parameters $H_0 = 71 {\rm km}$ $\rm s^{-1}$  $\rm Mpc^{-1}$, $\Omega_m = 0.3$, and $\Omega_{\Lambda} = 0.7$.

\section{GRB Radio Light Curves and Multifrequency Peak Evolution} \label{sec:theory}

\begin{figure*}
\plotone{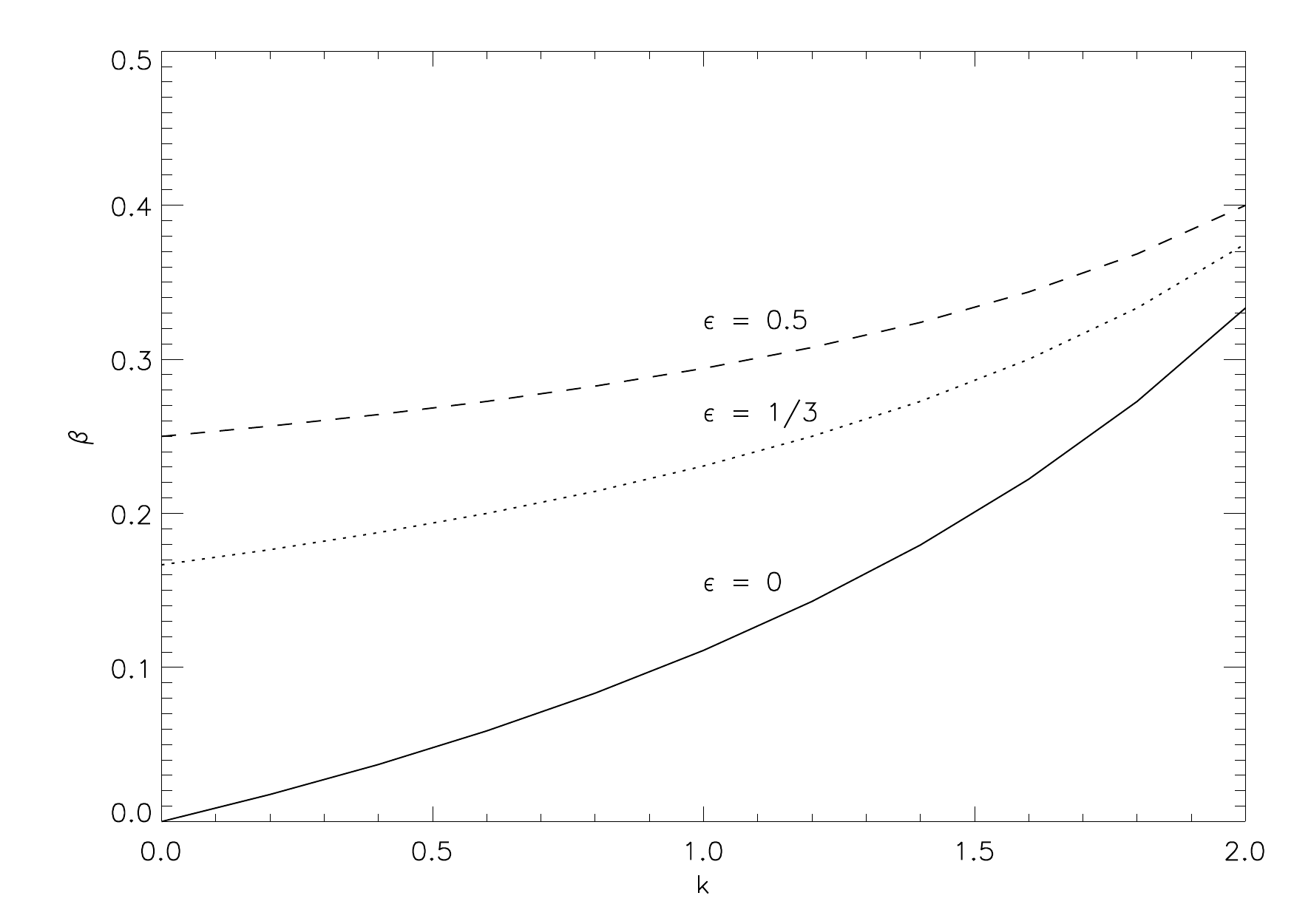}
\caption{\small{The relations between $k$ and $b$. In this figure, solid, dotted, and dashed lines represent the $\epsilon = 0$, $\epsilon = 1/3$, and $\epsilon = 1/2$ cases, respectively.}}\label{fig:k-beta}
\end{figure*}

In this section, our goal is to derive the relations between GRB radio afterglow peak times $t_{{\rm peak},\nu}$ versus $k$, as well as peak fluxes $F_{{\rm peak}, \nu}$ versus $k$, i.e., the $k$-dependent expressions of the power-law indices $a$ and $b$ for $t_{{\rm peak},\nu}\propto \nu^{-a}$, and $F_{{\rm peak}, \nu}\propto \nu^{b}$, respectively. We investigate the analytical behaviors of multiband afterglow with arbitrary circumburst density-distribution index $k$ in order to get the expressions of $a$ and $b$. Ignoring the early afterglow evolution due to forward-reverse-shock interactions, we consider the shock with radius $R > R_{{\rm dec}}$ only, in which $R_{{\rm dec}}$ is the deceleration radius of relativistic shock, with a corresponding deceleration time $t_{{\rm dec}} = R_{{\rm dec}} \left(1+z\right)/2\Gamma_0^2 c$, a cosmological redshift of $z$, and an initial bulk Lorentz factor of $\Gamma_0$. By this time, the GRB afterglow has already entered the self-similar phase. Here we do calculations for semiradiative shock waves for generality, although by the time of the emergence of radio peaks, i.e., several days after the initial GRB trigger, the shock waves are usually adiabatic. Assuming the circumburst density can be described as $n = A_\ast R^{-k} = A \left(R/R_*\right)^{-k}$ ($k=0$ for uniformly distributed interstellar medium, $k = 2$ for stellar wind, with a typical value of the characteristic length $R_*\sim 10^{17}$ cm; see \citealt{Wu2005}), the hydrodynamical evolution should be $\Gamma^2 \propto R^{-m}$, and $R \propto t^{1/\left(m+1\right)}$ (\citealt{Blandford1976}), with self-similar index $m$ as
\begin{equation}
m = \frac{3-k}{1-\epsilon},\label{eq:1}
\end{equation}
where $\epsilon$ denotes the radiative efficiency of the shock, $R$ the radius of the shock wave, $\Gamma$ the bulk Lorentz factor of the shock, and $t$ the observed time after the initial burst \citep{Wu2005}. The radiative efficiency $\epsilon$ can be considered as a constant at the early fast-cooling regime, and it evolves slowly after entering the slow-cooling regime at later times. However, as long as we have power-law-distributed electrons, $N_e \left(\gamma_e\right) \propto \gamma_e^{-p}$, with $\gamma_e$ as the Lorentz factor of the  electrons and $N_e$ the electron number, and the distribution index $p$ does not deviate too much from $2$, the effect of the evolving $\epsilon$ is not significant \citep{Wu2005}. Because the typical value of $p$ for relativistic-shock-accelerated electrons is $2.2$ (\citealt{Achterberg2001}, also see \citealt{Curran2009} for observational constraints of an average value of $k 
\sim 2.24$ with six \textit{Swift} GRBs), which is close to $2$, here we ignore the evolution of $\epsilon$.

In this work, we focus on radio peaks occurring at $\geqslant T_{90} + 0.5$ days in the rest frame only, thus the effects of the equal-arrival-time surface (EATS) can be ignored. That is because, for a typical GRB, the radio-peak time $t_{{\rm peak}}\sim T_{90} +  0.5$ days $< \Gamma^{-2} R/2c $, which is the maximum time delay due to the EATS effects. Also, because the major concern of this paper is the behavior of GRB radio afterglows, we ignore the inverse Compton (IC) scattering, for IC is not a dominant factor at lower frequencies \citep{Wu2005}.

Following the procedures of \cite{Wu2005} and \cite{vanderHorst2007a}, we first consider the evolution of spherical shock waves with various $k$ values. For synchrotron radiation caused by power-law-distributed electrons, a key time in the light curve is $t_{cm}$, which marks the transition point between the early fast-cooling phase to the late slow-cooling phase, with an electron minimum Lorentz factor $\gamma_{m}$ equal to the electron cooling Lorentz factor $\gamma_c$, and a minimum frequency $\nu_m$ equal to the cooling frequency $\nu_c$ (\citealt{Sari1998}; \citealt{Panaitescu2000}). We define $\nu_{cm} = \nu_m\left(t_{cm}\right) =\nu_c\left(t_{cm}\right)$. The transition occurs at 
\begin{eqnarray}
t_{cm} & = & A \sigma_T^{\left(3-k\right)/2} \frac{\left(1+z\right)\sigma_T^{1/2}}{2c} \left[\frac{\left(3-k\right) E_{cm}}{4\pi m_p c^2}\right]^{\left(2-k\right)/2}\nonumber\\
& & \left(\frac{2m_p}{3m_e} \varepsilon_e^{3/4} \varepsilon_B^{1/4}\zeta_{1/6}^{1/2} \right)^{4-k}. \label{eq:5} \\%2
\nu_{cm} & = & \frac{1}{16} \left( \frac{3}{2 \pi} \right)^{5/4}\frac{m_e c^3}{e^2\left(1+z\right)} \left(\frac{2 m_p}{3 m_e}\right)^{\left(3k-7\right)/2} \nonumber\\
& &\times\varepsilon_e^{\left(9k-20\right)/8} \varepsilon_B^{\left(3k-8\right)/8} \zeta_{1/6}^{\left(3k-4\right)/4} \nonumber \\
&  & \times \left(A \sigma_T^{\left(3-k\right)/2}\right)^{-3/2} \left[\frac{\left(3-k\right) E_{cm}}{4 \pi m_p c^2}\right]^{\left(3k-4\right)/4}, \label{eq:4} %3
\end{eqnarray}
where $m_e$ is the electron mass, $m_p$ the mass of a proton, $e$ the charge of an electron, $c$ the speed of light in vacuum, $\sigma_T$ the electron's Thomson scattering cross section, $E_{cm}$ the isotropic energy of the shock wave at $t_{cm}$, and $\zeta_{1/6} = 6\left(p-2\right)/\left(p-1\right)$. 

Following the assumptions of \cite{Wu2005}, we take typical values for parameters as follows: the isotropic energy $E_{cm} = 10^{53}$ erg, $\varepsilon_e = 1/3$, $\varepsilon_B = 10^{-2.5}$, and the electron distribution index $p=2.2$. It should be noted that although some later works, including \cite{Nava2014}, \cite{Beniamini2017}, as well as \cite{Santana2014}, hinted that $\varepsilon_e \sim 0.15$ and $\varepsilon_B \sim 10^{-8}-10^{-3}$ should be the case for some GRBs, the exact values of $\varepsilon_e$ and $\varepsilon_B$ do not change the overall shape of the afterglow light curves. Thus, our calculations still apply to their samples.

The characteristic flux density $F_{\nu,max}$ at any given time can be obtained with the characteristic flux density of synchrotron radiation at a certain frequency on $t_{cm}$ as $F_{\nu,max}\left(t_{cm}\right)$. (It should be noted that $F_{\nu,max}$ is different from $F_{{\rm peak}, \nu}$ because the latter denotes the peak flux of the radio light curve at a given frequency $\nu$, rather than the radiation peak across the radio spectrum). $F_{\nu,max}$ can be calculated as follows, regardless of fast or slow cooling 
\begin{equation}
F_{\nu,max} = F_{\nu,max}\left(t_{cm}\right) \left(\frac{t}{t_{cm}}\right)^{ \left(6-3k-2m\right)/ \left[2\left(m+1\right) \right]}.\label{eq:11} %4
\end{equation}

Also, at the radio band with low-enough frequencies, the synchrotron self-absorption (SSA) effects cannot be ignored. In this case, we have another characteristic time $t_a$, which denotes the time when $\nu = \nu_a$, where $\nu_a$ is the self-absorption frequency. Because in this work the later-time behavior of the radio light curve is our main focus, we only discuss the later-time slow-cooling regime, with $\nu_a = \min \left\{\nu_{as,<},\nu_{as,>} \right\}$, and
\begin{eqnarray}
\nu_{as,<} & = & \nu_m \left[ \frac{c_0 \left(p-1 \right)}{3-k}  \frac{enR}{B \gamma_{{\rm min}}^5} \right]^{3/5}, \nu_a<\nu_m, \label{eq:18}\\
\nu_{as,>} & = & \nu_m \left[ \frac{c_0 \left(p-1 \right)}{3-k}  \frac{enR}{B \gamma_{{\rm min}}^5} \right]^{2/\left(p+4\right)}, \nu_m<\nu_a<\nu_c.\label{eq:19} 
\end{eqnarray} 
where $c_0 \approx 10.4\left[\left(p+2\right) /\left(p+2/3 \right)\right] \approx 15.24$, and the strength of the magnetic field $B$ as well as shock radius $R$ can be calculated with their corresponding values at $t_{cm}$ (e.g., see Equations (13) and (14) in \citep{Wu2005}).

In the slow-cooling phase, the transition from $\nu_{as,<}$ to $\nu_{as,>}$ occurs at $t_{am}$, when $\nu_a = \nu_m$. For various combinations of physical parameters, typically we have $t_{am} \sim 10^3 t_{cm}$. For many cases, the GRB afterglow may have already finished the ultra-relativistic self-similar evolution and even began the transition to the Newtonian regime by the time of $t_{am}$, making \cite{Blandford1976} no longer applicable. And it is worth noting that in our numerical discussions of late-time afterglow, including the transrelativistic phase, in Section \ref{subsec:trans-relativistic}, the evolution of $\nu_{as,>}$ is included in calculations, with corresponding effects of this factor in this regime considered. Thus, we do not take $\nu = \nu_{as,>}$ into consideration in analytical calculations.

For typical parameter values, we have $\nu_{cm} \sim 10^{3} - 10^{4}$ GHz, which is much higher than in the radio band; hence in this work, we only take the $\nu < \nu_{cm}$ band into consideration. Firstly, we calculate the light curves for radio afterglows. If the flux $F_{\nu} \propto t^\alpha$ has a positive temporal index $\alpha$ before a turning point and then shows a negative $\alpha$, we define the turning point as a radio peak. We only consider the $0\leqslant k \leqslant 2$ case. The reason is that the self-similar solution proposed by \cite{Blandford1976} only stands as long as $k<4$. And in reality, it is difficult for the circumburst environment to show density distributions steeper than $k>2$, since the original medium distribution and the stellar wind from the progenitor are the only main factors affecting $k$. 

\begin{figure}
\gridline{\fig{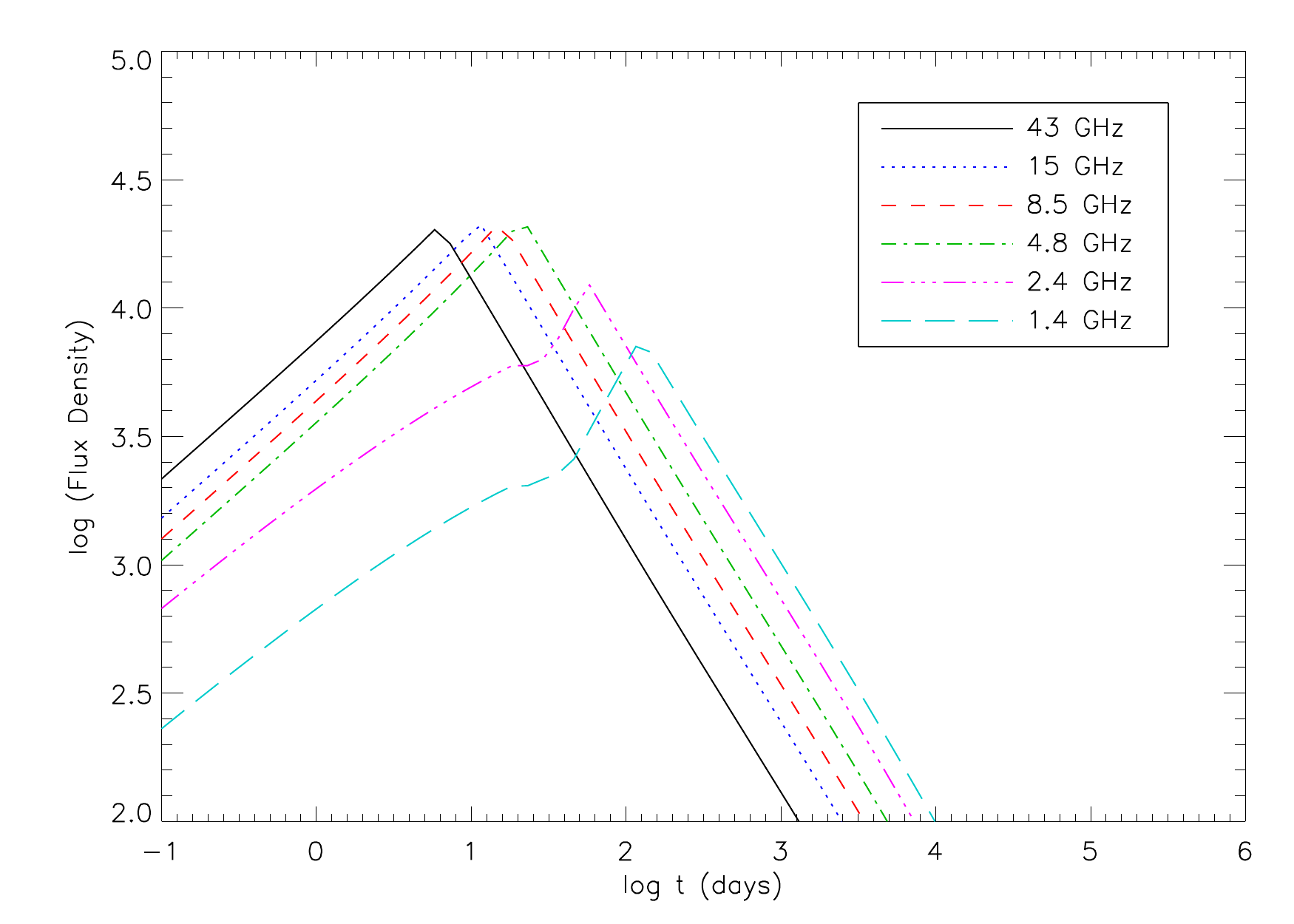}{0.4\textwidth}{}}
\gridline{\fig{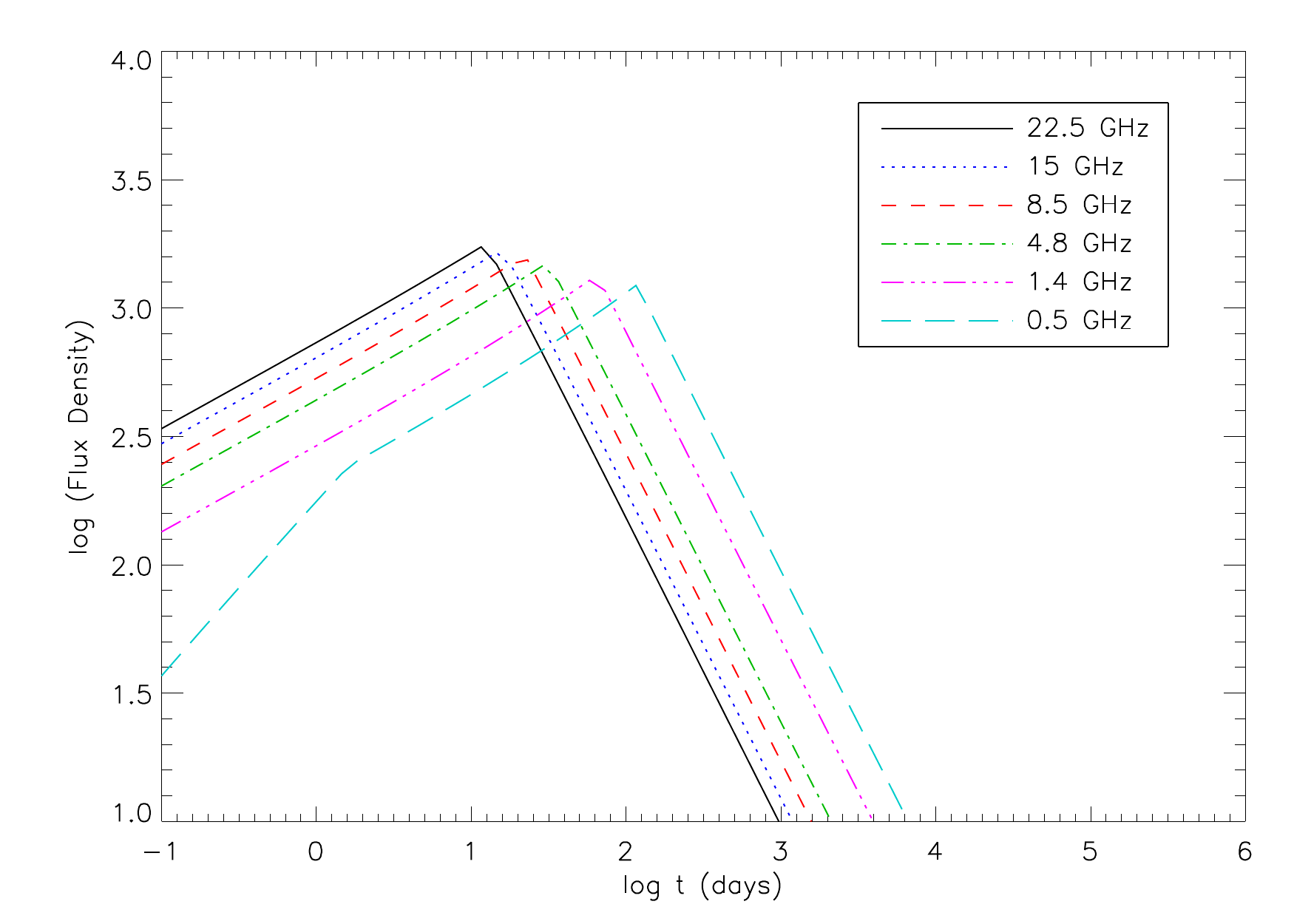}{0.4\textwidth}{}}
\gridline{\fig{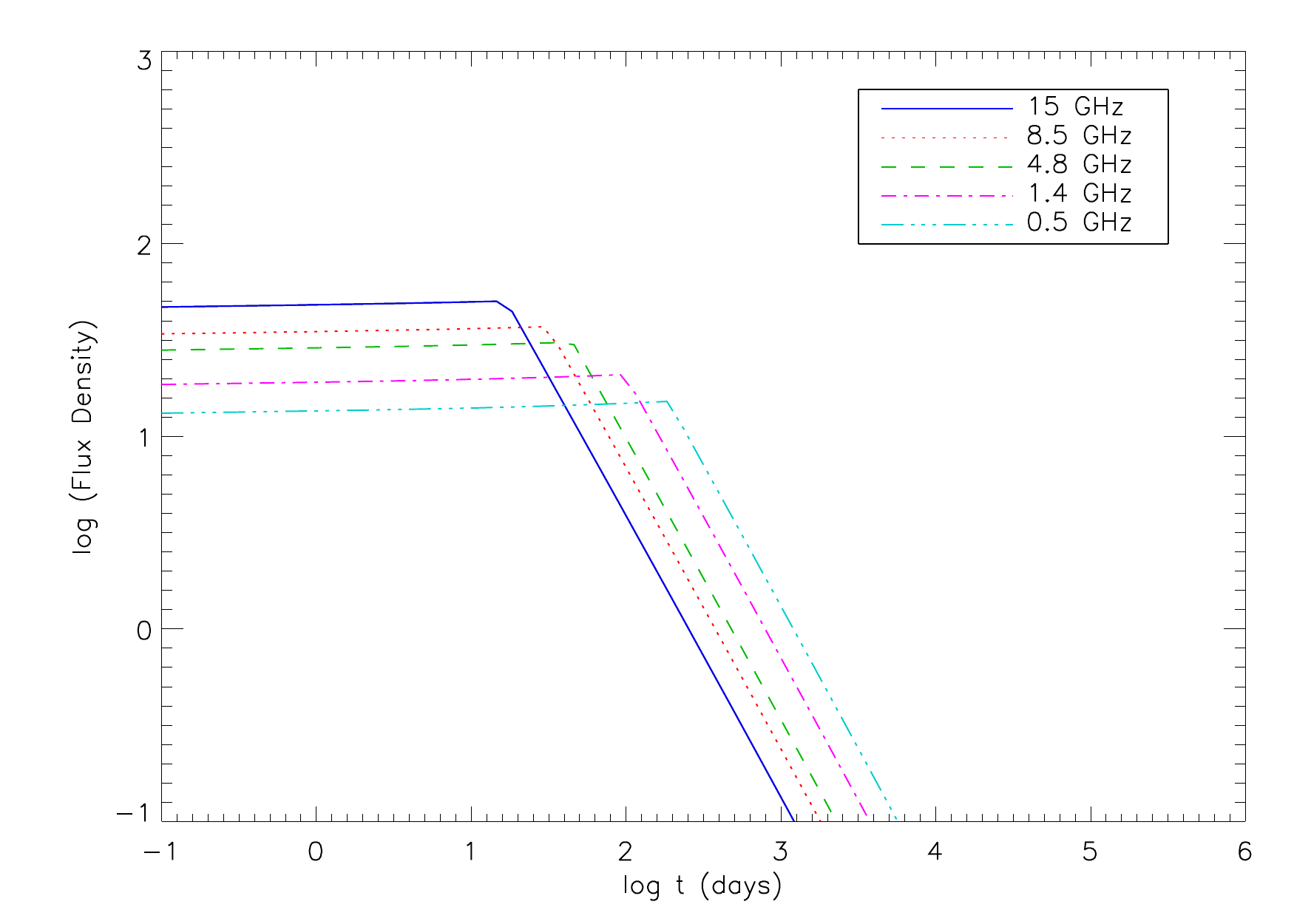}{0.4\textwidth}{}}
\caption{\small{The multiband GRB radio afterglow light curves for $k=0$ (upper panel), $k=1.1$ (middle panel), and $k=2$ (lower panel) calculated with the numerical model proposed by \cite{Huang1999}. Here we take the GRB isotropic energy $E_{\rm iso} = 10^{53}$ erg, characteristic density $A = 1$ ${\rm cm}^{-3}$, and adiabatic shock wave $\epsilon =0$. The y-axis are shown in arbitrary units. It can be seen that for the $k=0$ case, the low-frequency afterglows peak during the transrelativistic-to-Newtonian phase, and the relations between peak times/fluxes and frequencies are different from the ultrarelativistic regime, while for the $k=1.1$ and $2$ cases, the radio afterglows never enter the trans-relativistic phase during our calculations. }}\label{fig:lc_k}
\end{figure}

\begin{figure}
\gridline{\fig{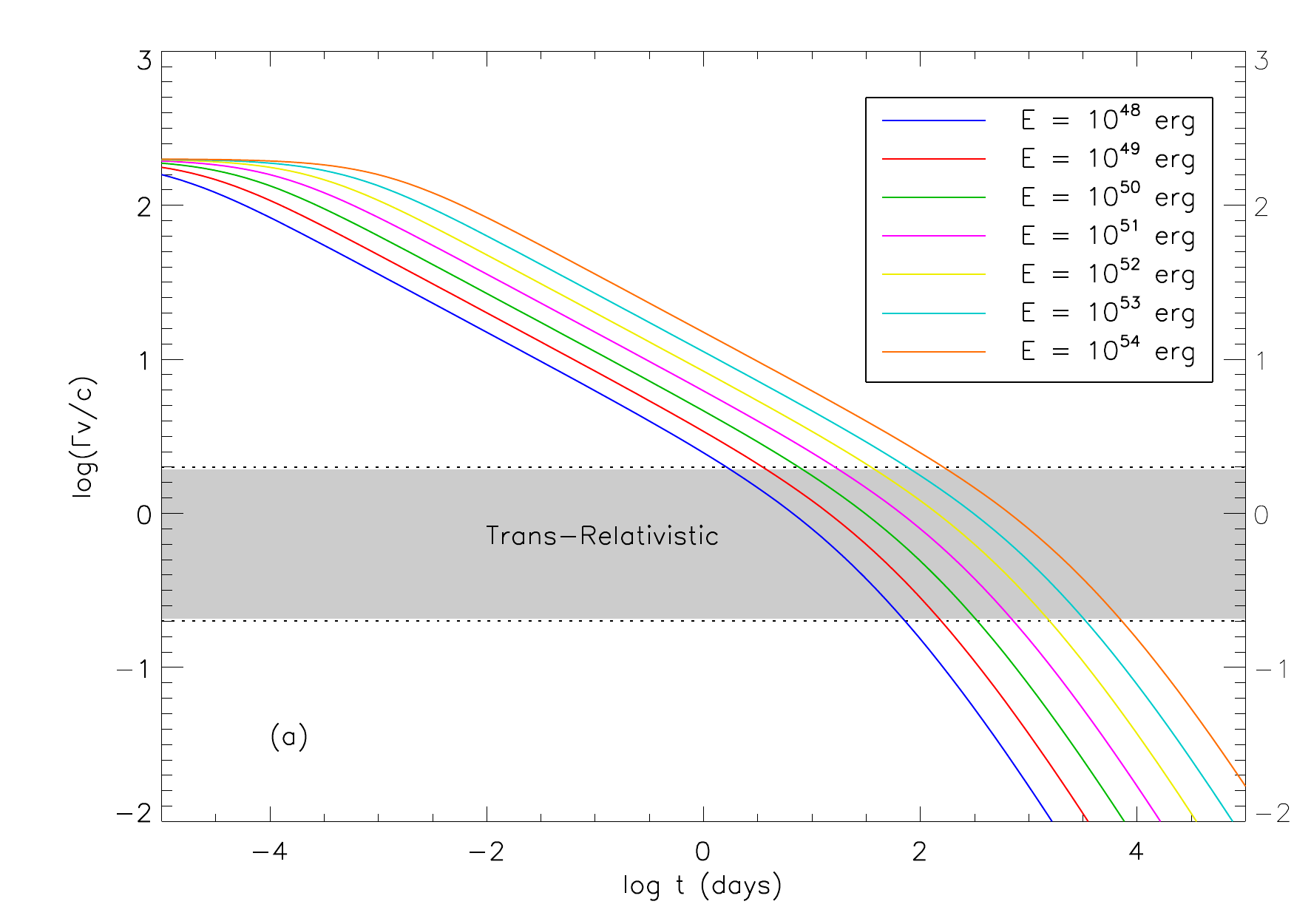}{0.4\textwidth}{}}
\gridline{\fig{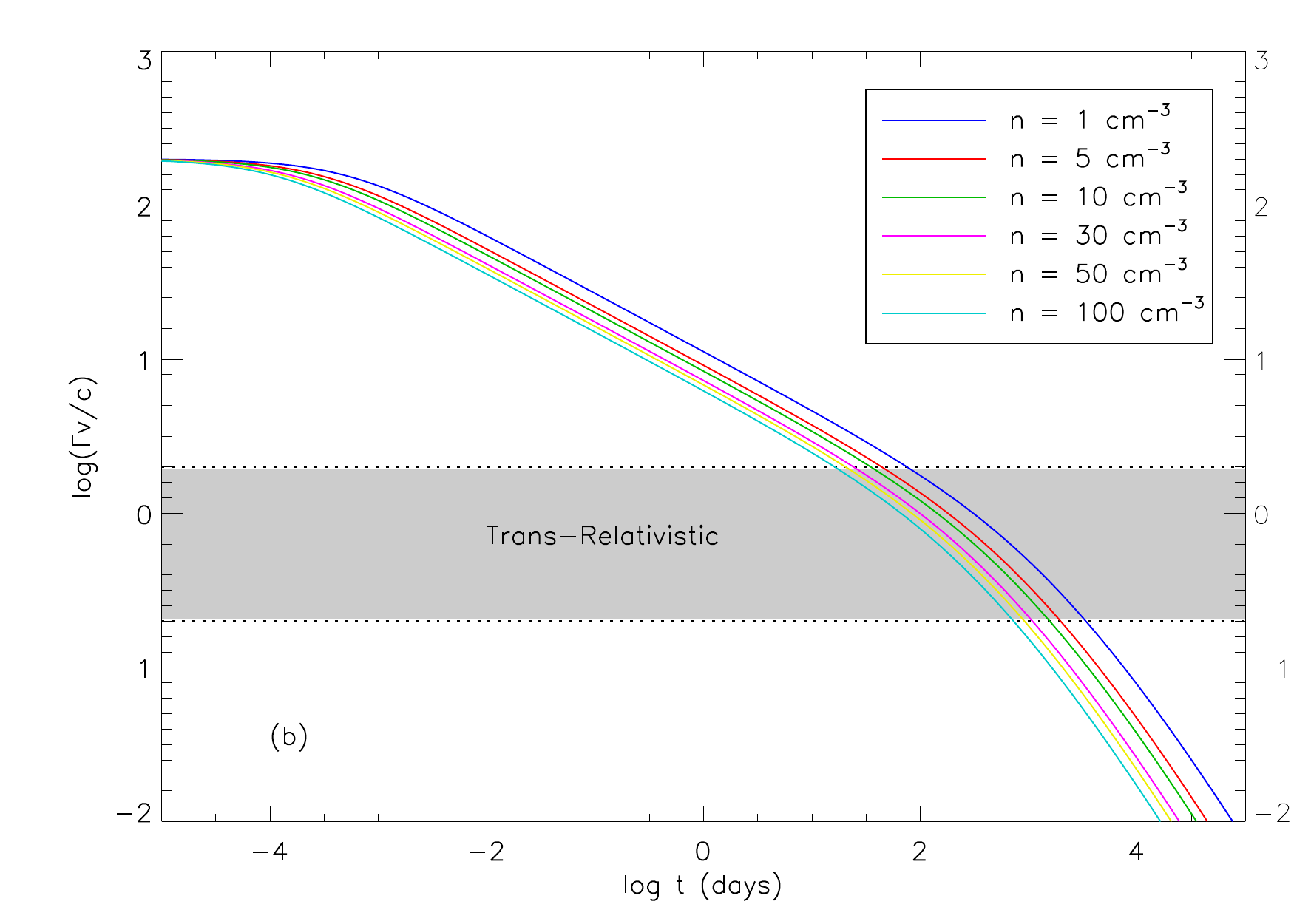}{0.4\textwidth}{}}
\gridline{\fig{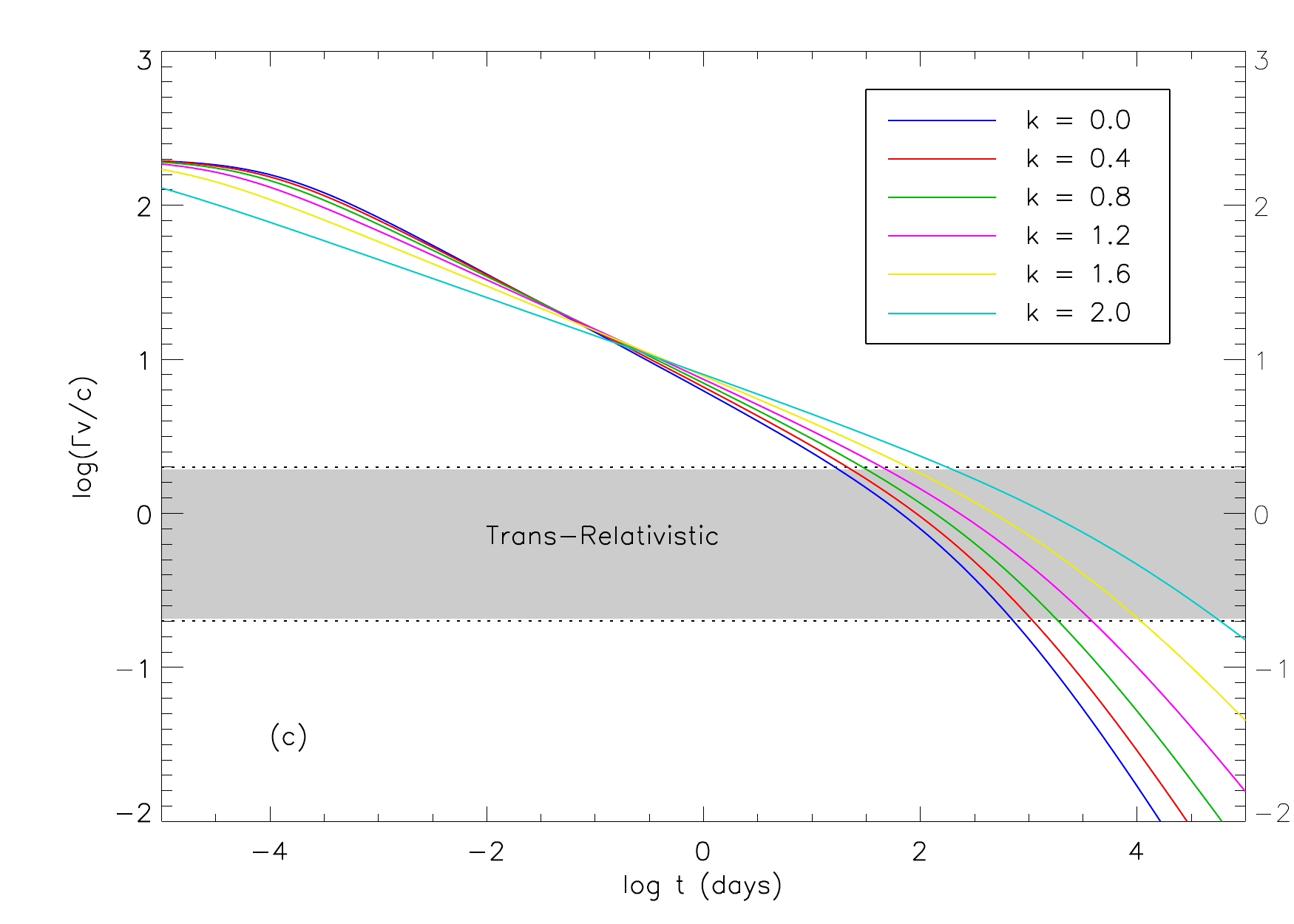}{0.4\textwidth}{}}
\caption{\small{The upper panel shows the temporal evolution of $\Gamma v/c$ with various isotropic energies. The transrelativistic evolution featuring $0.2<\Gamma v/c<2$ is marked with the gray-shaded region bounded by the dotted lines.} Here we take $k=0$ and $n = 1$ $\text{ cm}^{-3}$. It can be seen that the lower the $E_{\rm iso}$ is, the earlier the transrelativistic phase begins. The middle panel shows the temporal evolution of $\Gamma v/c$ for various circumburst densities, with $k=0$, $E_{\rm iso} = 10^{53}$ erg. It can be seen that the larger the density $n$, the earlier the transition begins. The lower panel shows the temporal evolution of $\Gamma v/c$ with various $k$, $E_{\rm iso} = 10^{53}$ erg, and characteristic density $A = 100$ ${\rm cm}^{-3}$. It can be seen that the smaller the $k$ is, the earlier the transition time.}\label{fig:gammabeta_parameter}
\end{figure}

Thus, similar to \cite{Wu2005} and \cite{vanderHorst2007a} by comparing the values of characteristic times $t_a$, $t_{cm}$, $t_c$, as well as $t_m$, with the latter two items corresponding to the time points when $\nu_c$ and $\nu_m$ equal to the observing frequency $\nu$, respectively, we can get the overall profile of GRB radio afterglow light curves. Depending on the values of $k$, three types of light curves can be obtained. For smaller $k$ with $0<k<\frac{32-12p-8 \epsilon}{12-3p+\left(p-4\right)\epsilon}$, when $\nu_a \left(t_{cm}\right) < \nu < \nu_{cm}$ , with $\nu_a \left(t_{cm}\right)$ means the SSA frequency at $t_{cm}$, we have $t_a < t_{cm}<t_m<t_c$, and the peak occurs at $t_m$, with flux before and after peak as
\begin{equation}
F_{\nu} \propto \left\{
\begin{array}{ll}
\nu^{1/3} t^{\left(9-4k-m\right) / \left[3\left(m+1\right)\right]}, & t_{cm} < t < t_m,\\
\nu^{-\left(p-1\right)/2 } t^{\left[12-5k-p\left(4m+k\right)\right] / \left[4\left(m+1\right)\right]}, & t_m < t < t_c,\\
\end{array}
\right.\label{eq:20}
\end{equation}
For $\nu <\nu_a\left(t_{cm}\right)$, the order changes to $t_{cm}<t_a <t_m<t_c$, and the flux peaks again at $t_m$. The light curve around peak time can be expressed as
\begin{equation}
F_{\nu} \propto \left\{
\begin{array}{ll}
\nu^{1/3} t^{\left(9-4k-m\right) / \left[3\left(m+1\right)\right]}, & t_a < t < t_{m},\\
\nu^{-\left(p-1\right)/2 } t^{\left[12-5k-p\left(4m+k\right)\right] / \left[4\left(m+1\right)\right]}, & t_m < t < t_c,\\
\end{array}
\right.\label{eq:21}
\end{equation}

It should be noted that, in this case, the upper limit of $k$, i.e., $\frac{32-12p-8 \epsilon}{12-3p+\left(p-4\right)\epsilon}$, depends on the electron distribution index $p$, as well as the shock radiation efficiency $\epsilon$, and is quite sensitive to both parameters. For $p \sim 2.2$, $\epsilon \sim 0$, this upper limit is $\sim 1.037$. 

When the value of $k$ increases to $\frac{32-12p-8 \epsilon}{12-3p+\left(p-4\right)\epsilon} < k < 4/3$, at higher frequencies, $\nu_a\left(t_{cm}\right) < \nu < \nu_{cm}$, we have $t_a < t_{cm}<t_m$, with $t_c$ disappearing in the radio band. In this case, the high-frequency light curve still peaks at $t_m$, and the radio flux around this time can be calculated as 
\begin{equation}
F_{\nu} \propto \left\{
\begin{array}{ll}
\nu^{1/3} t^{\left(9-4k-m\right) / \left[3\left(m+1\right)\right]}, & t_{cm} < t < t_m,\\
\nu^{-\left(p-1\right)/2 } t^{\left[12-5k-p\left(4m+k\right)\right] / \left[4\left(m+1\right)\right]}, & t_m < t. \\
\end{array}
\right.\label{eq:22}
\end{equation}
For $\nu < \nu_a\left(t_{cm}\right)$, $t_{cm} < t_a < t_m$, $t_m$ remains as $t_{\rm peak,\nu}$, and the peak-time light curve is
\begin{equation}
F_{\nu} \propto \left\{
\begin{array}{ll}
\nu^{1/3} t^{\left(9-4k-m\right) / \left[3\left(m+1\right)\right]}, & t_a < t < t_m,\\
\nu^{-\left(p-1\right)/2 } t^{\left[12-5k-p\left(4m+k\right)\right] / \left[4\left(m+1\right)\right]}, & t_m < t. \\
\end{array}
\right.\label{eq:23}
\end{equation}
And both the high- and low- frequency radio fluxes peak at $t_{m}$ again.

For a larger $k$ ($4/3<k<2$), the SSA frequency at $t_{ac}$ is lower than $\nu_{cm}$, that is, $\nu_{ac}< \nu_{cm}$, and the radio band should be divided into three sections. For $\nu_{ac} < \nu < \nu_{cm}$, $t_a < t_c< t_{cm}<t_m$, while we have $t_c<t_a<t_{cm}<t_m$ for $\nu<\nu_a\left(t_{cm}\right)$. In these two cases, the slope between $t_{cm}$ and $t_m$ is slightly negative as long as $\epsilon > 0$, only with an early peak at $t_a$. However, a significant break does exist at $t_m$, as can be seen from the equation below. Considering possible observational errors, such a turnover could still be taken as peak time:
\begin{equation}
F_{\nu} \propto \left\{
\begin{array}{ll}
\nu^{1/3} t^{\left(9-4k-m\right) / \left[3\left(m+1\right)\right]}, & t_{cm} < t < t_m,\\
\nu^{-\left(p-1\right)/2 } t^{\left[12-5k-p\left(4m+k\right)\right] / \left[4\left(m+1\right)\right]}, & t_m < t.
\end{array}
\right.\label{eq:24}
\end{equation}

While for $\nu<\nu_a\left(t_{cm}\right)$, we expect $t_c<t_{cm}<t_a<t_m$, again with $t_m$ serving as peak time. Here the light curve can be expressed as
\begin{equation}
F_{\nu} \propto \left\{
\begin{array}{ll}
\nu^{1/3} t^{\left(9-4k-m\right) / \left[3\left(m+1\right)\right]}, & t_a < t < t_m,\\
\nu^{-\left(p-1\right)/2 } t^{\left[12-5k-p\left(4m+k\right)\right] / \left[4\left(m+1\right)\right]}, & t_m < t.
\end{array}
\right.\label{eq:26}
\end{equation}

It can be seen that the GRB radio afterglow peaks at $t_{{\rm peak}} = t_m$, regardless of $k$ value or observing frequency. Because we have $t_m \propto \nu^{-2\left(m+1 \right)/\left(4m+k\right)}$, the peak times and peak fluxes can be calculated with Eqs. \ref{eq:20}-\ref{eq:26} as 
\begin{eqnarray}
t_{{\rm peak},\nu} &\propto& \nu^{-2\left(m+1 \right)/\left(4m+k\right)},\label{eq:27}\\
F_{{\rm peak},\nu} &\propto& \nu^{1/3} t_m^{\left(9-4k-m\right) / \left[3\left(m+1\right)\right]} \nonumber\\
& \propto & \nu^{\left(-6+3k+2m\right)/\left(4m+k\right)}.\label{eq:28}
\end{eqnarray}
That is, the $a$ and $b$ indices we seek should be
\begin{eqnarray}
a & = & -2\left(m+1 \right)/\left(4m+k\right),\label{eq:29}\\
b & = & \left(-6+3k+2m\right)/\left(4m+k\right).\label{eq:30}
\end{eqnarray}

For $\epsilon = 0$, it can be seen that $t_{{\rm peak},\nu} \propto \nu^{-2/3}$. Even for $\epsilon>0$ cases, as long as the shock radiation efficiency is low enough, the $t_{{\rm peak},\nu}\propto \nu^{-a}$ index $a$ should not deviate too much from $2/3$. For the $F_{{\rm peak},\nu}\propto \nu^{b}$ index $b$ with $\epsilon = 0$, $b$ increases with $k$, that is, $b = 0$ when $k=0$, and $b = 1/3$ when $k = 2$. If $\epsilon>0$, the $b$ value is larger than that in the $\epsilon = 0$ case, although such a difference is not significant for larger $k$. In Figure \ref{fig:k-beta}, the relations between $k$ and $b$ with various $\epsilon$ are shown, according to which the $k$ value can be drawn from $b$. And it should be noted that for later peaks occurring at $T_{90} + $several days, $\epsilon=0$ usually applies, and thus a single $k$ value can be constrained with $b$, as shown in Eq. \ref{eq:30}.

Eqs \ref{eq:29} and \ref{eq:30} cannot be applied for high-frequency ($\nu_{ac}<\nu<\nu_{cm}$) radio afterglows with $k>4/3$, the flux of which peaks at $t_{a} \ll 1$ day. Because for real GRBs radio follow-up observations usually do not begin at such early times, and the behaviors of afterglows are often contaminated by the early reverse shocks, such a high-frequency early peak with a large $k$ have little value for our purpose, and its behavior is beyond the scope of this work. 

\subsection{Late-time Afterglow Behaviors and Transition from Relativistic to Newtonian Phase} \label{subsec:trans-relativistic}

\begin{figure*}
\plotone{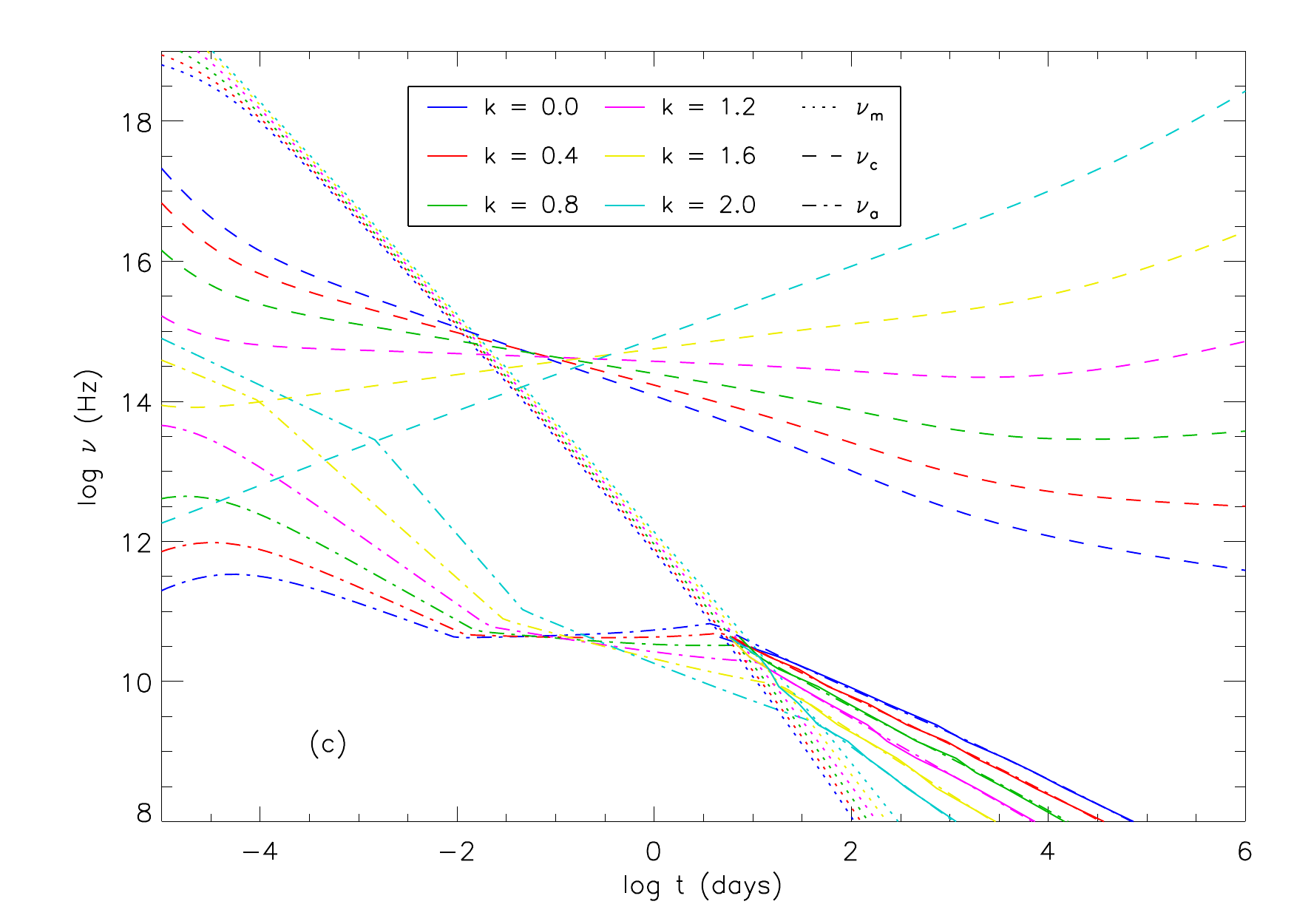}
\caption{\small{The temporal evolution of characteristic frequencies for various $k$ (lower panel). Here the dashed, dotted and dash-dotted lines represent the evolution of cooling frequency $\nu_c$, minimum frequency $\nu_m$, as well as SSA frequency $\nu_a$. The solid lines show the evolution of multi-band radio peak frequencies, as well as peak times. It can be seen that considering trans-relativistic effects and evolution of $\nu_a$}, the peak time $t_{{\rm peak},\nu}$ for these cases should be $\approx t_a$. However, other possibilities for peak also exist for various physical parameters.} \label{fig:nu_parameter} %$E_{\rm iso}$ (upper panel), $n$ (middle panel), and  }}
\end{figure*}

The GRB afterglow begins the transition from relativistic to Newtonian phase (the so-called ``transrelativistic'' stage) when the shock wave expands to a larger radius. The evolutionary phase with Lorentz factor $0.2 \leqslant\Gamma v/c \leqslant 2$ can be considered as ``transrelativistic'', because by this time, the shock velocity has been significantly decreased and the amount of swept-up circumburst medium becomes large enough, while the afterglow evolution in this stage can be described by neither the relativistic self-similar solution proposed by \cite{Blandford1976}, nor the Newtonian Sedov-Taylor solution (\citealt{Taylor1950}; \citealt{Sedov1969}). Here, numerical calculations based upon dynamical equations proposed by \cite{Huang1999}, which provide a unified framework from relativistic to deep Newtonian stages of shock-wave evolution, are required, with examples shown in Figure \ref{fig:lc_k}. And because in late phases GRB afterglows should be adiabatic, it is safe to assume $\epsilon = 0$ in our transrelativistic calculations. Figure \ref{fig:gammabeta_parameter} shows the temporal evolution of $\Gamma v/c$ for adiabatic shocks exhibiting various physical parameters, with the trans-relativistic phase occurring during $T_{90} + $dozens to hundreds of days. And considering the transrelativistic phase only begins when the swept-up mass is comparable with the mass of the shock wave itself, it is clearly seen that a larger $E_{\rm iso}$, a smaller $A$, and a larger $k$ can lead to a later transition. 

According to numerical calculations, for spherical ejecta, the radio flux peak no longer appears at $t_m$ during such late times, due to the change of blast-wave dynamics, as well as the transition from $\nu_a = \nu_{as,<}$ to $\nu_a = \nu_{as,>}$, which may occur before or after the beginning of transrelativistic phase, depending on the circumburst density profiles and $E_{\rm iso}$. For example, as can be seen from Figure \ref{fig:nu_parameter}, $t_a$ acts as the peak time for our typical parameters, though other possibilities also exist. Hence, the values of $a$ and $b$ in this phase differ significantly with predictions made by the self-similar shock-wave theory.

\begin{figure*}
\plotone{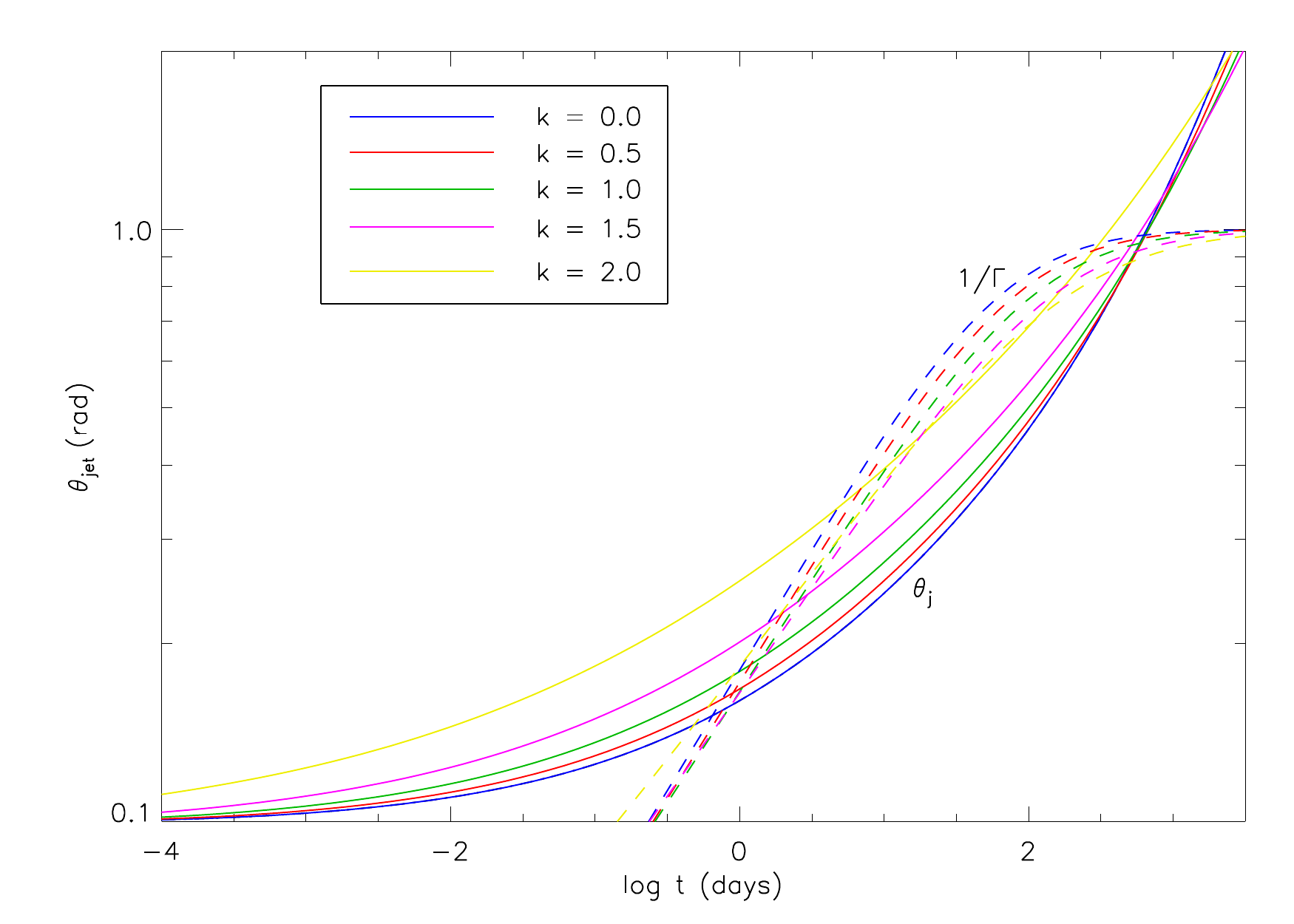}
\caption{\small{The temporal evolution of jet opening angles $\theta_j$ for sideway-expanding jets under various circumburst medium distribution index $k$. Here different colors represent various $k$-values, the solid line the evolution of $\theta_j$, and the dashed line the evolution of $1/\Gamma$. It can be seen that during late times, the condition of $\Gamma > \theta_j^{-1}$ can be satisfied once more, making the afterglow flux to rise again, providing a constant expanding speed.}} \label{fig:theta_k}
\end{figure*}

In late times covering the transrelativistic regime, the analytical expressions to describe the evolution of the bulk Lorentz factor $\Gamma$ and other quantities no longer exist. However, the relations between $t_{{\rm peak},\nu}$, $F_{{\rm peak},\nu}$ and $\nu$ can be analyzed numerically. We fitted $a$ and $b$ with various parameter values and found that both $a$ and $b$ show no strong dependence on $E_{\rm iso}$ or $A$. While these indices do change more significantly with $k$. For typical parameter values, the larger the $k$ value, the flatter the late time $a$, and the smaller the $b$. Generally speaking, for shocks with $E_{\rm iso} = 10^{53}$ erg, $A = 100 {\rm cm}^{-3}$, $\epsilon =0$, $p=2.2$, $\varepsilon_e = 1/3$, and $\varepsilon_B = 10^{-2.5}$, we have $a \sim 1.6$, and $b \sim 0.95$ for the $k=0$ case. While $a$ decreases from $\sim 1.6$ to $\sim 1.0$, $b$ decreases from $\sim 1.0$ to $\sim 0.89$ when $k$ increases from $0$ to $2$. By comparing Fig. \ref{fig:nu_parameter} with Fig. \ref{fig:gammabeta_parameter}(c), it can be seen that the late peaks due to the transition of $\nu_a$ could appear earlier than the beginning of the transrelativistic phase for the circumburst environment with $k \geqslant 1$ and small $A$. However, because our numerical calculations cover the complete from as early as $< T_{90} + 0.1$  days, the general trend mentioned above still applies. 

We also investigate the effects on transrelativistic $a$ and $b$ caused by the electron distribution index $p$, the proportion of electron and magnetic field energies $\varepsilon_e$ and $\varepsilon_B$, as well as shock radiation efficiency $\epsilon$. We find that for electrons with distribution index $2.0<p<2.5$, the values of $a$ and $b$ depend weakly on $p$, and a larger $p$ can lead up to a steeper $b$, as well as a shallower $a$. However, such a trend is not as clear as the $k$ dependence mentioned above, and the change in $a$ is hardly noticeable. For example, $a$ decreases from 1.6 to 1.55 when $p$ increases from $2.1$ to $2.4$. Meanwhile, a larger $p$ can give rise to a $b$ value as large as $\sim 1.1$. Besides, both $a$ and $b$ depend weakly on $\varepsilon_B$, and a smaller $\varepsilon_B$ can lead to flatter $a$ and $b$. For example, $a$ decreases from $\sim 1.6$ to $\sim 1.4$ when $\varepsilon_B$ decreases from $0.01$ to $0.001$, while neither $\varepsilon_e$ nor $\epsilon$ shows a strong influence on $a$ and $b$. In short, the variation of $b$ within a wide parameter range is relatively small, while the value of $a$ is largely determined by $k$. With the value of $a$ in the transrelativistic regime, one can put a stringent constraint on $k$.

For wideband observations covering $\sim 1 - 10^2$ GHz frequency range, the values of the $t_{\rm peak,\nu}-\nu$ and $F_{\rm peak,\nu}-\nu$ indices $a$ and $b$ fitted from peak data should fall between $a \sim 2/3$, $b \sim 0-1/3$ for the ultrarelativistic regime, and $a \sim 1-0.9$ as well as $b \sim 1.6-1$ for the transrelativistic regimes, depending on $k \sim 0-2$. Similar to Eqs. \ref{eq:29} and \ref{eq:30}, such a transrelativistic trend with steeper $a$ and $b$ compared with the ultrarelativistic case can also be used to put constraints on $k$ qualitatively, using $a$ and $b$ derived with multiband light-curve peaks.

\subsection{Corrections for the Jet} \label{subsec:jet}

\begin{figure*}
\gridline{\fig{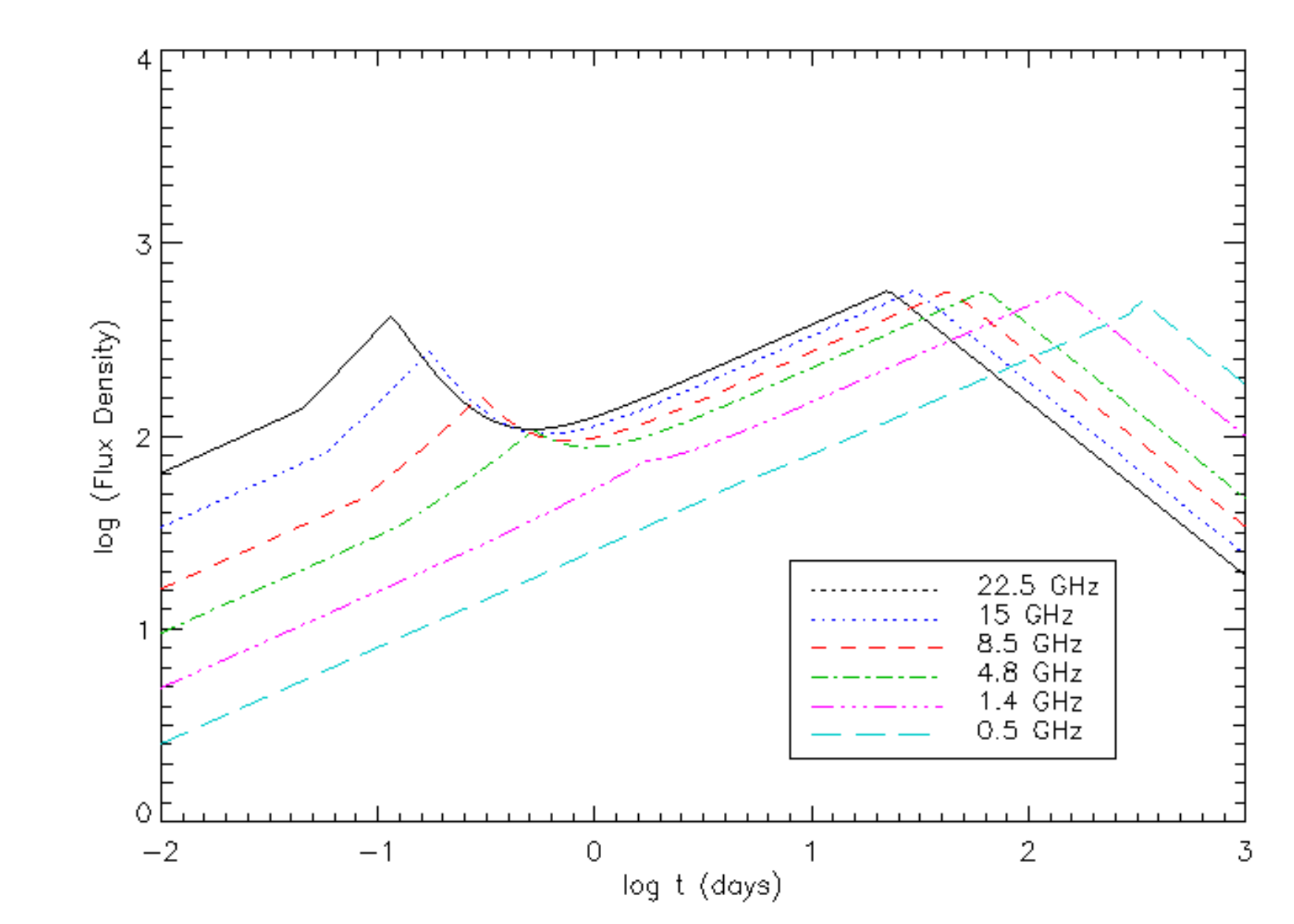}{0.45\textwidth}{}
          \fig{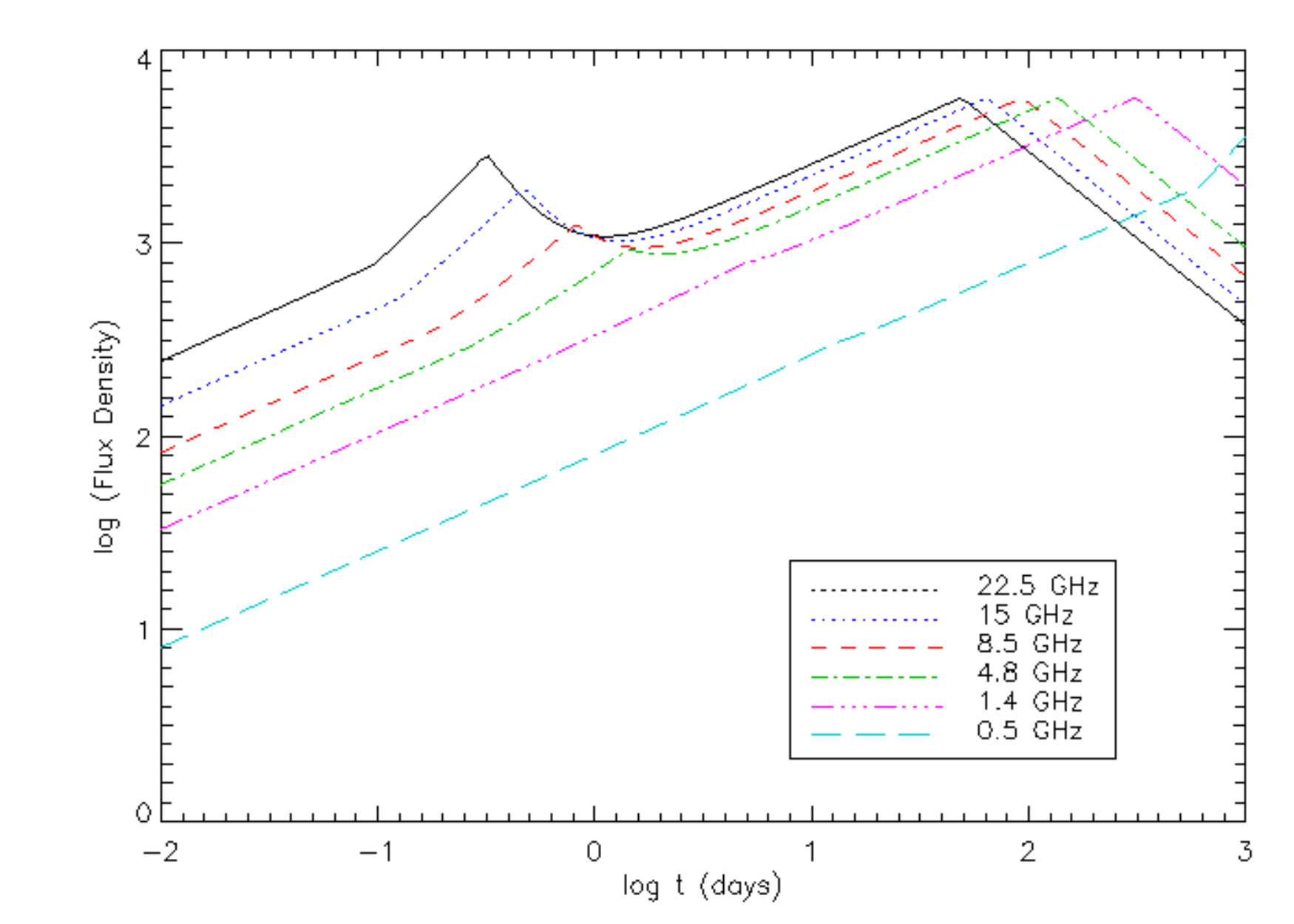}{0.45\textwidth}{}
          }
\caption{\small{Theoretical GRB radio afterglow light curves considering forward-reverse shock interactions with $E_{\rm iso} = 10^{52}$ erg(left) and $10^{53}$ (right) erg, respectively. The other parameters are adopted as the typical values listed in Section \ref{sec:theory}. It can be seen that such interactions mainly affect higher frequency light curves at early times ($\leqslant 1$ day), with a smaller $E_{\rm iso}$ can give rise to an earlier FS/RS peak. Also, For frequencies $< 4.8$ GHz, the FS/RS peak is hardly noticeable, with the only prominent structure as the later peak shown in Fig. \ref{fig:lc_k}.}} \label{fig:RSFS}
\end{figure*}

Many GRBs have already entered postjet break evolution when reaching radio afterglow peak times, especially for low observing frequencies, that is,  $t_{{\rm peak}, \nu} > t_b$, where $t_b$ denotes the jet break time \citep{Chandra2012}. It is necessary to discuss the effects brought by a GRB jet in order to get a full understanding of multiband afterglows. Assuming the GRB jet-opening angle is $\theta_j$, with an initial opening angle $\theta_0$. When the bulk Lorentz factor of GRB shock wave decreases to $\Gamma < \theta_j^{-1}$, the jet-opening angle becomes larger than the relativistic beaming angle $1/\Gamma$, and thus, approximately speaking, the flux density from the jet is $\sim \Gamma^2 \theta_0^2$ times smaller than that from spherical shock.

Firstly, we consider a jet without sideways expansions, that is, $\theta_j = \theta_0$. Here we perform a similar analysis to \cite{Meszaros1999}. Assuming the flux density from spherical shock evolves as $F_{\nu,1} \propto t^{a_1}$, a jet with the same parameters can give rise to a flux density $F_{\nu,2} \propto t^{a_2}$. Because in the ultrarelativistic phase the bulk Lorentz factor evolves as $\Gamma \propto t^{\left(3-k\right)/\left(8-2k\right)}$, and $F_{\nu,1}/F_{\nu,2} \propto t^{a_1 - a_2} \propto \Gamma^{-2} \theta_0^{-2}$, it can be seen that
\begin{equation}
a_2 = a_1 - \frac{3-k}{4-k},\label{eq:31}
\end{equation}
That is, analytically speaking, for a relativistic jet without sideways expansions, the slope of the afterglow light curves during $t>t_j$ is the same as the slopes presented by Eqs. \ref{eq:20}-\ref{eq:26} minus a correction factor $\frac{3-k}{4-k}$. It can be seen that no radio peak exists after $t_j$, no matter the $k$ value. If a jet exists during the radio afterglow evolution, observationally the radio-peak time $t_{{\rm peak},\nu}$ should equal the jet break time $t_j$, if no peak occurs before $t_j$, and $t_{{\rm peak},\nu}$ no longer depends on the observing frequency $\nu$ in this case, making $a = 0$. If we assume that $t_j<t_m$, and that $t_j$ is not far from $t_m$, we can calculate that $b = 1/3$. Such a $b$ value cannot be distinguished from peaks caused by spherical shocks. Although \cite{Huang2000} show that the shock dynamics can be affected by the jet, according to the numerical calculation, the afterglow flux decreases smoothly after jet break when sideways expansions of the jet can be ignored even with evolving temporal index, and later emission peaks do not appear. Thus, the analytical conclusions on $a$ and $b$ still apply.

Analytical expressions for jets with sideways expansions only exist when $k=0$, according to \cite{Sari1999}. In this case, we also have $a_2 < a_1$, and $a_2 \leqslant 0$. That is, no emission peak exists after the jet break. Hence, similar to jets without expansions, the possible peak marks the jet break time $t_j$, with $a = 0$, $b = 1/3$ (providing that $t_j$ occurs not too early). For arbitrary $k$ values, a numerical calculation based upon dynamical equations provided by \cite{Huang2000} is needed. We take the local speed of sound $c_s = c/\sqrt{3}$ as the speed of jet expansion to calculate the dynamical evolution and find that the existence of a jet has a greater impact on late-time afterglow behaviors. During this time, the condition $\Gamma > \theta_j^{-1}$ can be satisfied once more, making the afterglow flux rise again, forming a new emission peak. As early peaks are produced by spherical shocks, the late peaks due to jet effects appear earlier at higher frequencies, too. Because for typical parameters the late peak appears at several hundred days after GRB prompt emission, the analytical solutions proposed by \cite{Sari1999} or \cite{Rhoads1999} do not apply, even for the $k=0$ ISM environment. The flux level of such a late peak is much lower than that of early peaks, for the afterglow has already entered transrelativistic phase during this time, making the relativistic beaming effect much less significant. And it should be noted that when $k \sim 2$, we can almost have $\Gamma > \theta_j^{-1}$ satisfied all the way. As a result, it is possible that no jet break exists in the stellar-wind environment. Figure \ref{fig:theta_k} shows the temporal evolution of the jet-opening angle for various $k$, and its relation with $1/\Gamma$.

According to numerical results, the $a$ value of the late emission peaks for sideways-expanding jets decreases with increasing $k$, which is similar to the case with spherical shock waves. The value of $a$ decreases from $\sim 1.1$ to $\sim 0.88$ when $k$ increases from $0$ to $2$. Meanwhile, the value of $b$ for jets increases with larger $k$, which is opposite to the spherical case. And $b$ is quite small this time, even smaller than the ultrarelativistic case. For example, our results show that $b \sim 0.15$ for $k=0$, and $b \sim 0.45$ for $k=2$. However, it should be noted that all such results are based upon one single assumption, that is, the jet expands at a constant speed of $c/\sqrt{3}$. In reality, this assumption may not be satisfied after the afterglow enters the transrelativistic regime. Late-time afterglows may expand quite slowly, making the jet-opening angle $\theta_j<1/\Gamma$, thus no late peak can be formed. Also, earlier numerical simulation shows that jets do not have significant sideways expansions (\citealt{Kumar2003}; \citealt{Granot2003}). In one word, effects caused by jet may be more complex in real GRBs.

\section{Observational Implications} \label{sec:obs}

\subsection{Sample Selection}

We collected data on GRBs with multiband radio light-curve observations (that is, GRBs with radio light curves in no fewer than two bands) from the literature, in order to put constraints on the circumburst density profiles based on Eqs. \ref{eq:27} and \ref{eq:28}. Only the observed flux and corresponding epochs, rather than fitted peak values, such as the ones shown in Table 4 of \cite{Chandra2012} or Table 1 of \cite{Li2015}, were utilized in this analysis. Because it can be seen from Eqs. \ref{eq:27} and \ref{eq:28}, as well as in Section \ref{subsec:trans-relativistic} that theoretically speaking, both relativistic and transrelativistic cases give rise to nonnegative $a$ and $b$, no matter of the $k$ value, high-frequency GRB radio afterglow light curves should exhibit a higher $F_{\rm peak}$, as well as a $t_{\rm peak}$ no later than lower-frequency peaks, a general trend of later and weaker peaks at lower frequencies is required during our sample selection, in order to comply with the theoretical framework of this analysis. For this reason, GRB 980519 (\citealt{Frail2000a}) was excluded, due to the late appearance of higher-frequency peaks compared with the lower-frequency ones, which is hard to explain with the GRB afterglow model based upon synchrotron radiation, unless we take large observation errors and light-curve modulation due to interstellar scintillation into consideration. Besides, GRBs 050416A (\citealt{Soderberg2007}), 020813, and 050713B (\citealt{Li2015}) were omitted, due to their higher peak fluxes compared with their lower-frequency ones. Although our strategy is somewhat biased, such a bias could be considered as originating from model constraints, especially for our radio-peak-focused approach to get the $k$ value.

\startlongtable
\begin{deluxetable*}{l l l c c l}
\tablecaption{Characteristics of GRB Multi-Band Radio Afterglow Emission Peaks}\label{table:data}
\tablenum{1}
\tablewidth{0pt}
\tabletypesize{\scriptsize}
\tablehead{
\colhead{GRB} & \colhead{Telescope$^a$} & 
\colhead{Frequency} & \colhead{$F_{\rm peak,\nu}$} & 
\colhead{$t_{\rm peak,\nu}$} & \colhead{Reference}\\ 
\colhead{} & \colhead{} & \colhead{(GHz)} & \colhead{($T_{90}$ + days)} & 
\colhead{($\mu$Jy)($T_{90}$ + days)} & \colhead{}
} 
\startdata
970508	&  VLA  &  $1.43$  &  $81.03^{+21.08}_{-10.49}$  & $485 \pm 45$   & \cite{Frail2000b} \\ 
      	&  VLA  &  $4.86$  &  $20.2^{+1.84}_{-1.43}$  & $1350 \pm 55$ & \cite{Frail1997} \\
      	&  VLA  &  $8.46$  &  $20.2^{+1.84}_{-1.43}$  & $1200 \pm 47$ & \cite{Frail1997} \\
\hline
970828  &  VLA  &  $4.86$  &  $5.6^{+20.8}_{-2.1}$  & $99 \pm 50$ & \cite{Djorgovski2001} \\
      	&  VLA  &  $8.46$  &  $3.5^{+2.1}_{-1.0}$ & $147 \pm 33$ & \cite{Djorgovski2001} \\
\hline
980329	&  VLA  &  $4.9$   &   $28.86^{+3.08}_{-3.92}$ $^e$ & $128 \pm 42$ & \cite{Taylor1998}\\
	&  VLA  &  $8.3$   &  $12.93^{+6.95}_{-3.03}$ & $274 \pm 34$ & \cite{Taylor1998}\\
\hline
980425	&  ATCA &  $1.38$  & $45.7^{+6.0}_{-5.7}$ &  $28900 \pm 500$  & \cite{Kulkarni1998}\\
	&  ATCA &  $2.49$  & $34.7^{+2.1}_{-1.8}$ &  $33500 \pm 500$  & \cite{Kulkarni1998}\\
	&  ATCA &  $4.86$  & $11.7^{+2.9}_{-1.8}$ &  $44600 \pm 300$  & \cite{Kulkarni1998}\\
	&  ATCA &  $8.46$  & $11.7^{+2.9}_{-1.8}$ &  $49400 \pm 200$  & \cite{Kulkarni1998}\\
\hline
980703	&  VLA  &  $1.43$  & $49.10^{+0.94}_{-9.99}$ &  $148 \pm 37$  & \cite{Frail2003}\\
	&  VLA  &  $4.86$  & $14.38^{+2.77}_{-2.15}$ &  $1200 \pm 48$  & \cite{Frail2003}\\
	&  VLA  &  $8.46$  & $12.93^{+2.15}_{-6.92}$ &  $1050 \pm 35$  & \cite{Frail2003}\\
\hline
990510	&  ATCA &  $4.8$   & $9.22$ $\left(>0.72\right)^{b}$  & $177 \pm 36$  &  \cite{Harrison1999}\\
	&  ATCA &  $8.7$   & $3.31^{+1.93}_{-2.59}$ &  $227 \pm 30$  & \cite{Harrison1999}\\
\hline
991208$^f$&  VLA  &  $4.86$  & $21.96$ $\left(>12.04\right)^{b}$ & $820 \pm 52$  & \cite{Galama2000}\\
	&  VLA  &  $8.46$  & $11.73^{+1.00}_{-1.11}$ &  $1990 \pm 33$  & \cite{Galama2000}\\
	&  Ryle &  $15.0$  & $5.36^{+2.06}_{-1.11}$ &  $2100 \pm 600$  & \cite{Galama2000}\\
\hline
000301C &  VLA  &  $4.86$  & $48.06^{+16.07}_{-6.00}$ &  $226 \pm 51$  & \cite{Berger2000}\\
	&  VLA  &  $8.46$  & $12.17^{+2.00}_{-7.19}$ &  $483 \pm 26$  & \cite{Berger2000}\\
	&  Ryle &  $15.0$  & $3.57$ $\left(<8.8\right)^c$  &  $660 \pm 160$  & \cite{Berger2000}\\
\hline
000418$^g$&  VLA  &  $4.86$ & $14.63^{+1.93}_{-1.98}$  & $1120 \pm 52$  & \cite{Berger2001}\\
	&  VLA  &  $8.46$  & $14.63^{+1.93}_{-1.98}$ &  $1240 \pm 46$  & \cite{Berger2001}\\
	&  Ryle &  $15.0$  & $12.32^{+2.20}_{-1.98}$ &  $1350 \pm 480$  & \cite{Berger2001}\\
\hline
000911	&  VLA  &  $4.86$  & $11.05^{+6.15}_{-4.86}$ &  $71 \pm 23$  & \cite{Price2002a}\\
	&  VLA  &  $8.46$  & $3.06^{+1.00}_{-0.87}$ &  $278 \pm 36$  & \cite{Price2002a}\\
\hline
000926 &  VLA  &  $4.86$  & $10.75^{+0.96}_{-3.55}$ &  $395 \pm 52$  & \cite{Harrison2001}\\
	&  VLA  &  $8.46$  & $8.22^{+2.56}_{-1.03}$ &  $566 \pm 34^h$  & \cite{Harrison2001}\\
	&  Ryle &  $15.0$  & $4.70^{+4.06}_{-0.78}$ &  $820 \pm 390$  & \cite{Harrison2001}\\
	&  VLA  &  $22.5$  & $7.19$ $\left(<8.22\right)^c$ & $1415 \pm 185$  & \cite{Harrison2001}\\
\hline
010921	&  VLA  &  $4.86$  & $37.77$ $\left(>34.82\right)^d$ & $140 \pm 28$ & \cite{Price2002b}\\
	&  VLA  &  $8.46$  & $27.01^{+10.76}_{-1.08}$ & $229 \pm 22$  & \cite{Price2002b}\\
	&  VLA  &  $22.5$  & $25.93$ $\left(<32.80\right)^c$  &  $330 \pm 90$  & \cite{Price2002b}\\
\hline
011121  &  ATCA & $4.80$  & $6.86^{+8.16}_{-3.44}$ & $510 \pm 38$  & \cite{Price2002c}\\
	&  ATCA &  $8.70$  & $6.86^{+8.16}_{-3.56}$ &  $610 \pm 39$  & \cite{Price2002c}\\
\hline
020903	&  VLA  &  $1.5$   & $36.7^{+37.3}_{-8.10}$ &  $294 \pm 91$  & \cite{Soderberg2004a}\\
	&  VLA  &  $4.9$   & $36.7^{+37.3}_{-8.10}$ &  $832 \pm 47$  & \cite{Soderberg2004a}\\
	&  VLA  &  $8.5$   & $23.8$ $\left(<25.7\right)^c$ & $1058 \pm 19$ & \cite{Soderberg2004a}\\
\hline
021004	&  VLA  &  $4.86$  & $27.2^{+4.5}_{-7.0}$ &  $476 \pm 38$  & \cite{dePostigo2005}\\
	&  VLA  &  $8.46$  & $16.3^{+3.9}_{-3.7}$ &  $720 \pm 49$  & \cite{dePostigo2005}\\
	&  VLA  &  $22.5$  & $6.2^{+4.0}_{-0.8}$ &  $1282 \pm 85$  & \cite{dePostigo2005}\\
\hline
030329	&  GMRT &  $0.61$ & $334.15^{+43.86}_{-52.66}$ & $1140 \pm 250$ & \cite{vanderHorst2008}\\
	&  WSRT &  $0.84$ &  $318.555$ $\left(<379.267\right)^c$ & $2332 \pm 288$  & \cite{vanderHorst2008}\\
	&  WSRT &  $1.4$ & $170.739^{+25.152}_{-44.507}$ & $2290 \pm 190$ &  \cite{vanderHorst2005}\\
	&  WSRT &  $2.3$ & $72.356^{+49.741}_{-31.013}$ &  $3840 \pm 70$  & \cite{vanderHorst2005}\\
	&  VLA  &  $4.86$  & $33.58^{+2.94}_{-2.07}$ & $11620 \pm 80$  & \cite{Berger2003}\\
	&  VLA  &  $8.46$  & $14.87^{+1.79}_{-2.18}$ &  $19150 \pm 80$  & \cite{Berger2003}\\
	&  VLA  &  $15.0$  & $11.90^{+0.79}_{-2.41}$ &  $31400 \pm 250$  & \cite{Berger2003}\\
	&  VLA  &  $22.5$  & $9.49^{+2.41}_{-1.81}$ &  $48160 \pm 230$  & \cite{Berger2003}\\
	&  VLA  &  $43.3$  & $6.89^{+0.79}_{-2.13}$ &  $55330 \pm 430$  & \cite{Berger2003}\\
\hline
031203$^i$&  VLA  &  $4.86$  & $65.32^{+7.98}_{-11.90}$ & $751 \pm 45^j$ &  \cite{Soderberg2004b}\\
	&  VLA  &  $8.46$  & $811 \pm 40$ & $19.45^{+3.03}_{-2.02}$ & \cite{Soderberg2004b}\\ 
\hline
050820A &  VLA  &  $4.86$  & $2.19^{+1.96}_{-2.034}$ & $256 \pm 78$ & \cite{Cenko2006}\\
	&  VLA  &  $8.46$  & $0.93^{+1.23}_{-0.814}$ &  $634 \pm 62$  & \cite{Cenko2006}\\
\hline
051022	&  WSRT &  $4.9$   & $2.22^{+1.08}_{-1.03}$ &  $342 \pm 34$  & \cite{Rol2007}\\ 
	&  VLA  &  $8.5$   & $1.64$ $\left(<13.37\right)$ & $585 \pm 49^c$  & \cite{Cameron2005}\\ 
\hline
060218$^k$ &  VLA  &  $1.43$  & $6.97^{+5.11}_{-5.12}$ & $134 \pm 145$ & \cite{Soderberg2006}\\
	&  VLA  &  $4.86$  & $4.85^{+2.12}_{-1.02}$ & $328 \pm 61$  & \cite{Soderberg2006}\\
	&  VLA  &  $8.46$  & $1.87$ $\left(<3.00\right)^c$ & $453 \pm 77$ & \cite{Soderberg2006}\\
\hline
070125	&  VLA  &  $4.86$  & $27.90^{+7.01}_{-5.17}$ & $308 \pm 78$  & \cite{Chandra2008}\\
	&  VLA  &  $8.46$  & $18.86^{+1.01}_{-1.01}$ & $660 \pm 29$  & \cite{Chandra2008}\\
	&  VLA  &  $14.94$  & $16.06^{+1.83}_{-1.09}$ & $1410 \pm 137$  & \cite{Chandra2008}\\
	&  VLA  &  $22.5$  & $12.80^{+1.29}_{-2.07}$ & $1603 \pm 235$  & \cite{Chandra2008}\\
\hline
071003	&  VLA  &  $4.86$  & $3.84$ $\left(<21.68\right)^c$ & $220 \pm 54$ & \cite{Perley2008}\\
	&  VLA  &  $8.46$  & $3.82^{+4.90}_{-2.06}$ &  $430 \pm 50^l$  & \cite{Perley2008}\\
\hline
100418A$^m$&  JVLA &  $4.95$  & $72^{+23}_{-16}$ &  $513 \pm 44$  & \cite{Moin2013}\\
	&  JVLA &  $8.46$  & $48^{+8}_{-15}$ &  $1454 \pm 21$  & \cite{Moin2013}\\
\hline
100814A$^n$&  JVLA &  $4.7$  & $11.3^{+12.9}_{-3.9}$ &  $461.13$  & \cite{dePasquale2012}\\
	&  JVLA &  $7.9$   & $11.3^{+12.9}_{-3.9}$ &  $560.43$  & \cite{dePasquale2012}\\	
\hline
111215A$^o$&  WSRT &  $1.4$   & $31.01^{+16.99}_{-8.98}$ & $373 \pm 170$ & \cite{vanderHorst2015}\\
	&  WSRT &  $4.8$   & $22.03^{+8.98}_{-6.94}$ & $1079 \pm 51$  & \cite{vanderHorst2015}\\
	&  JVLA &  $6.7$   & $15.41^{+10.99}_{-1.97}$ & $1180 \pm 30$  & \cite{Zauderer2013}\\
	&  JVLA &  $8.4$   & $15.41$ $\left(<26.40\right)^c$ & $1340 \pm 63$  & \cite{Zauderer2013}\\
	&  JVLA &  $19.1$  & $11.49^{+3.92}_{-8.14}$ &  $1670 \pm 47$  & \cite{Zauderer2013}\\
	&  JVLA &  $24.4$  & $3.35$ $\left(<11.49\right)^c$ & $1990 \pm 47$ & \cite{Zauderer2013}\\
\hline
130427A	&  JVLA &  $5.1$   & $2.04^{+2.71}_{-1.36}$ &  $1760 \pm 88$  & \cite{Perley2014}\\
	&  JVLA &  $6.8$   & $0.68$ $\left(<2.04\right)^c$ &  $2570 \pm 160$  & \cite{Perley2014}\\
	&  AMI  &  $15.7$  & $0.64^{+0.91}_{-0.28}$ &  $4160 \pm 220$  & \cite{Anderson2014}\\
\hline
140304A	&  JVLA &  $4.90$  & $8.55^{+9.71}_{-4.00}$ & $112 \pm 16$  & \cite{Laskar2018}\\
	&  JVLA &  $7.00$  & $8.55^{+9.71}_{-4.00}$ & $166 \pm 12$  & \cite{Laskar2018}\\
	&  JVLA &  $8.55$  & $8.55^{+9.70}_{-4.00}$ & $187 \pm 16$  & \cite{Laskar2018}\\
	&  JVLA &  $11.0$  & $4.55^{+4.00}_{-3.01}$ &  $241 \pm 17$  & \cite{Laskar2018}\\
	&  JVLA &  $13.5$  & $4.55^{+4.00}_{-3.01}$ &  $280 \pm 13$  & \cite{Laskar2018}\\
	&  JVLA &  $16.0$  & $4.55^{+4.00}_{-3.01}$ &  $323 \pm 13$  & \cite{Laskar2018}\\
	&  JVLA &  $19.2$  & $4.55^{+4.00}_{-3.00}$ &  $333 \pm 18$  & \cite{Laskar2018}\\
	&  JVLA &  $24.5$  & $1.55$ $\left(<3.00\right)^c$ &  $307 \pm 28$  & \cite{Laskar2018}\\
	&  JVLA &  $30.0$  & $1.55$ $\left(<3.00\right)^c$ &  $384 \pm 40$  & \cite{Laskar2018}\\
	&  JVLA &  $37.0$  & $1.55$ $\left(<3.00\right)^c$ &  $469 \pm 55$  & \cite{Laskar2018}\\
\hline
141121A$^p$&  JVLA &  $3$ & $21.350^{+2.979}_{-4.965}$ & $156 \pm 21$  & \cite{Cucchiara2015}\\
	&  JVLA &  $5$     & $21.350^{+2.979}_{-4.965}$ & $184 \pm 15$  &\cite{Cucchiara2015}\\
	&  JVLA &  $7$     & $16.385^{+4.965}_{-4.983}$ & $270 \pm 18$  & \cite{Cucchiara2015}\\
	&  AMI  &  $15$    & $3.037^{+5.315}_{-2.127}$ & $460 \pm 50$  & \cite{Cucchiara2015}\\
\hline
160509A$^q$&  JVLA & $1.644$ & $10.03$ $\left(>6.00\right)^d$ & $170.2 \pm 102.6$ & \cite{Laskar2016}\\
	&  JVLA &  $2.679$ & $6.00^{+4.03}_{-0.69}$ & $334.8 \pm 40.1$  & \cite{Laskar2016}\\
	&  JVLA &  $3.523$  & $4.06^{+1.25}_{-4.06}$ & $384.4 \pm 27.1$  & \cite{Laskar2016}\\
	&  JVLA &  $5.0$   & $2.1479^{+0.8458}_{-0.9797}$ & $509.1 \pm 18.9$  & \cite{Laskar2016}\\
	&  JVLA &  $7.4$   & $4.06^{+1.94}_{-1.0663}$ & $1389.1 \pm 20.2$  & \cite{Laskar2016}\\
	&  JVLA &  $8.5$   & $2.9781^{+1.0819}_{-1.8253}$ & $1049.0 \pm 17.1$  & \cite{Laskar2016}\\
	&  JVLA &  $11.0$  & $2.1479^{+1.94}_{-1.0819}$ & $1261.0 \pm 17.9$  & \cite{Laskar2016}\\
	&  JVLA &  $13.5$  & $1.1369$ $\left(<2.9623\right)^c$ & $946.0 \pm 27.2$ & \cite{Laskar2016}\\
	&  JVLA &  $16.0$  & $1.1369$ $\left(<2.9623\right)^c$ & $1025.9 \pm 31.7$  & \cite{Laskar2016}\\
	&  JVLA &  $19.2$  & $1.1163$ $\left(<2.1259\right)^c$ & $1341.0 \pm 33.7$  & \cite{Laskar2016}\\
	&  JVLA &  $24.5$  & $1.1163$ $\left(<2.1259\right)^c$ & $1891.0 \pm 56.3$  & \cite{Laskar2016}\\
	&  JVLA &  $30.0$  & $1.0911$ $\left(<2.9165\right)^c$ & $2117.8 \pm 47.9$  & \cite{Laskar2016}\\
\hline
160625B$^r$&  JVLA &  $1.77$ & $22.52^{+25.86}_{-10.02}$ & $346 \pm 62$ & \cite{Alexander2017}\\
	&  JVLA &  $2.68$  & $12.94^{+10.02}_{-6.19}$ &  $621 \pm 31$  & \cite{Alexander2017}\\
	&  JVLA &  $3.52$  & $12.94^{+10.02}_{-6.19}$ &  $475 \pm 40$  & \cite{Alexander2017}\\
	&  JVLA &  $5.0$   & $2.50^{+3.79}_{-1.13}$ &  $932 \pm 24$  & \cite{Alexander2017}\\
	&  JVLA &  $7.1$   & $2.50^{+3.79}_{-1.13}$ &  $1310 \pm 20$  & \cite{Alexander2017}\\
	&  JVLA &  $8.5$   & $2.49^{+3.79}_{-1.14}$ &  $1135 \pm 28$  & \cite{Alexander2017}\\
	&  JVLA &  $11.0$  & $2.49^{+3.79}_{-1.14}$ &  $946 \pm 25$  & \cite{Alexander2017}\\
\hline
170817A	&  JVLA &  $2.6$   & $162.9^{+35.8}_{-57.8}$ &  $69.5 \pm 16.3$  & \cite{Margutti2018}\\
	&  ATCA &  $5.5$   & $149.6^{+13.3}_{-24.0}$ &  $99.1 \pm 8.4$  & \cite{Dobie2018}\\
	&  ATCA &  $9.0$   & $125.6^{+24.0}_{-10.2}$ &  $84.5 \pm 10.9$  & \cite{Dobie2018}\\
\hline
171010A	&  JVLA &  $4.5$   & $9.66^{+18.93}_{-7.09}$ & $929 \pm 42$  & \cite{Bright2019}\\
	&  JVLA &  $5.5$   & $9.66^{+18.93}_{-7.09}$ & $978 \pm 33$  & \cite{Bright2019}\\
	&  JVLA &  $6.5$   & $9.66^{+18.93}_{-7.09}$ & $1025 \pm 35$  & \cite{Bright2019}\\
	&  JVLA &  $7.5$   & $9.66^{+18.93}_{-7.09}$ & $1090 \pm 35$  & \cite{Bright2019}\\
	&  JVLA &  $8.5$   & $2.56$ $\left(<9.27\right)^c$ & $1024 \pm 40$  & \cite{Bright2019}\\
	&  JVLA &  $9.5$   & $2.56$ $\left(<9.27\right)^c$ & $1580 \pm 54$  & \cite{Bright2019}\\
	&  JVLA &  $10.5$  & $2.56$ $\left(<9.27\right)^c$ & $1858 \pm 64$  & \cite{Bright2019}\\
	&  JVLA &  $11.5$  & $2.56$ $\left(<9.27\right)^c$ & $2009 \pm 71$  & \cite{Bright2019}\\
	&  JVLA &  $15.5$  & $2.37^{+0.94}_{-1.02}$ & $2520 \pm 140$  & \cite{Bright2019}
\enddata
\tablecomments{\\
$^a$ ``VLA'' for Very Large Array; ``JVLA'' for the upgraded Karl G. Jansky Very Large Array; ``ATCA'' for Australia Telescope Compact Array; ``GMRT'' for Giant Meterwave Radio Telescope; ``WSRT'' for Westerbork Synthesis Radio Telescope; ``Ryle'' for Ryle Telescope; ``AMI'' for Arcminute Mircrokelvin Imager.\\
$^b$ Error bars of $t_{\rm peak,\nu}$ denotes the times on which the adjacent measurements are taken around the corresponding peak of the light curve.\\
$^c$ Flux decreases monotonically during the complete observing campaign without a definite peak; we took the highest $S_{\nu}$ as peak flux.\\
$^d$ Flux increases monotonically during the complete observing campaign without a definite peak; we took the highest $S_{\nu}$ as peak flux.\\
$^e$ The 4.9 MHz light curve of GRB 980329 exhibits two peaks, according to \cite{Taylor1998}. Since the first peak appears earlier than the 8.3 MHz peak, which is in contradiction with the standard afterglow model, here we adopt the later peak for our analysis.\\
$^f$ The 30 GHz data acquired by the Owens Valley Radio Observatory's 40 m telescope from \cite{Galama2000} were discarded due to scintillated light curve without a definite peak, and the later appearance of the highest flux compared to the 15 GHz data; the 1.43 GHz data acquired by VLA from \cite{Galama2000} were discarded due to the earlier peak time compared to higher frequencies.\\
$^g$ The 22.46 GHz data acquired by VLA from \cite{Berger2001} were discarded due monotonic light curve and later appearance of the highest flux compared to the 15 GHz data.\\
$^h$ The 8.46 GHz data point with $S_{\nu} = 566 \pm 34$ $\mu$Jy acquired at 2000 Octpber 5.216 (UTC) by VLA from \cite{Harrison2001} was adopted as the peak flux; a later rebrightening at October 12.771 with $S_{\nu} = 644 \pm 126$ $\mu$Jy was discarded due to the later appearance compared to the 4.86 GHz peak, as well as the larger error bar.\\
%$^i$ No error for peak time provided by \cite{Li2015}.\\
$^i$ The 1.43 GHz data acquired by VLA from \cite{Berger2003} were discarded due to scintillated light curves and earlier appearance of the highest flux measurement compared to the 4.86 GHz data; the 22.5 GHz data were discarded due to monotonic light curve and a maximum flux significantly lower than the 8.46 GHz data.\\
$^j$ Two flux measurements with similar flux levels, $S_{\nu} = 728 \pm 55$ and $749 \pm 63$ $\mu$Jy, were performed at 2004 January 4.33 and January 15.35 (UTC), respectively, as shown in \cite{Soderberg2004b}. It seems that the radio afterglow of this burst suffered significantly from interstellar scintillation, according to data presented in \cite{Soderberg2004b}, thus making the exact peak time somewhat hard to determine.\\
$^k$ The 15 GHz measurements acquired by the Ryle Telescope from \cite{Soderberg2006} were discarded due to large measurement errors and lack of data points; the 22.5 GHz measurements acquired by VLA from \cite{Soderberg2006} were discarded due to lack of data points.\\
$^l$ An 8.46 GHz measurement performed at 2007 October 12.04 (UTC) shows a flux level of $S_{\nu} = 431 \pm 51$ $\mu$Jy (\citealt{Perley2008}), which is slightly larger than the peak data listed in this table. The earlier $S_{\nu} = 430 \pm 50$ $\mu$Jy was chosen as a peak considering the theoretical predictions for earlier peaks at higher frequencies. However, due to no 8.46 GHz data has been taken between the two measurements, it is difficult to know the exact peak time for the GRB 071003 radio afterglow at this frequency.\\
$^m$ ACTA 5.5 GHz data for GRB 100418A listed in \cite{Moin2013} were discarded, due to the monotonic light-curve behavior, which is inconsistent with the VLA data.\\
$^n$ No errors in the flux measurements shown in \cite{dePasquale2012}.\\
$^o$ The 15 GHz measurements acquired by AMI from \cite{vanderHorst2015} were discarded due to the lower flux and later peak compared to the 4.8 GHz data.\\
$^p$ The 13 and 15 GHz measurements acquired by JVLA from \cite{Cucchiara2015} were discarded due to the lower flux compared to the 7 GHz data.\\
$^q$ The 1.255 GHz measurements acquired by JVLA from \cite{Laskar2016} were discarded due to a peak flux that appeared earlier than the 1.644 GHz peak, with a flux higher than the 5.0 GHz peak.\\
$^r$ The 1.45 GHz measurements acquired by JVLA from \cite{Alexander2017} were discarded due to an earlier peak compared to 1.77 GHz; all data with observed frequency $> 13.5$ GHz were discarded due to monotonic decline with maximum $S_{\nu}$ lower than 11.0 GHz peak, due to the lack of early observations. \\
}
\end{deluxetable*}

In total, we selected 32 GRBs, including 31 long bursts and 1 short one (GRB 170817A), as our sample. For observations at certain frequencies in our sample, GRBs do not comply with the theoretical trend of higher $F_{\rm peak}$, earlier $t_{\rm peak}$ at higher $\nu$ -- such data points have been omitted during our analysis. Meanwhile, we largely ignored the light-curve peaks occurring earlier than $\sim 1-2$ days in the observer's frame to avoid possible contamination from early reverse shocks, because as can be seen in Fig. \ref{fig:RSFS}, the typical peak time due to forward-reverse-shock interactions should be at $\sim T_{90} +$ several$\times 10^{-1}$ days, which corresponds to $\sim T_{90} + 1-2$ days in the observer's frame because the typical redshift value of GRB is $z\sim 3$ (\citealt{Jakobsson2005}). Also, $\geqslant 50$ GHz data are not taken into consideration, as in such high frequencies, the $t_{\rm peak} = t_m$ conclusion does not apply (see Section \ref{sec:theory}). Considering the sparsely sampled nature of GRB radio observations, as well as the not-so-steep slopes of temporal evolution for radio afterglow light-curve fluxes (see Fig. \ref{fig:lc_k}), we took an approximate approach to make our analysis: The maximum observed flux is recognized as of the peak flux $F_{\rm peak,\nu}$ at a certain observing frequency, the corresponding observing time as the estimated peak time $t_{\rm peak,\nu}$, and the epochs of adjacent observations give the upper and lower limits of $t_{\rm peak,\nu}$. Besides, in order to make constraints with as many observing frequencies as possible, we took another strategy to utilize the monotonically evolved light curves: the reading of the last data point of a monotonically increasing light curve was taken as the lower limit of $F_{\rm peak}$, and the second-to-last epoch as the possible lower range of error for $t_{\rm peak}$. Similarly, for a constantly declining light curve, the first data point sets the lower limit of peak flux, and the second observing epoch the upper limit of peak time. 

Some GRBs have been observed by various telescopes at similar frequencies. For example, GRB 030329 has been observed by WSRT at 1.4 GHz (\cite{vanderHorst2005}, and GMRT at 1.28 GHz (\cite{vanderHorst2008}). Also,  GRB 100418A has been observed by VLA and ACTA at 4.95 and 5.5 GHz, respectively \cite{Moin2013}. Because calibrations between different instruments could lead to extra complications, in this case, we select peak data for analysis from as few telescopes as possible. Hence, our main uncertainties arise from errors in $t_{\rm peak,\nu}$ estimation, especially for GRB light curves with fewer observed data points. Information on the $t_{\rm peak,\nu}$ and $F_{\rm peak,\nu}$ for each band of our GRB samples, as well as descriptions of the discarded outlier data, is listed in Table \ref{table:data}. It is worth noting that as seen in Fig. \ref{fig:gammabeta_parameter}, considering the long-lasting process of transition between relativistic to deep Newtonian shock waves, nearly all of our samples should be still in the relativistic or transrelativistic phase when the light-curve peaks occur.

\subsection{Constraints on Circumburst Environment}

\begin{figure*}
\plotone{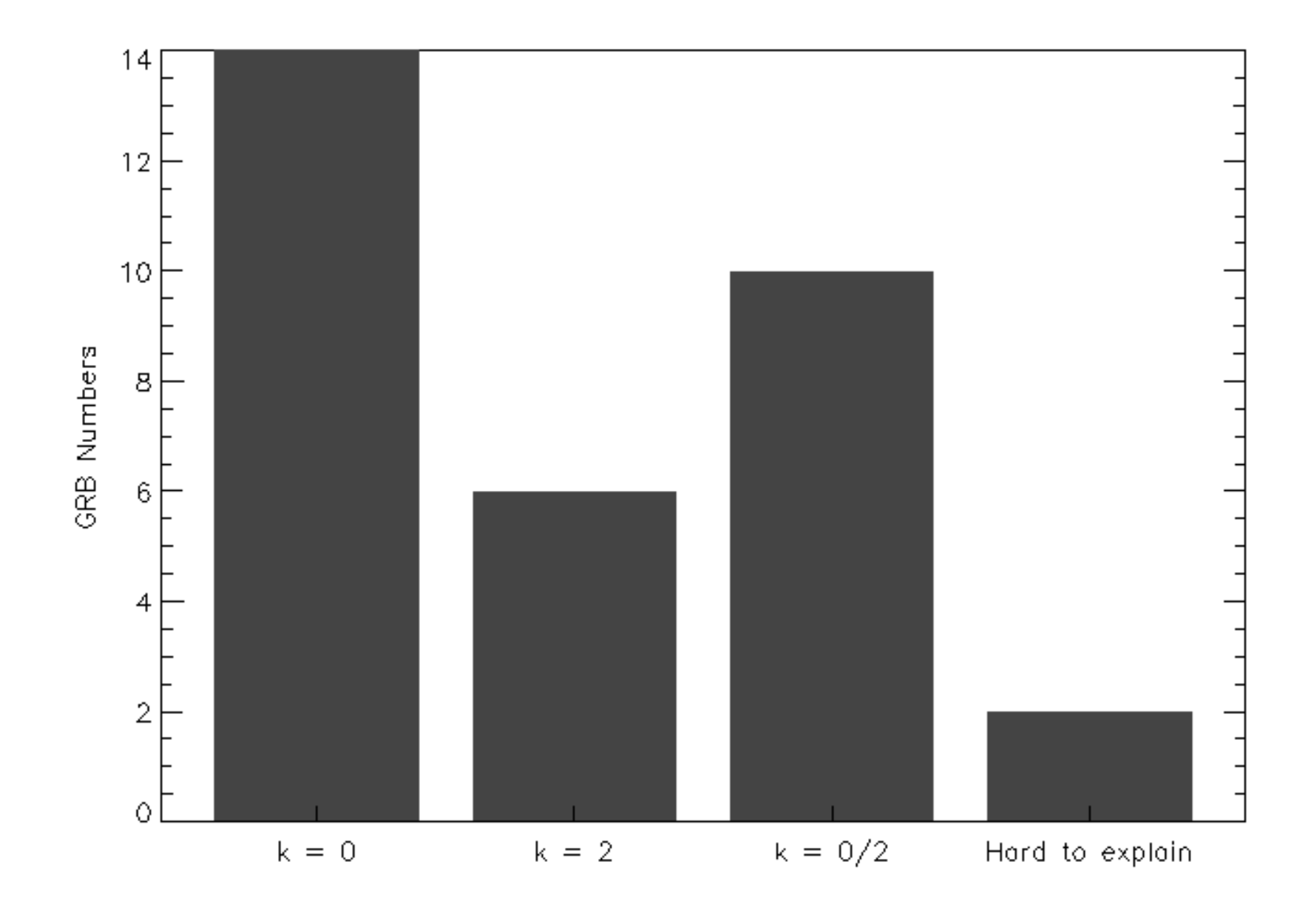}
\caption{\small{A statistic on the $k$ distributions for the 32 GRB sample. It can be seen that the radio peaks of 14 GRBs can be explained by the $k=0$ density profile (with six samples showing possible contamination by the reverse shock). Another six samples exhibit behaviors compatible with predictions for the $k\sim 2$ case. Also, we have 10 GRBs that can be explained by density distributions between the ISM and stellar-wind environment, along with another 2 bursts occurring in environment hard to constrain, due to observation errors.}}\label{fig:stat}
\end{figure*}

The values of the $a$ and $b$ indices of 32 sample GRBs are listed in Table \ref{table:peak}. Figure \ref{fig:tpeak_nu} shows the $a$ values, observing frequencies, as well as peak times in light curves of each sample GRB. Figure \ref{fig:fpeak_nu} shows the corresponding $b$ value, the observing bands, along with peak fluxes. And a statistic of the $k$ value in the circumburst environment for our complete sample can be found in Figure \ref{fig:stat}. We compare these fitted indices with theoretical predictions from Eqs. \ref{eq:29} and \ref{eq:30} and Section \ref{subsec:trans-relativistic}, taking the circumburst density data provided by \cite{Chandra2012} as well as possible observational errors into consideration. It can be seen that  more than 40\% of our samples can be explained with uniform ISM distribution, while nearly one-fifth samples can be described with stellar-wind environment. And due to various observational complications, both $k=0$ and $k=2$, or any $0<k<2$ circumburst density distributions work for about 30\% of the samples, while two GRBs cannot be properly constrained with our analysis using multiband radio peaks.

\startlongtable
\begin{deluxetable*}{l c c c c l}
\tablecaption{Results of Multi-Band Light Curve Peak Fitting\label{table:peak}}
\tablenum{2}
\tablewidth{0pt}
\tabletypesize{\scriptsize}
\tablehead{
\colhead{GRB} & \colhead{Band Coverage} & 
\colhead{$a$} & \colhead{$b$} & 
\colhead{Explanations$^a$}\\ 
\colhead{} & \colhead{(GHz)} & \colhead{} & \colhead{} & \colhead{}
} 
\startdata
970508	&	$1.43 - 8.46$	& $0.84^{+0.09}_{-0.06}$ & $0.56_{-0.06}^{+0.06}$ & (1,2)\\
970828	&	$4.86 - 8.46$	& $0.84^{+1.99}_{-0.68}$ & $0.71_{-0.80}^{+1.19}$ & (1,2)$^c$\\
980329	&	$4.9 - 8.3$	& $1.52^{+0.28}_{-0.60}$ & $1.44_{-0.56}^{+0.73}$ & (1)\\
980425	&	$1.38 - 8.46$	& $0.86^{+0.06}_{-0.08}$ & $0.31_{-0.01}^{+0.01}$ & (2)\\
980703	&	$1.43 - 8.46$	& $0.86^{+0.19}_{-0.10}$ & $1.21_{-0.14}^{+0.15}$ & (1)\\
990510	&	$4.8 - 8.7$	& $1.72^{+1.73}_{-1.42}$ & $0.42_{-0.36}^{+0.41}$ & (1,2)\\
991208	&	$4.86 - 15.0$	& $1.25^{+0.22}_{-0.34}$ & $1.51_{-0.42}^{+0.05}$ & (1,2)\\
000301C	&	$4.86 - 15.0$	& $2.31^{+0.79}_{-0.54}$ & $0.95_{-0.30}^{+0.28}$ & (1)\\
000418	&	$4.86 - 15.0$	& $0.15^{+0.11}_{-0.11}$ & $0.17_{-0.37}^{+0.25}$ $^d$ & (3,4)\\
000911	&	$4.86 - 8.46$	& $2.32^{+0.56}_{-0.58}$ & $2.46_{-0.53}^{+0.68}$ & (5)\\
000926	&	$4.86 - 22.5$	& $0.37^{+0.57}_{-0.17}$ & $0.80_{-0.18}^{+0.16}$ & (1,2)\\
010921$^b$ &	$4.86 - 22.5$	& $0.22^{+0.75}_{-0.09}$ & $0.54_{-0.24}^{+0.20}$ & (1,2)\\
011121	&	$4.80 - 8.70$	& $0.00^{+0.95}_{-0.85}$ & $0.30_{-0.15}^{+0.15}$ & (2,4)\\%$^c$
020903$^b$ &	$1.5 - 8.5$	& $0.21^{+0.70}_{-0.11}$ & $0.76_{-0.17}^{+0.20}$ & (1)\\
021004	&	$4.86 - 22.5$	& $0.97^{+0.07}_{-0.27}$ & $0.64_{-0.07}^{+0.06}$ & (1,2)\\
030329$^b$ &	$0.61 - 43.3$	& $1.00^{+0.03}_{-0.11}$ & $0.96_{-0.02}^{+0.03}$ & (2)\\
031203	&	$4.86 - 8.46$	& $2.19^{+0.16}_{-0.25}$ & $0.14_{-0.13}^{+0.13}$ & (1)\\
050820A & $4.86 - 8.46$	& $1.51^{+1.50}_{-1.72}$ & $1.63_{-0.49}^{+0.62}$ & (1)$^c$\\
051022	&	$4.9 - 8.5$	& $0.53^{+0.19}_{-3.07}$ & $0.97_{-0.22}^{+0.21}$ & (1)$^c$\\
060218	&	$1.43 - 8.46$	& $0.67^{+0.50}_{-0.20}$ & $0.69_{-0.46}^{+0.21}$ $^e$& (1,2)\\
070125$^b$ &	$4.86 - 22.5$	& $0.48^{+0.09}_{-0.07}$ & $1.12_{-0.15}^{+0.19}$ & (1)\\
071003	&	$4.86 - 8.46$	& $0.01^{+2.08}_{-1.79}$ & $1.21_{-0.42}^{+0.49}$ & (5)\\
100418A	&	$4.95 - 8.46$	& $0.76^{+0.51}_{-0.29}$ & $1.94_{-0.15}^{+0.16}$ & (1)\\
100814A	&	$4.7 - 7.9$	& $0.00^{+0.86}_{-0.80}$ & 0.38 & (3,4)\\	
111215A$^f$ &	$1.4 - 24.4$	& $0.64^{+0.26}_{-0.17}$ & $0.55_{-0.12}^{+0.17}$ & (1,2)\\
130427A$^b$ &	$5.1 - 15.7$	& $0.81^{+0.49}_{-0.64}$ & $0.77_{-0.06}^{+0.06}$ & (1)$^c$\\
140304A	&	$4.90 - 37.0$	& $1.00^{+0.37}_{-0.91}$ & $0.64_{-0.06}^{+0.06}$ & (1,2)\\
141121A	&	$3 - 15$	& $1.26^{+0.37}_{-0.52}$ & $0.70_{-0.09}^{+0.10}$ & (1)$^c$\\
160509A & $1.644 - 30.0$ & $0.76^{+0.27}_{-0.16}$ & $0.79_{-0.09}^{+0.15}$ & (1)$^c$\\
160625B	&	$1.77 - 11.0$	& $1.38^{+0.17}_{-0.30}$ & $0.65_{-0.06}^{+0.07}$ & (1)$^c$\\
170817A &	$2.6 - 9.0$	& $0.20^{+0.09}_{-0.24}$ & $0.18_{-0.21}^{+0.21}$ & (4)\\
171010A	&	$4.5 - 15.5$	& $1.56^{+0.55}_{-0.44}$ & $0.90_{-0.04}^{+0.04}$ & (1)\\
\enddata
\tablecomments{\\
$^a$ (1) Transrelativistic afterglows under a $k\sim0$ ISM environment. (2) Transrelativistic afterglows under a $k\sim 2$ wind environment. (3) Ultrarelativistic afterglows with larger $k$. (4) Afterglows produced by expanding jets with larger $k$. (5) Hard to explain, unless taking larger observational errors into consideration.\\
$^b$ Show signs of transrelativistic break; both $a$ and $b$ become steeper at later times.\\
$^c$ Possible reverse-shock contamination due to early peak time and/or large $E_{\rm iso}$.\\
$^d$ Large fitting errors for $b$ due to large observation errors of 15 GHz data; fitted as $b = 0.18_{-0.10}^{+0.10}$ without 15 GHz flux.\\
$^e$ The 1.43 GHz peak flux as shown in \cite{Soderberg2006} exhibits large errors comparable with the flux level. The fitting results without the 1.43 GHz data point is $a = 1.72^{+2.00}_{-0.55}$, $b = 0.58_{-0.43}^{0.42}$, which are still compatible with the transrelativistic afterglows in the ISM explanation.\\
$^f$ Fitted with the 1.4 GHz data, which exhibits large measured errors ($S_{\nu} = 373 \pm 170$ $\mu$Jy). The fitting result should be $a = 0.9^{+0.48}_{-0.35}$, $b = 0.36_{-0.02}^{+0.02}$ without 1.43 GHz data, which is compatible with explanations (1, 2, 4).}
\end{deluxetable*}

\subsubsection{\textbf{GRBs with ISM Density Profiles}}

According to our method, a total number of 14 GRBs out of the 32 samples can be solely explained with a $k=0$ ISM environment. As listed in Table \ref{table:peak}, we find that the values of $a$ and $b$ for multiband radio afterglow peaks of GRBs 980329, 980703, 000301C, 020903, 031203, 070125, 100418A, and 171010A lie between the ultra- and transrelativistic cases, and thus can be explained with transrelativistic afterglows under an ISM environment. And as can be seen in Figs \ref{fig:tpeak_nu} and \ref{fig:fpeak_nu}, although due to observational uncertainties, the time-frequency index $a$ shows large fitting errors for GRBs980329, 020903, and 100418A, which may lead to multiple possible $k$ values, the flux-$\nu$ index $b$ can be constrained stringently enough. In fact, as noted in Section \ref{subsec:trans-relativistic}, considering the late occurrence of the transrelativistic phase for $k=1.1$ or 2 cases, a $b\sim 1$ value alone can still lead to a confident conclusion of $k=0$, no matter of the $a$ value. Besides, tt is worth noting that the radio peaks of GRBs 020903 and 070125 show ``breaks'' of $a$ and $b$ values at a certain frequency, which could imply a transition from ultra- to transrelativistic phases, as the high-frequency peaks could appear during the ultrarelativistic phase with flatter $a$ and $b$ values, while the low-frequency ones occur later with steeper transrelativistic indices.

Compared with existing works using multiband modeling, it should be noted that \cite{Berger2000}, \cite{Soderberg2004b}, and \cite{Chandra2008} have drawn similar conclusions for GRBs 000301C, 031203, and 070125, respectively, that is, all these three GRBs occur in ISM environments. And \cite{Starling2008} found a possible ISM explanation for GRB 980329 and demonstrated that GRB 980703 can be explained by both $k=0$ and $2$ environments. \cite{Frail2003} also showed that the wind model cannot be ruled out (although not favored) to fit the light curve of GRB 980703, which is still compatible with our constraints. An exception is GRB 171010A, which has been explained by \cite{Bright2019} using a steep density profile, although some hard-to-explain features do exist. This conclusion is somewhat inconsistent with our constraints, which could be due to the relatively large fitting errors for the $b$ index, thus leaving large room for alternative explanations. 

However, although the rest six GRBs in this category, including GRBs 050820A, 051022, 130427A, 141121A, 160509A, and 160625B generally follow predictions of the $k=0$ ISM environment, \citealt{Cenko2006}, \citealt{Rol2007}, \cite{Perley2014}, \cite{Cucchiara2015}, along with \cite{Alexander2017} all pointed out a possible reverse-shock contribution of the radio afterglow peaks in these bursts. With the existence of reverse-shock contamination, the multiband peak method of analysis may no longer apply, thus the reliability of our related results could be compromized.

\subsubsection{GRBs Compatible with a Stellar-Wind Environment}

GRBs occurring in $k=2$ density profiles as implied by multiband radio peaks include GRBs 980425, 000418, 011121, 030329, and 100814A, along with the only short GRB in our sample, GRB 170817A. Among these, the behavior of the GRBs 980425 and 030329 light-curve peaks follows predictions for late-time afterglow ($\nu_a = \nu_{a,>}$) in the wind medium. And \cite{Berger2003} and \cite{vanderHorst2005} have fitted the afterglow light curves of GRB 030329 with both ISM and wind models, whose results are compatible with our conclusions.

Meanwhile, the $b$ indices of GRBs 000418 and 031203 are relatively shallow and are consistent with expanding jets in the ISM environment. However, their $a$ indices are too small to be explained with jets. It also should be noted that for several bursts, the medium- and low-frequency radio peaks may be explained with jet breaks, although the details depend much on the jet-opening angles (hence the jet break times). Analysis with multiband radio peaks may only hint at multiple explanations, and data from other bands (e.g., optical) should be utilized to clarify the existence of jets. Besides, it is quite possible that the speed of the jet's sideways expansion during the transrelativistic phase should be slowed down, making the results drawn in Section \ref{subsec:jet} no longer stand. 

The value of $a$ index of GRB 011121 is consistent with the existence of a jet, while the $b$ index falls within predictions for the $k=2$ case. Considering the simultaneous appearance of the multiband peak and the large errors in peak-time estimations that can be attributed to observational effects, a burst occurring in a wind density profile is a more suitable conclusion, which is supported by \cite{Price2002c}. A similar case applies for GRB 100814A, for which \cite{dePasquale2012} has pinned down to the existence of a jet.

For the gravitational-wave-associated GRB 170817A, the only short burst in our sample, its multiband radio afterglow exhibits a shallow $a$ with a shallow $b$. Such a case is more consistence with relativistic jet, which is compatible with the off-axis jet identified by previous works (e.g., see \citealt{Fong2019}).

\subsubsection{GRBs Explained by \textit{k} between 0 and 2}

This category consists of 10 samples, which are GRBs 970508, 970828, 990510, 991208, 000926, 010921, 021004, 060218, 111215, and 140304A. The uncertainties here mainly arise from the relatively large errors for peak-time/flux estimates. For example, the peak behavior of GRBs 970508, 990510, and 010921 lies somewhere in between the $k=0$ trans-relativistic and $k=2$ late time ($\nu_a = \nu_{a,>}$) cases, when taking fitting errors into consideration, making any $0\leqslant k \leqslant 2$ possible. This is compatible with the conclusion drawn by \cite{Starling2008}, who found a homogeneous circumburst medium for GRBs 970508 and 990510 with X-ray, optical, and IR afterglows. However, the exact $k$ for each sample is hard to pin down precisely, due to large uncertainties in $a$ and $b$.

For GRBs with high circumburst densities, including GRBs 991208, 000926, 021004, 060218, and 111215A, the indices of $a$ and $b$ can be explained with later transrelativistic shocks in the $k\sim 0$ environment. However, because a larger $A$ can lead to a later $t_{cm}$ (thus a later peak time $t_m$ in radio afterglow), the larger $k$ possibility, which can produce a steeper $b$ during the relativistic phase, cannot be excluded for these bursts. When compared with existing works, \cite{Harrison2001} showed that the temporal behavior of GRB 000926's light curve can be better modeled with a Comptonized uniform medium; \cite{Soderberg2006} fitted GRB 060218 with both $k=2$ and $k=0$ density profiles, as well as low an $E_{\rm iso}$; and \cite{Zauderer2013} has pointed out that the circumburst medium of GRB 111215A should be in the form of stellar wind, which are all consistent with our analysis. 

For GRBs 970828 and 140304A, although the $a$ and $b$ indices can be explained with both $k=0$ and $2$ possibilities when taking fitting errors into consideration, it should be noted that the afterglows of these bursts were significantly affected by reverse shock, as shown by \cite{Djorgovski2001} and \cite{Laskar2018}. And the radio light curve of GRB 140304A might even be contaminated by interstellar scintillation \citep{Laskar2018}. All of these factors can complicate our analysis, and thus compromise the robustness of the results.

\subsubsection{GRBs Hard to Explain with Multiband Light-curve Peaks}

Finally, we have two outliers that cannot be well constrained by the multiband peak method. One of them is GRB 000911, whose $a$ and $b$ values are much steeper than theoretical predictions for both trans- or ultra- relativistic shock waves, thus making its afterglow behavior hard to explain, even considering the fitting errors. Another hard-to-explain case is GRB 071003, which is located near the Galactic plane, with its radio data severely affected by scintillation \citep{Perley2008}, thus leaving a large room for errors in the $a$ and $b$ indices, therefore making multiple explanations, including both relativistic or transrelativistic ejecta propagating in the $k=0$ and $k=2$ circumburst environment, possible.

\section{Summary and Discussion} \label{sec:summary}

In this paper, we investigated the radio afterglows of gamma-ray bursts occurring in a power-law-distributed circumburst medium with $n\propto R^{-k}$, and an arbitrary $k$. We show that one can use multiband radio afterglow peak time $t_{{\rm peak},\nu}$ and $F_{{\rm peak},\nu}$ data to put a constraint on the density-distribution index $k$. We find that in the relativistic phase of the afterglow evolution, the peak time $t_{{\rm peak},\nu}$ corresponds to $t_m$, that is, the time when the minimum frequency of electrons equals to the observing frequency. And we have $t_{{\rm peak},\nu}\propto \nu^{-a}$, $a \sim 2/3$, which is independent of the shock radiation efficiency, while for $F_{{\rm peak},\nu} \propto \nu^{b}$, the value of $b$ depends more strongly on $k$. For adiabatic shocks, $b$ increases from $0$ to $1/3$, if we change $k$ from $0$ to $2$. And in the transrelativistic phase, similar dependencies between peak time/flux and frequency exist, with steeper $a$ and $b$ values than in the ultrarelativistic phase.

\begin{figure*}
\gridline{\fig{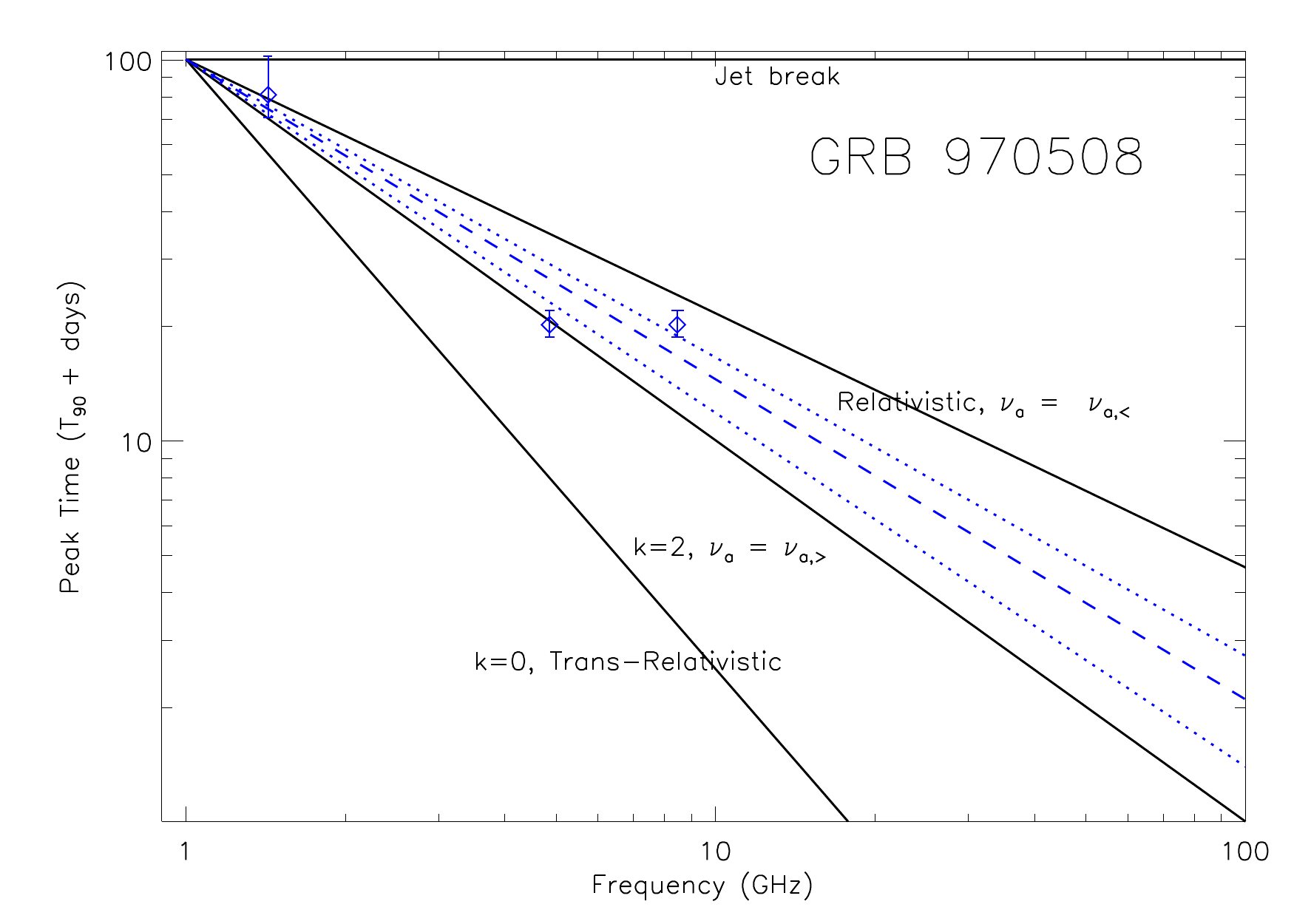}{0.31\textwidth}{}
          \fig{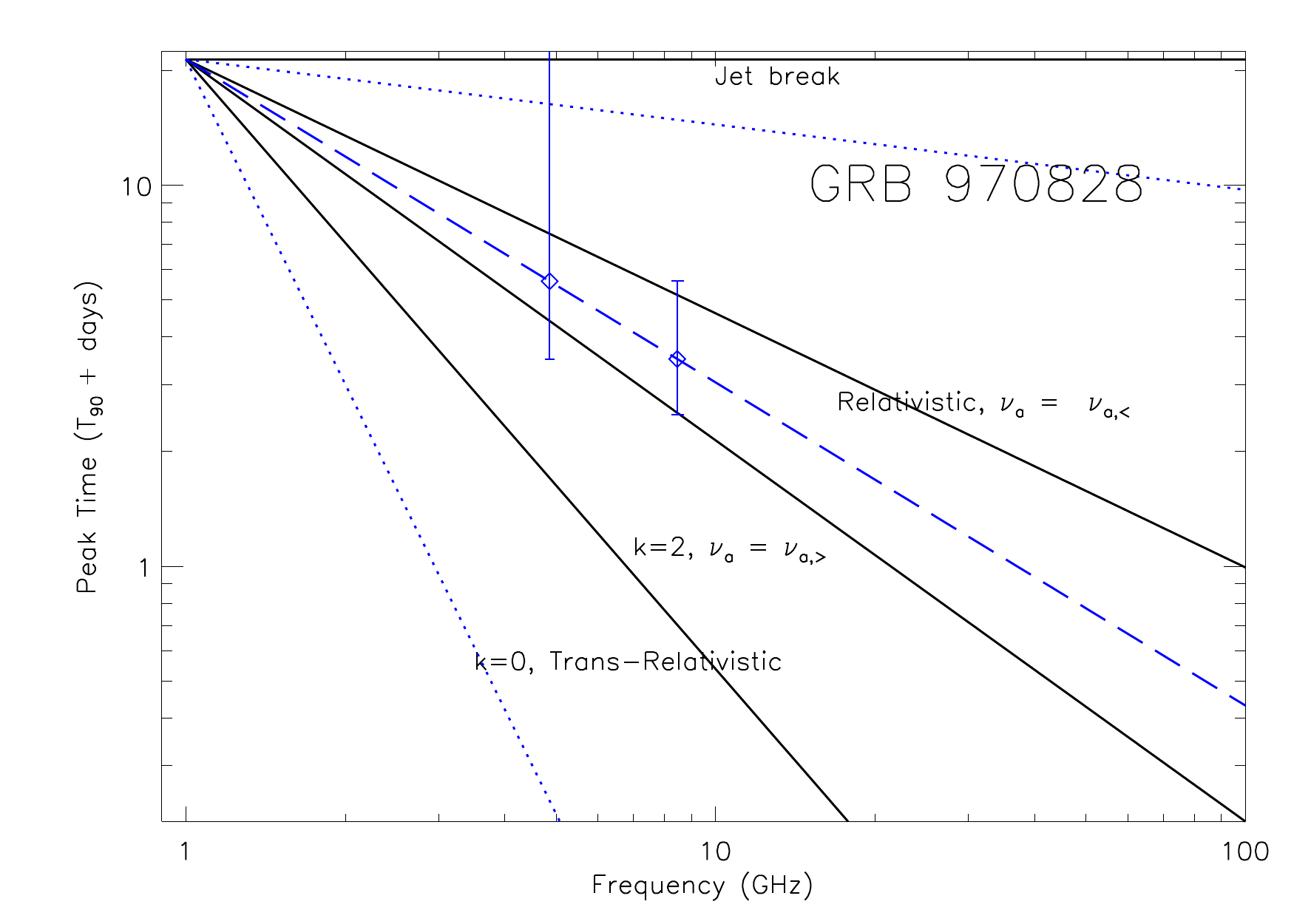}{0.31\textwidth}{}
          \fig{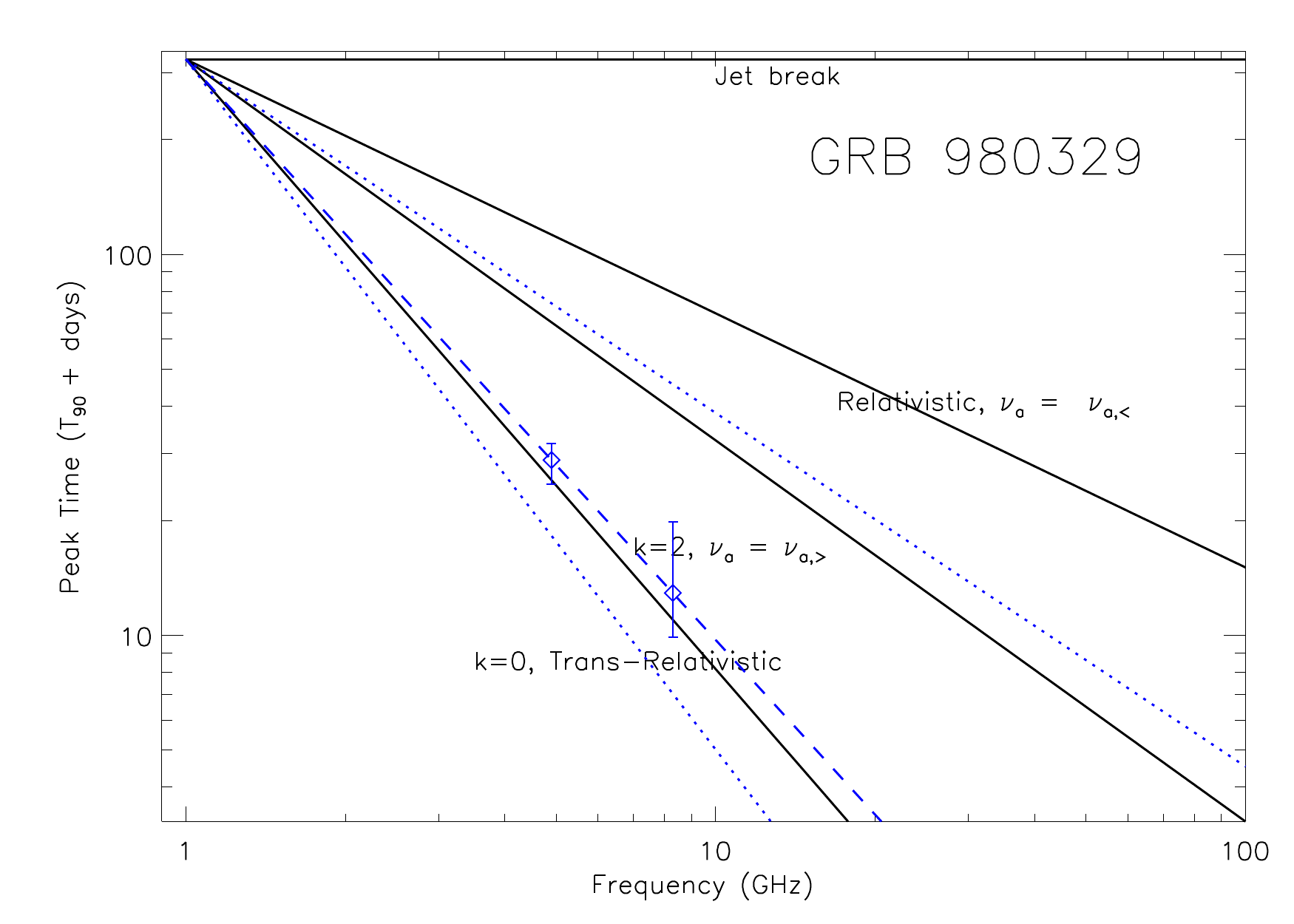}{0.31\textwidth}{}
          }
\gridline{\fig{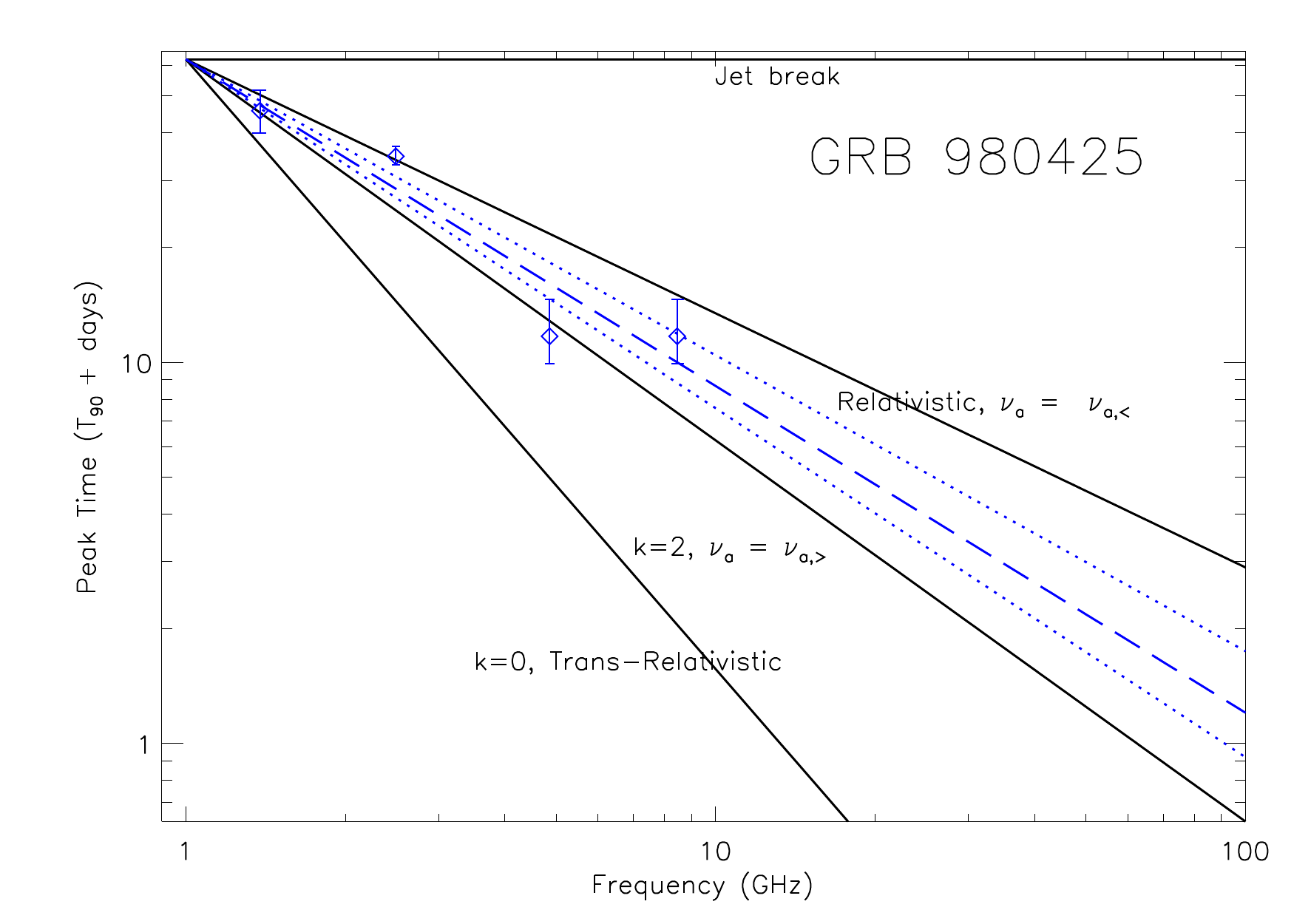}{0.31\textwidth}{}
          \fig{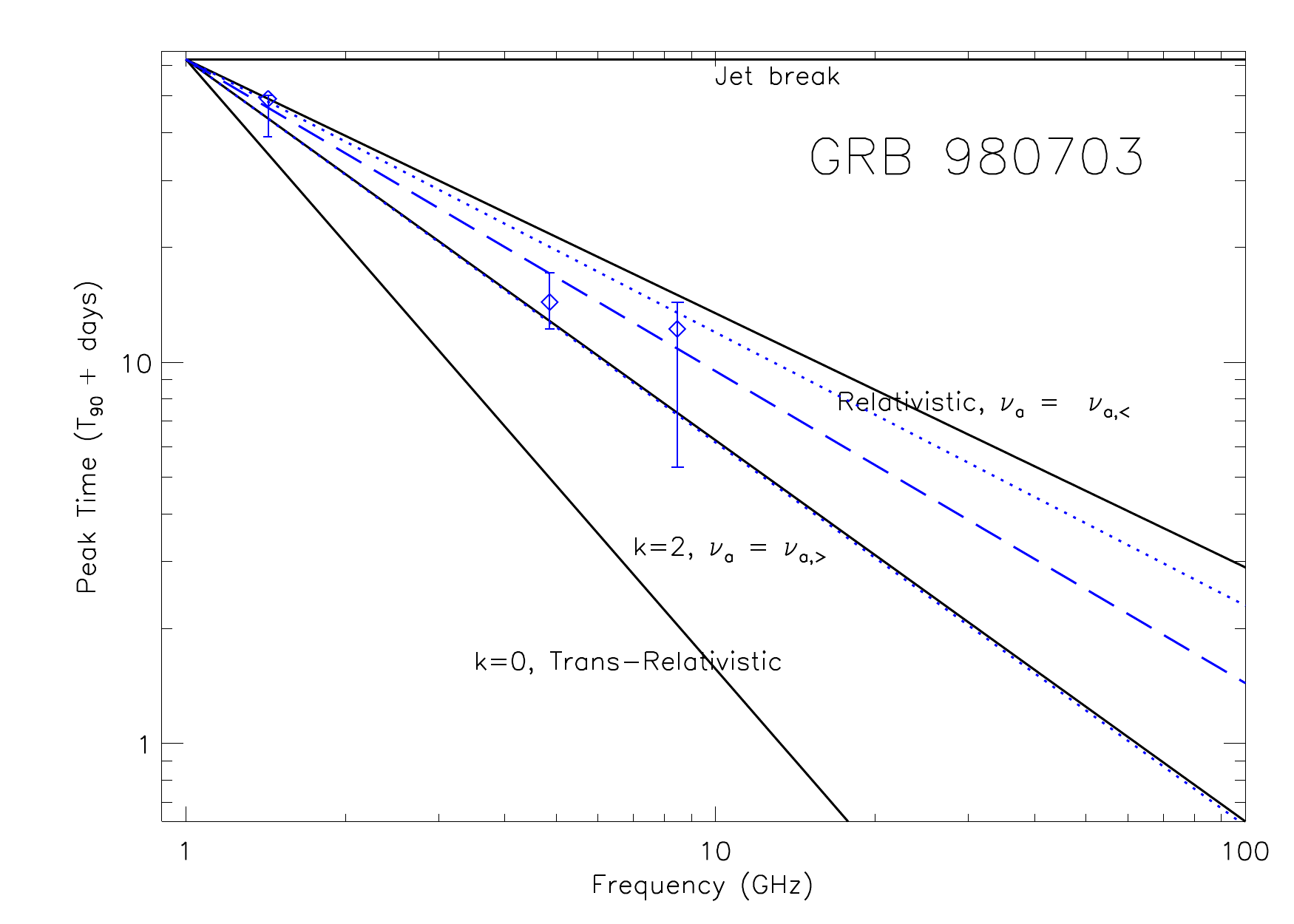}{0.31\textwidth}{}
          \fig{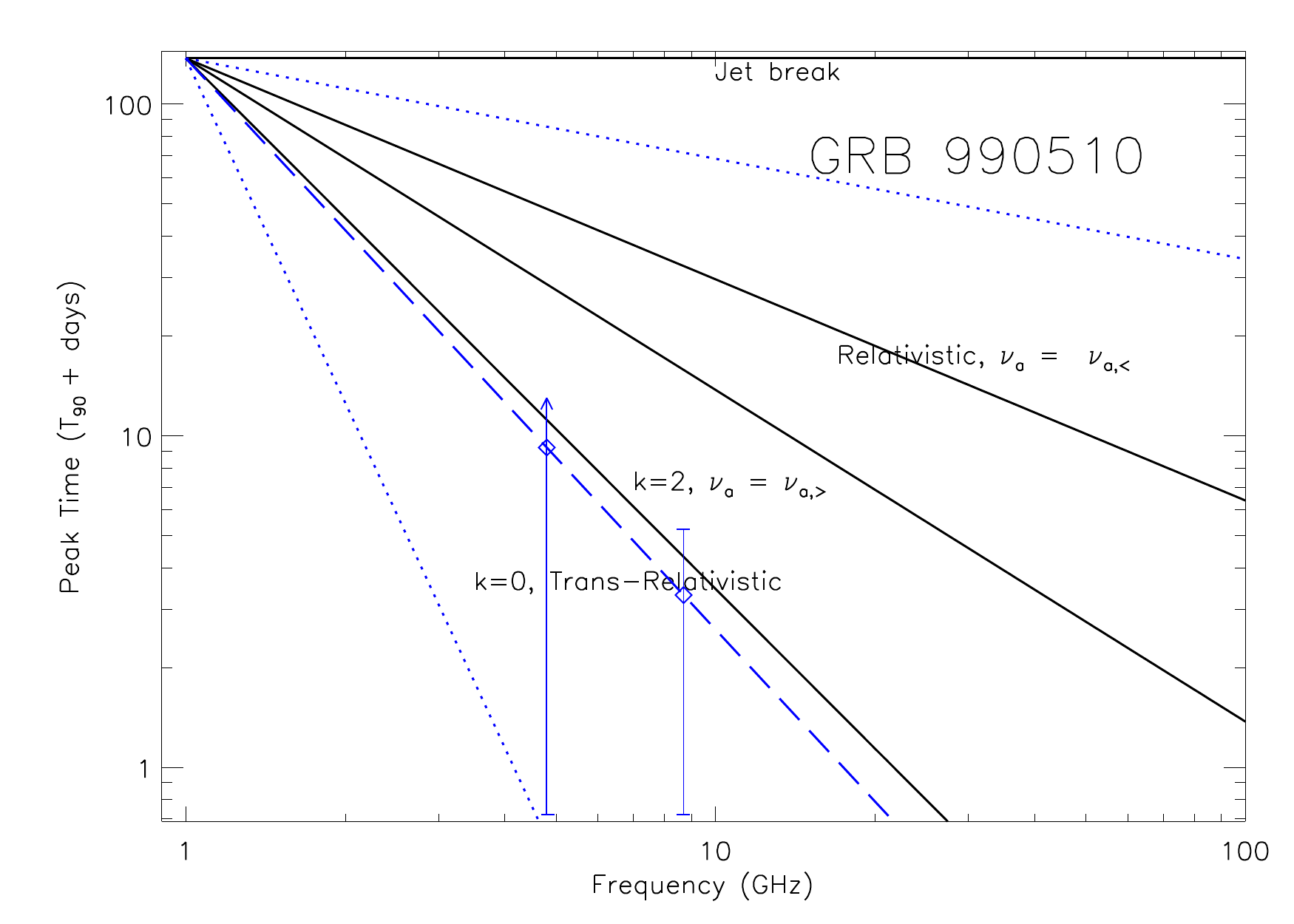}{0.31\textwidth}{}
          }
\gridline{\fig{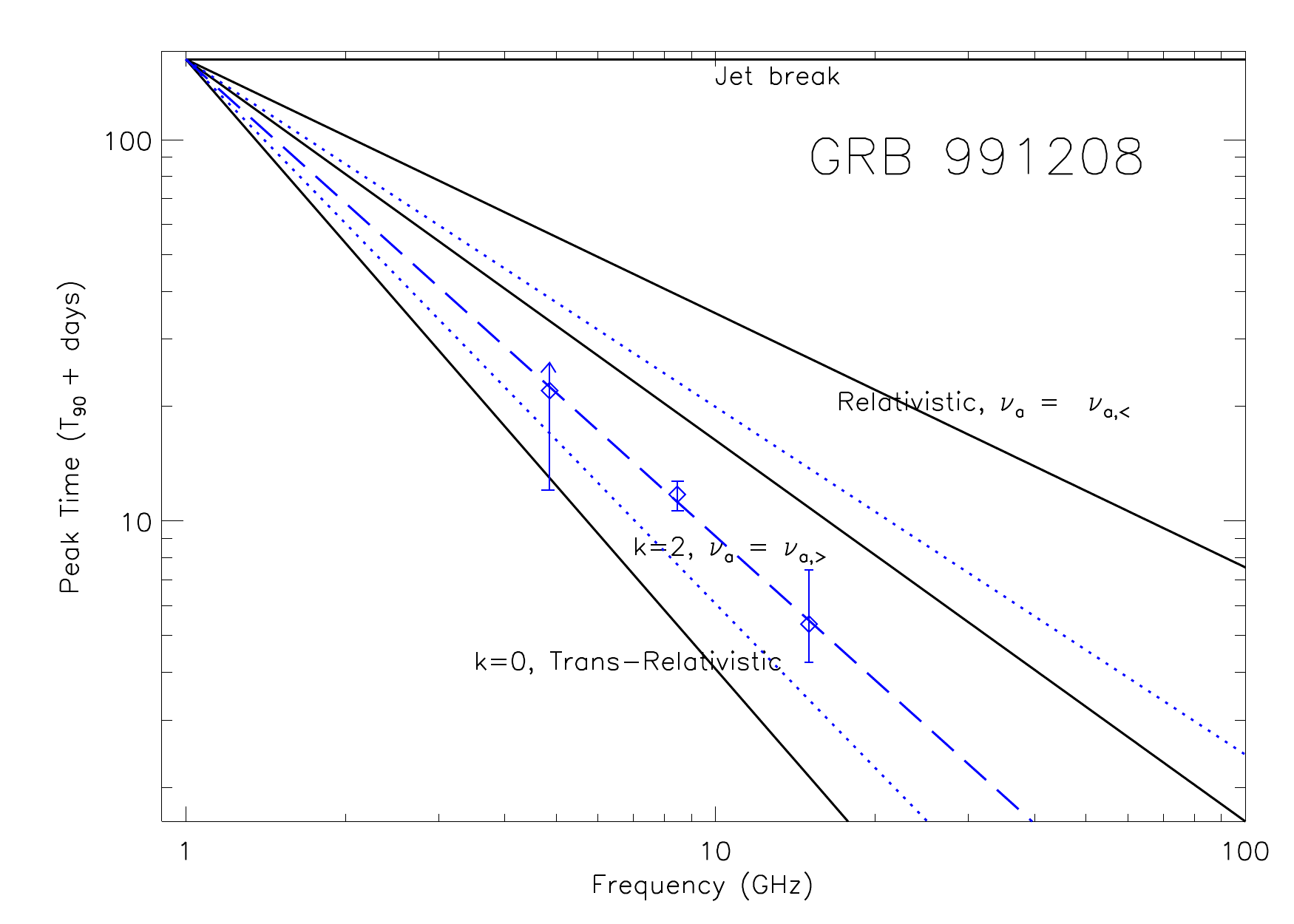}{0.31\textwidth}{}
          \fig{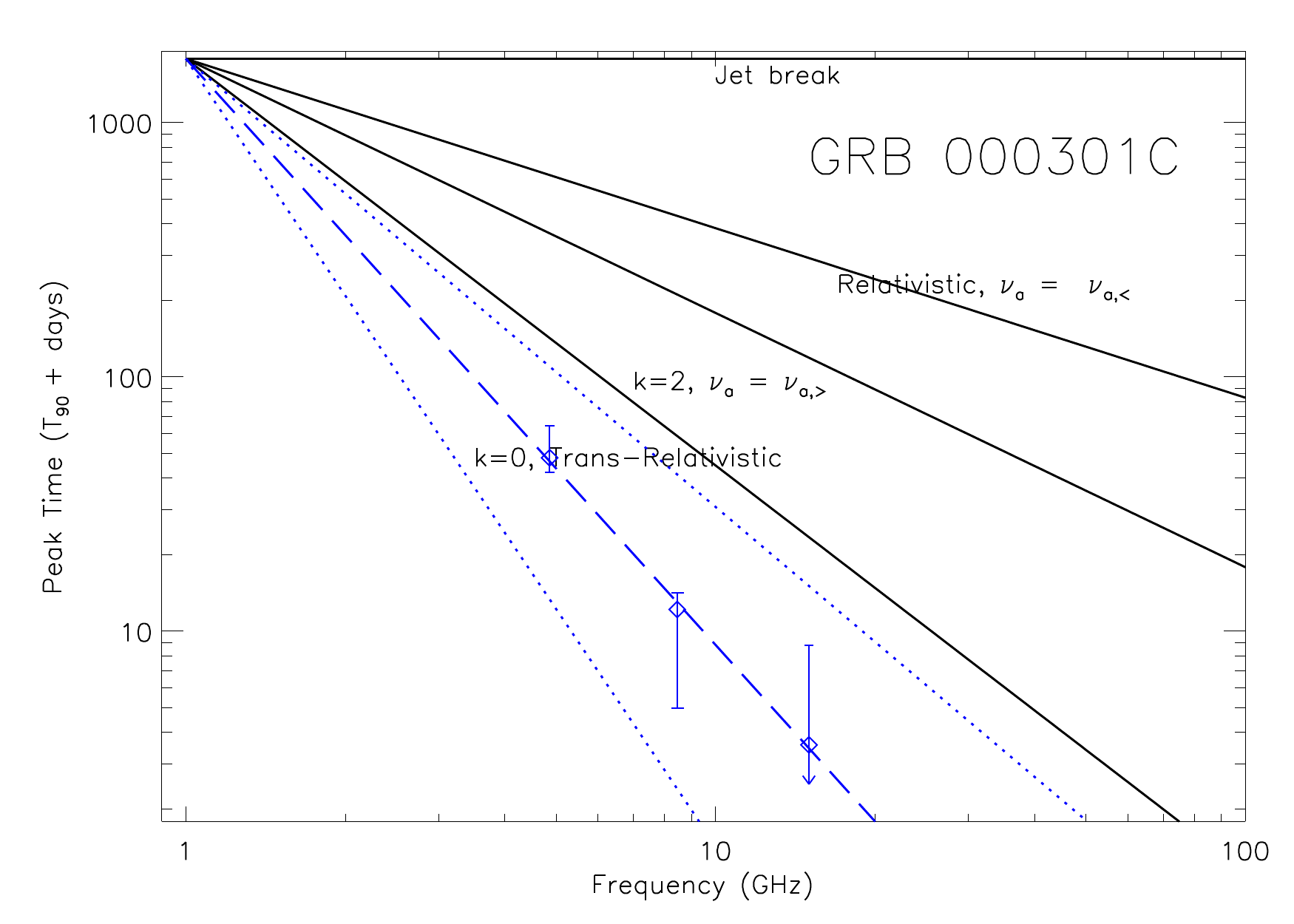}{0.31\textwidth}{}
          \fig{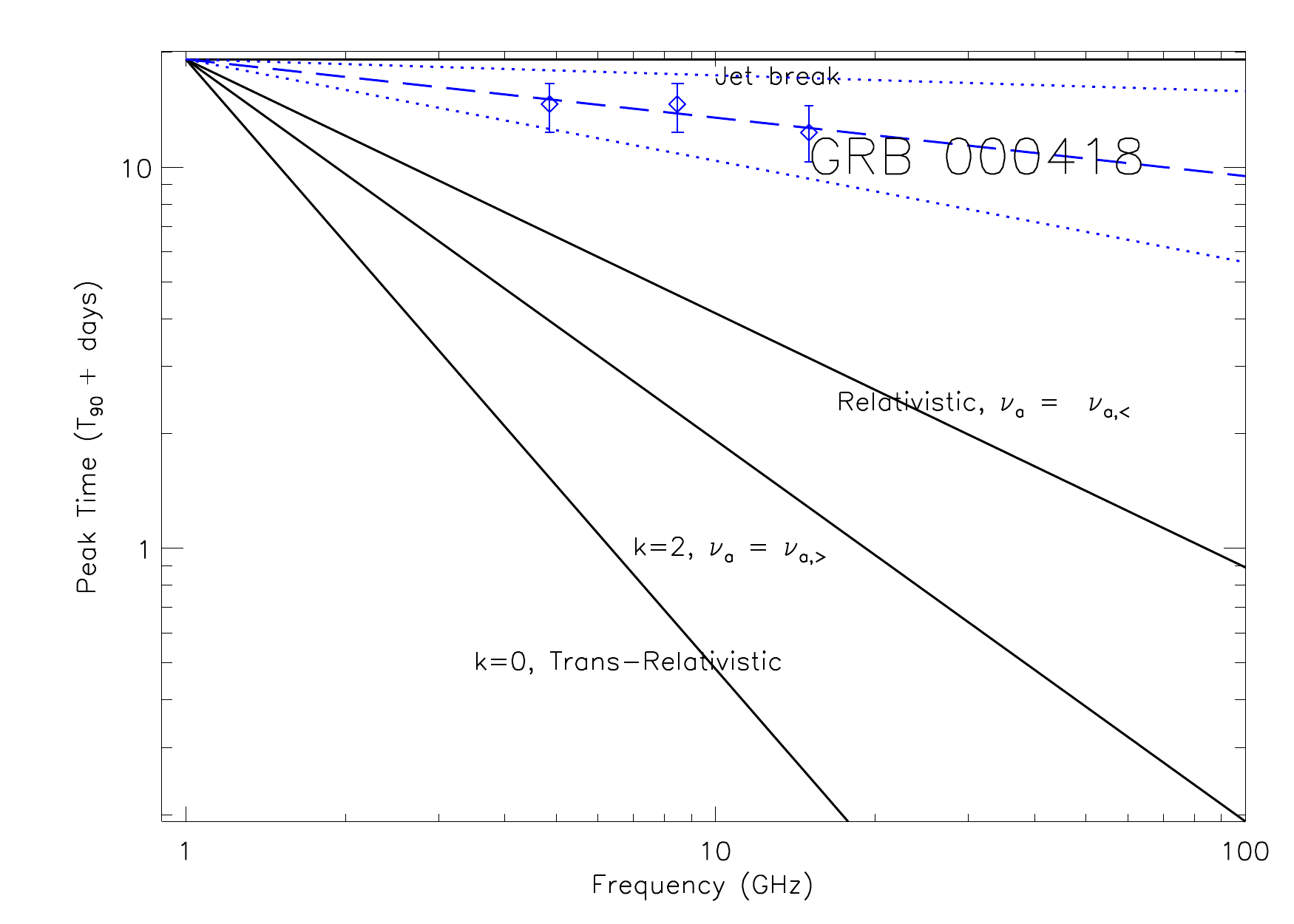}{0.31\textwidth}{}
          }
\gridline{\fig{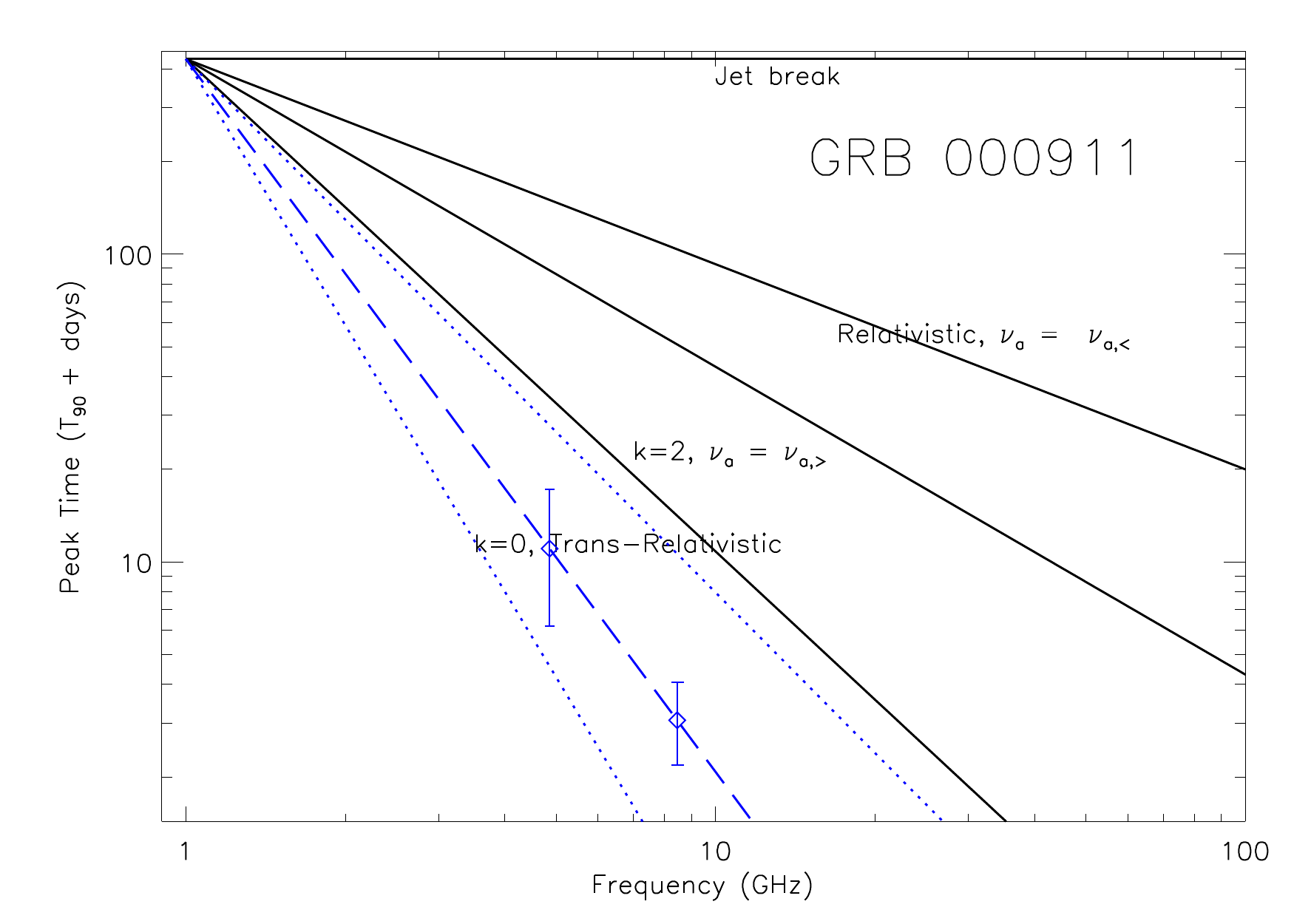}{0.31\textwidth}{}
          \fig{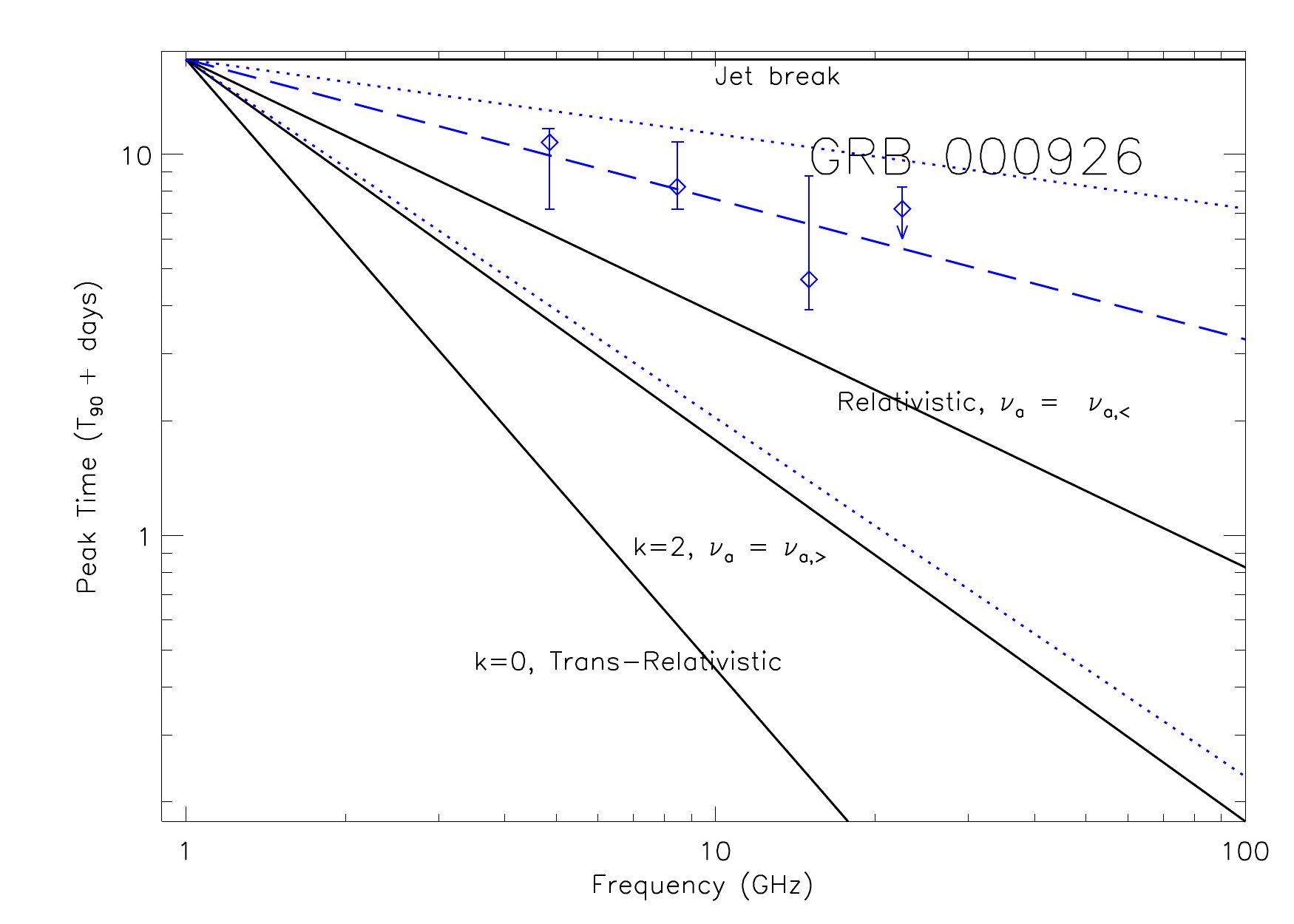}{0.31\textwidth}{}
          \fig{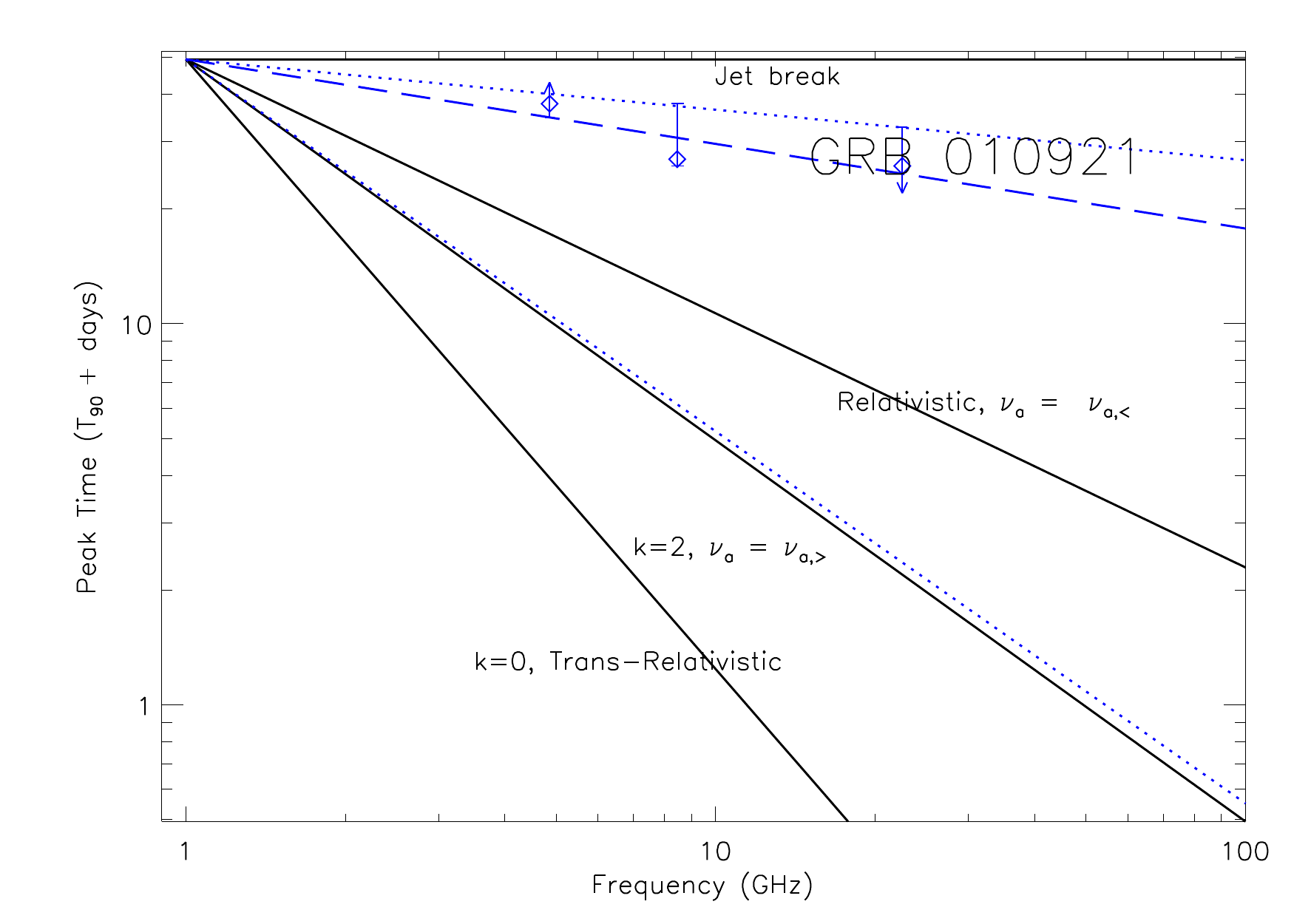}{0.31\textwidth}{}
          }
\caption{\small{The relations between radio afterglow peak fluxes $t_{{\rm peak},\nu}$ and observing frequencies $\nu/\nu_{{\rm min}}$ for our sample GRBs. The dashed line with data points in each figure shows the fitting results from observations, the blue dotted lines denote the possible range of fitting errors, while the black solid lines from top to bottom represent the theoretical predictions for relativistic jet break, relativistic spherical shock wave, $k=2$ late time ($\nu_a = \nu_{a,>}$), and $k=0$ transrelativistic cases, respectively.}}\label{fig:tpeak_nu}
%\end{center}
\end{figure*}

\clearpage
\gridline{\fig{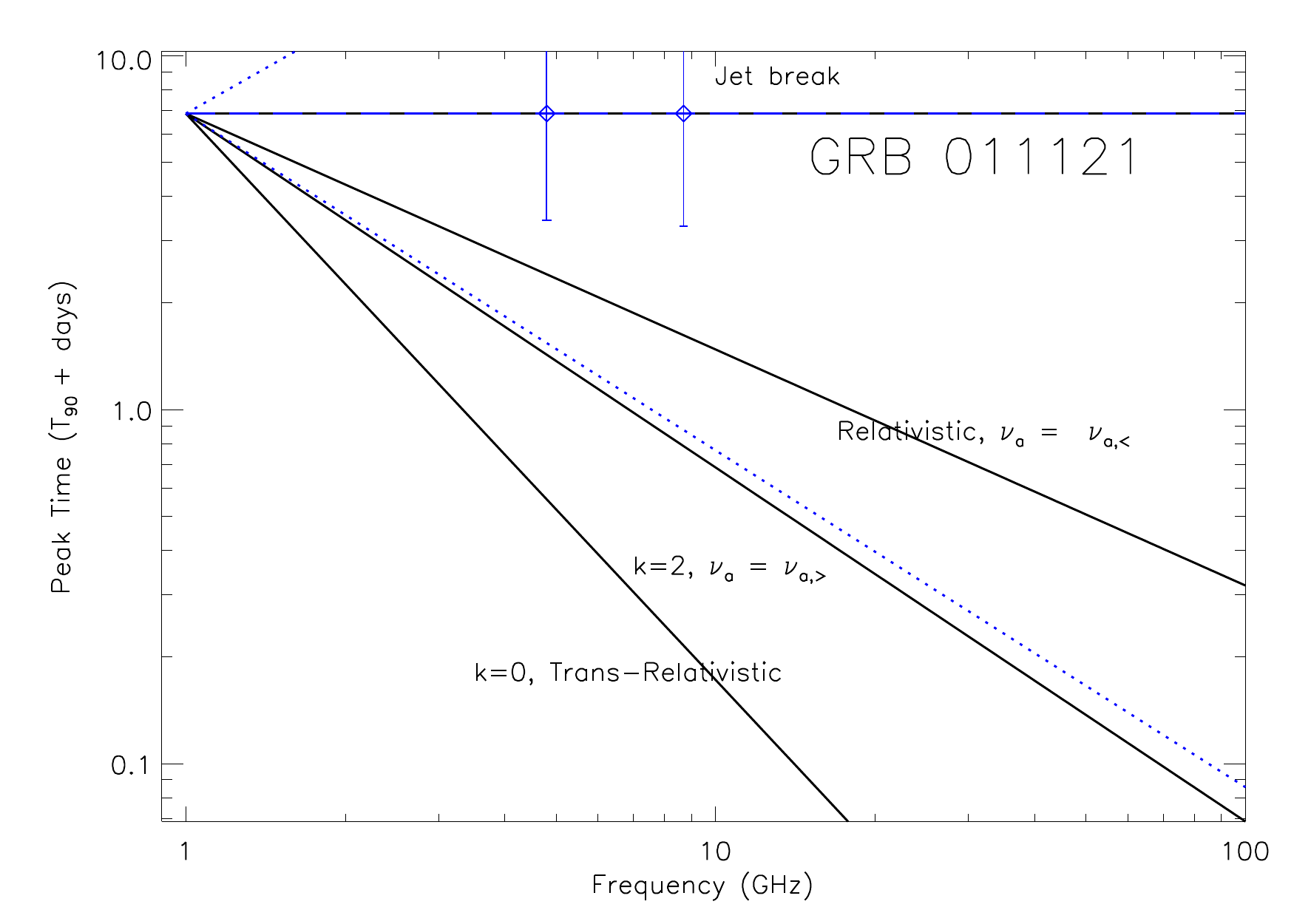}{0.31\textwidth}{}
          \fig{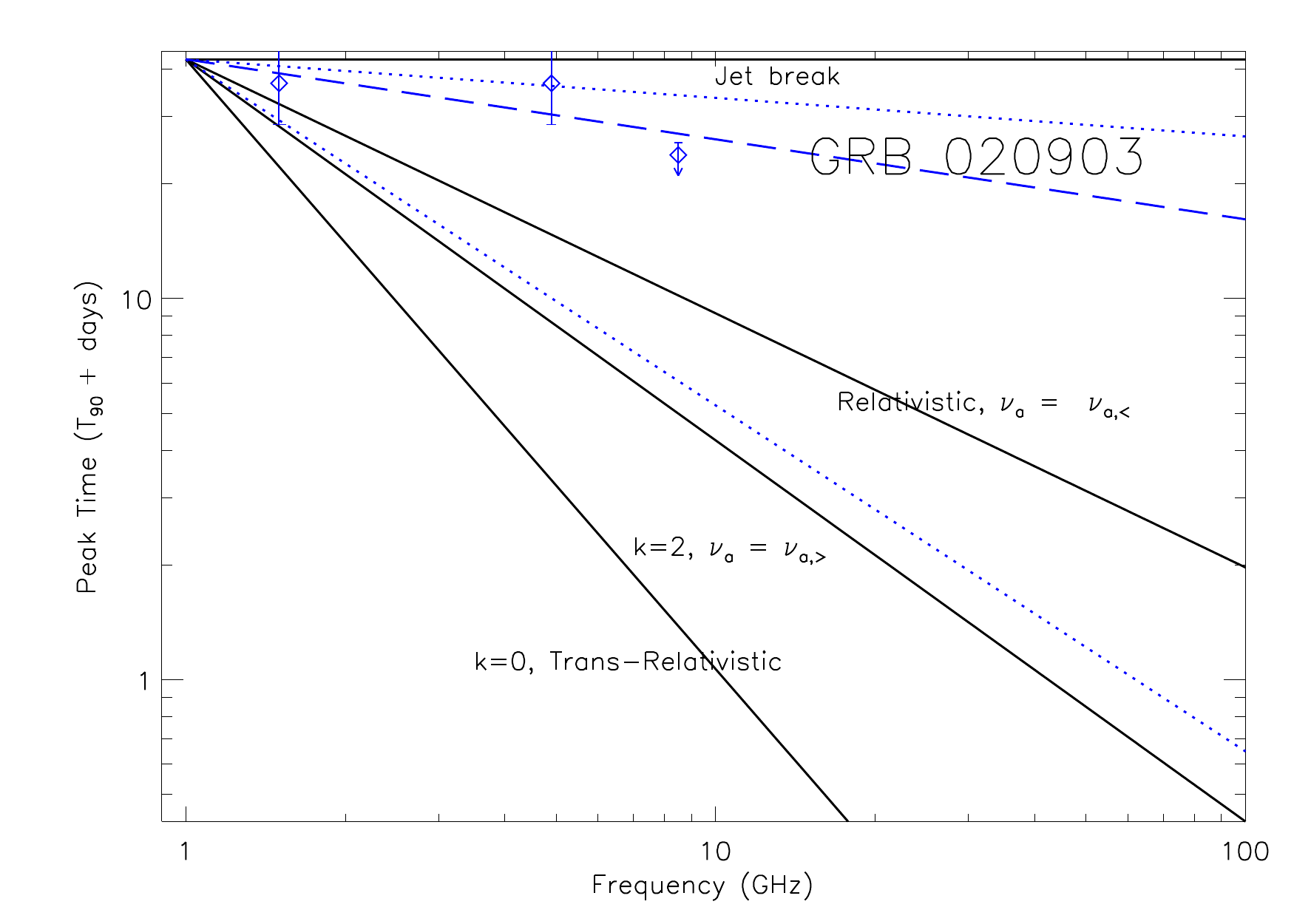}{0.31\textwidth}{}
          \fig{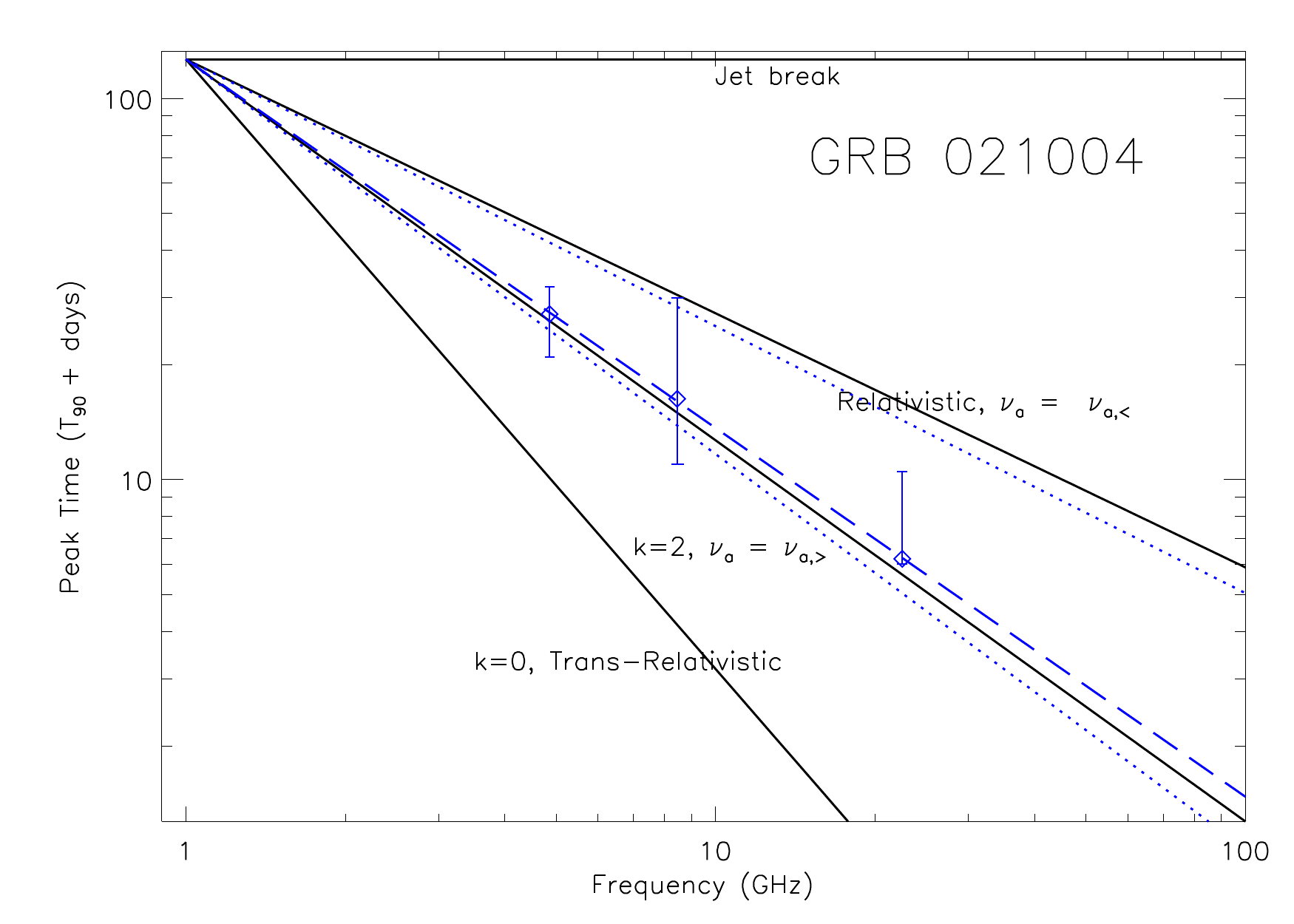}{0.31\textwidth}{}
          }
\gridline{\fig{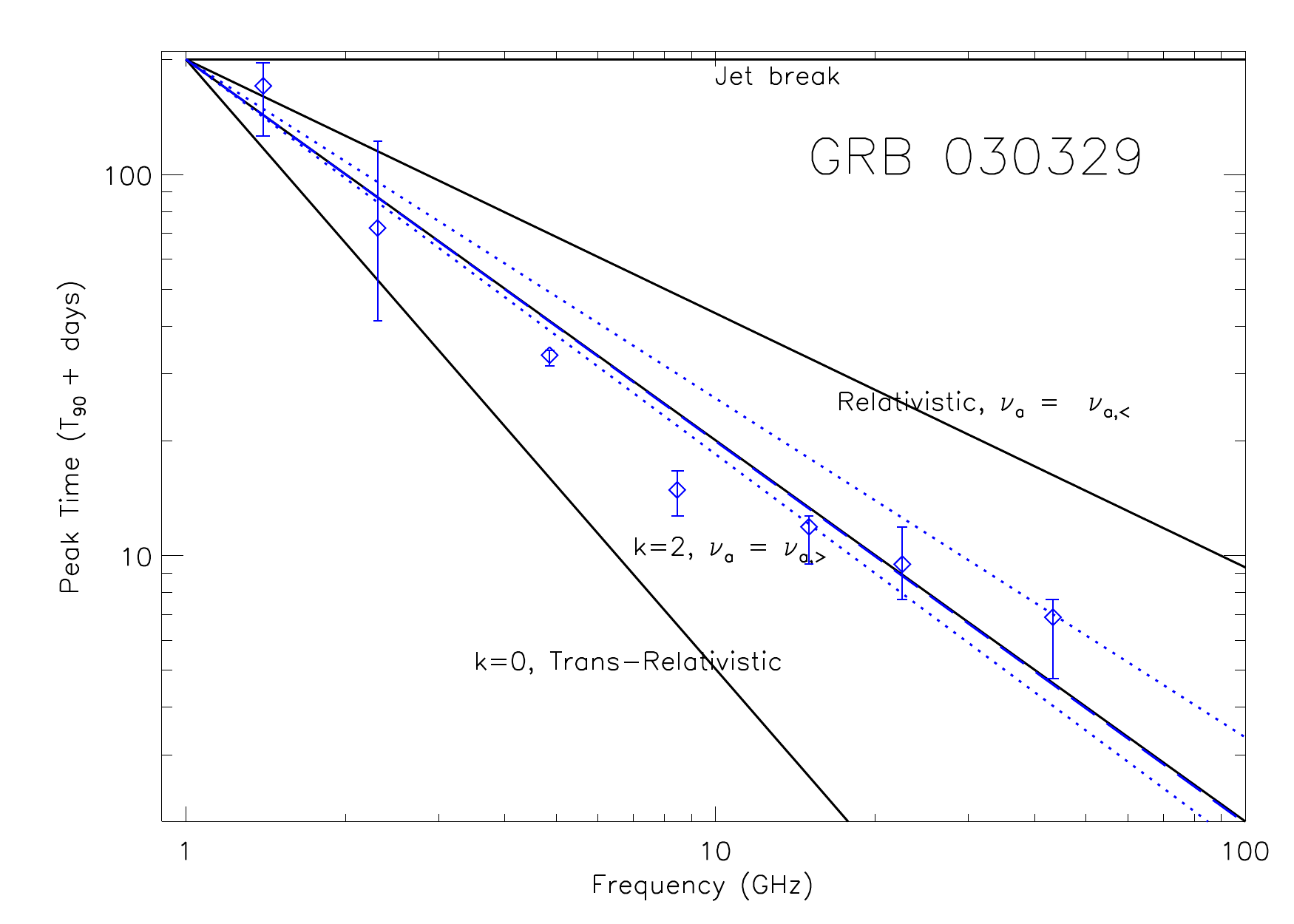}{0.31\textwidth}{}
          \fig{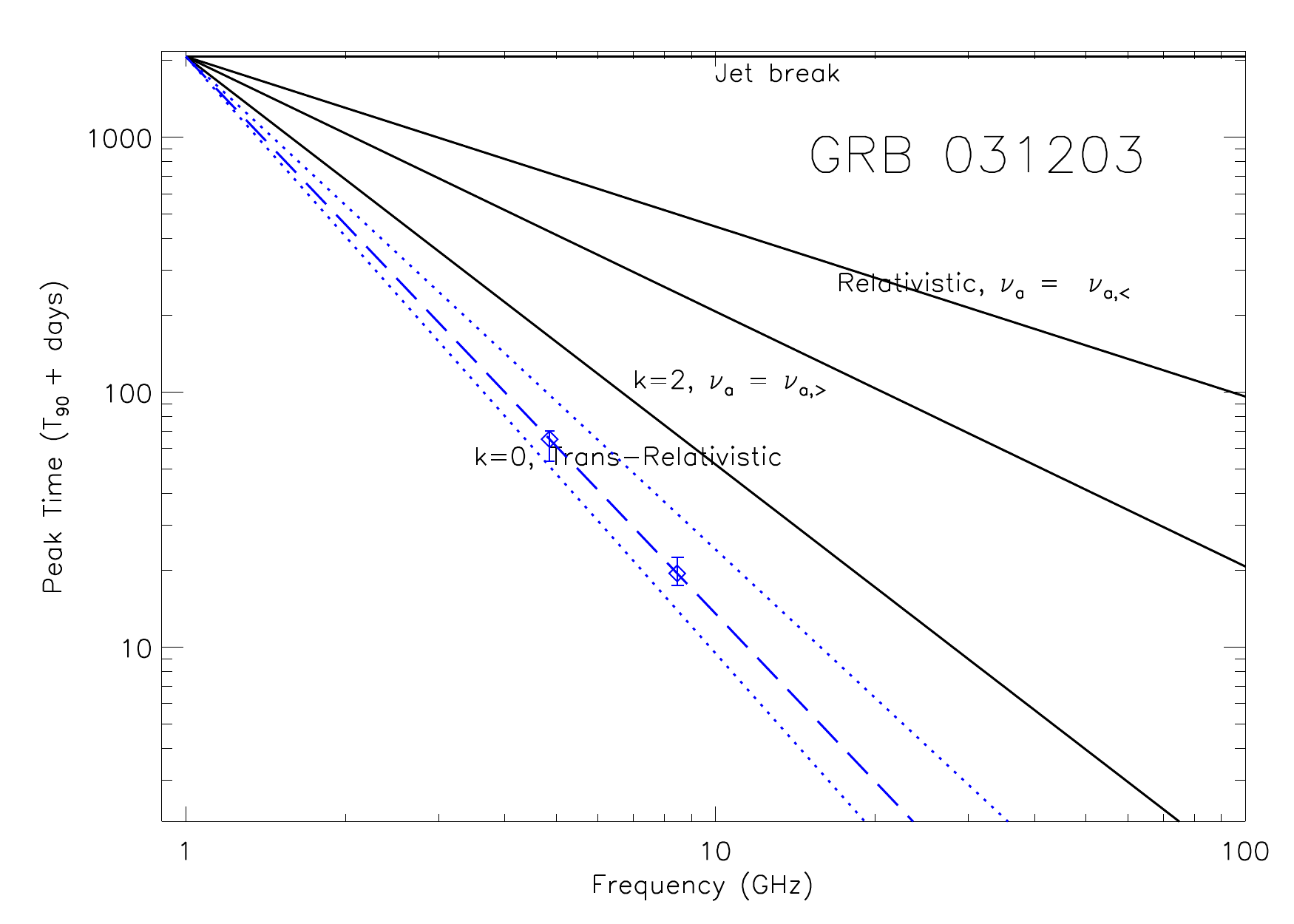}{0.31\textwidth}{}
          \fig{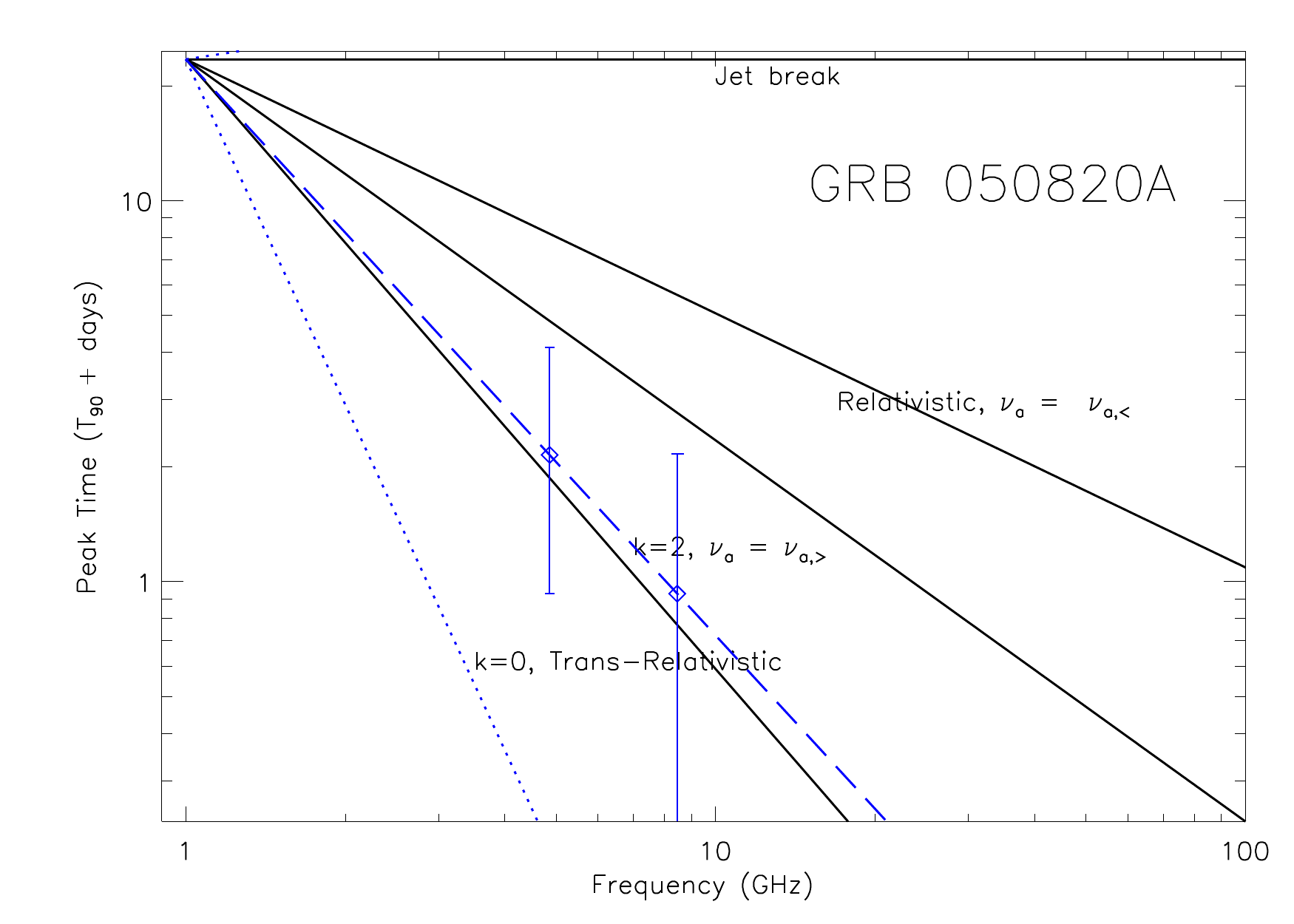}{0.31\textwidth}{}
          }
\gridline{
          \fig{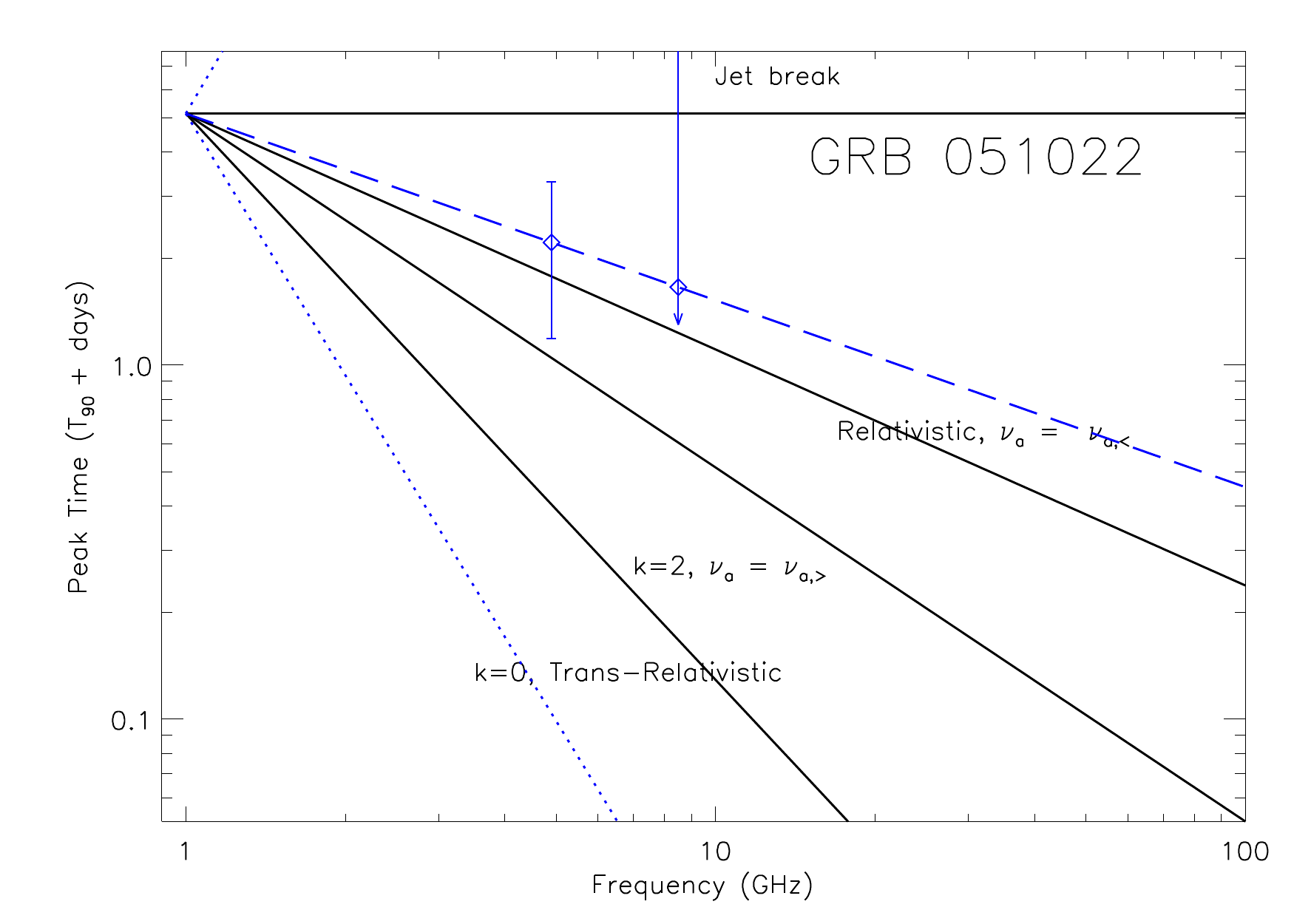}{0.31\textwidth}{}
          \fig{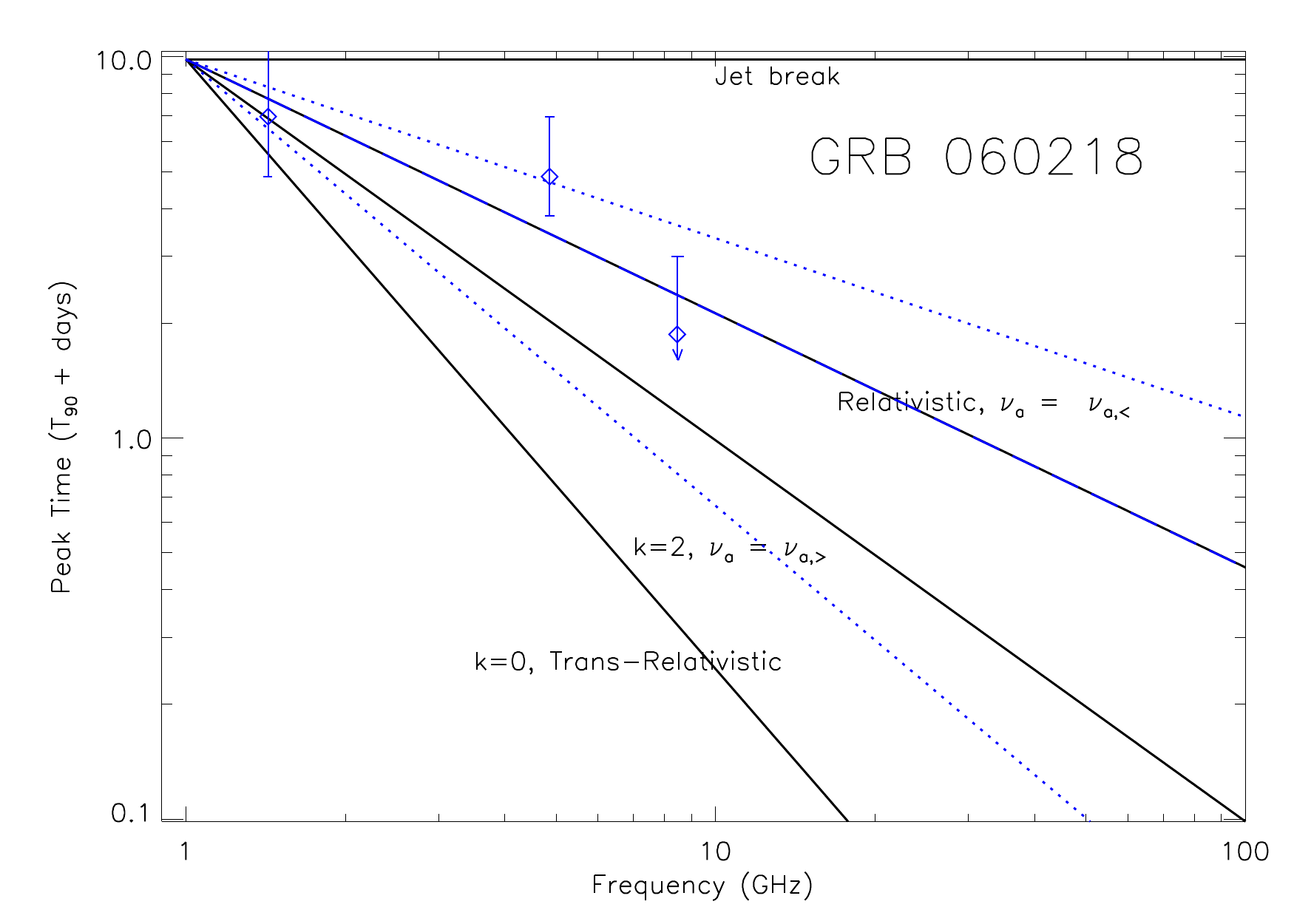}{0.31\textwidth}{}
          \fig{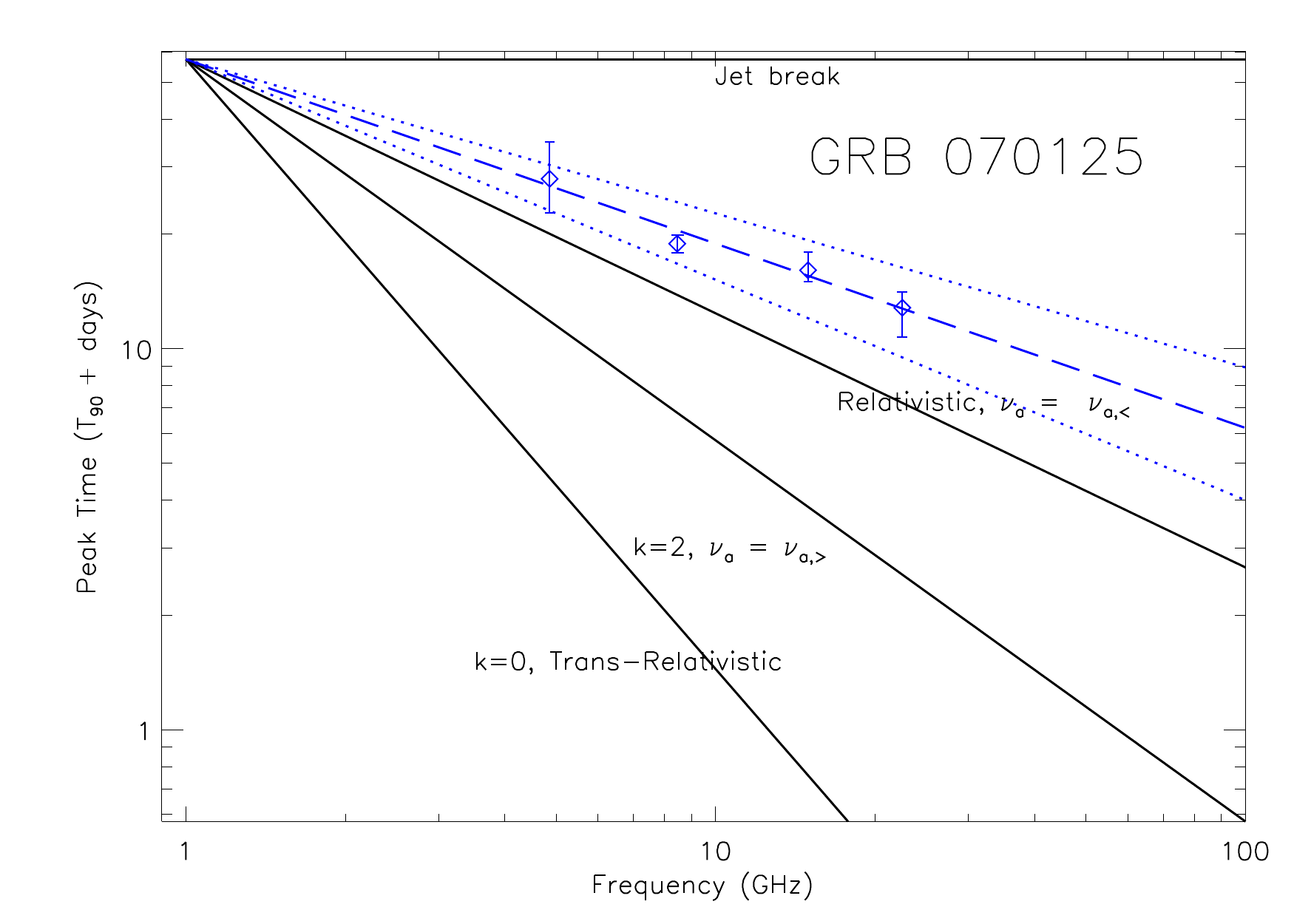}{0.31\textwidth}{}
          }
\gridline{\fig{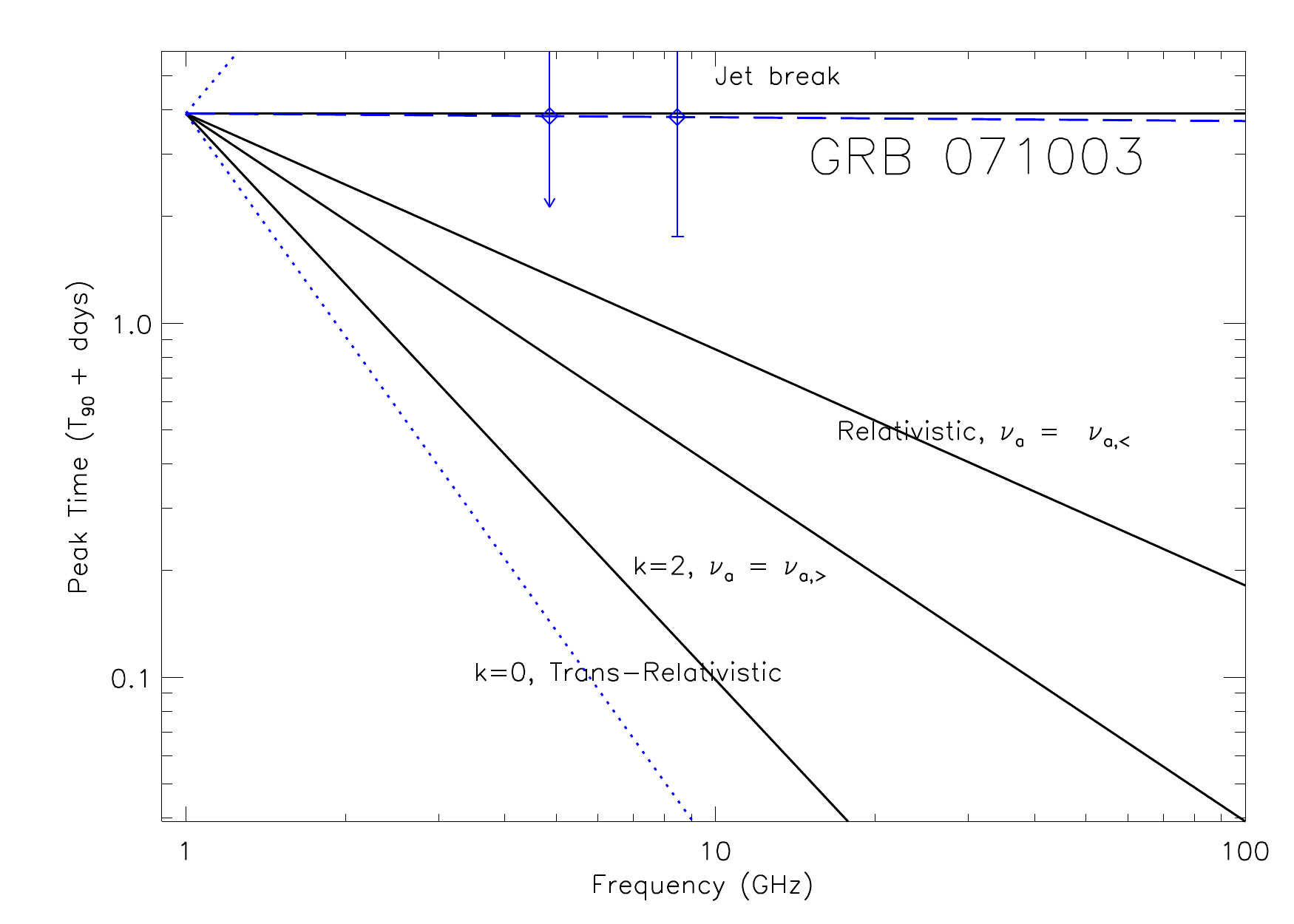}{0.31\textwidth}{}
          \fig{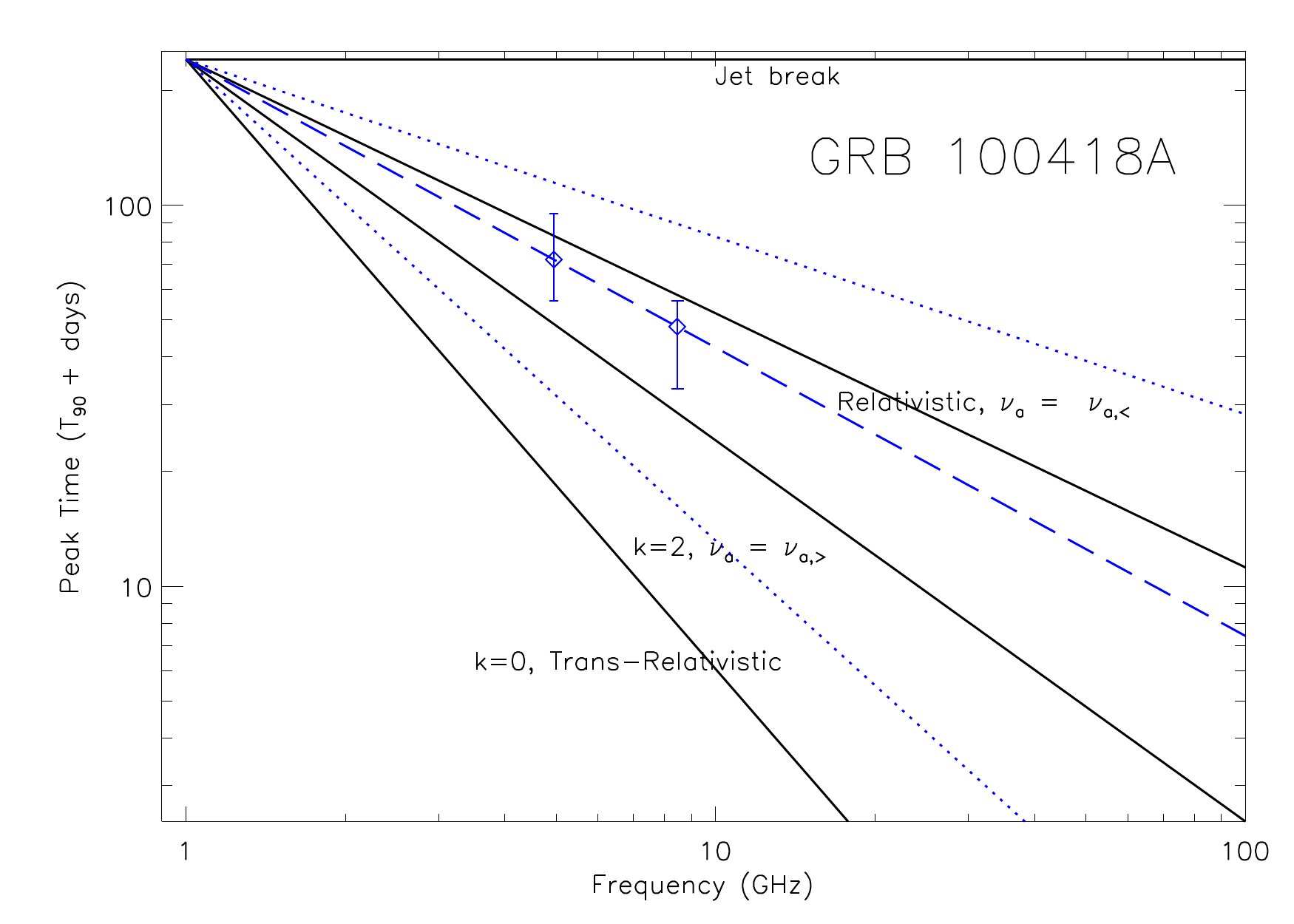}{0.31\textwidth}{}
          \fig{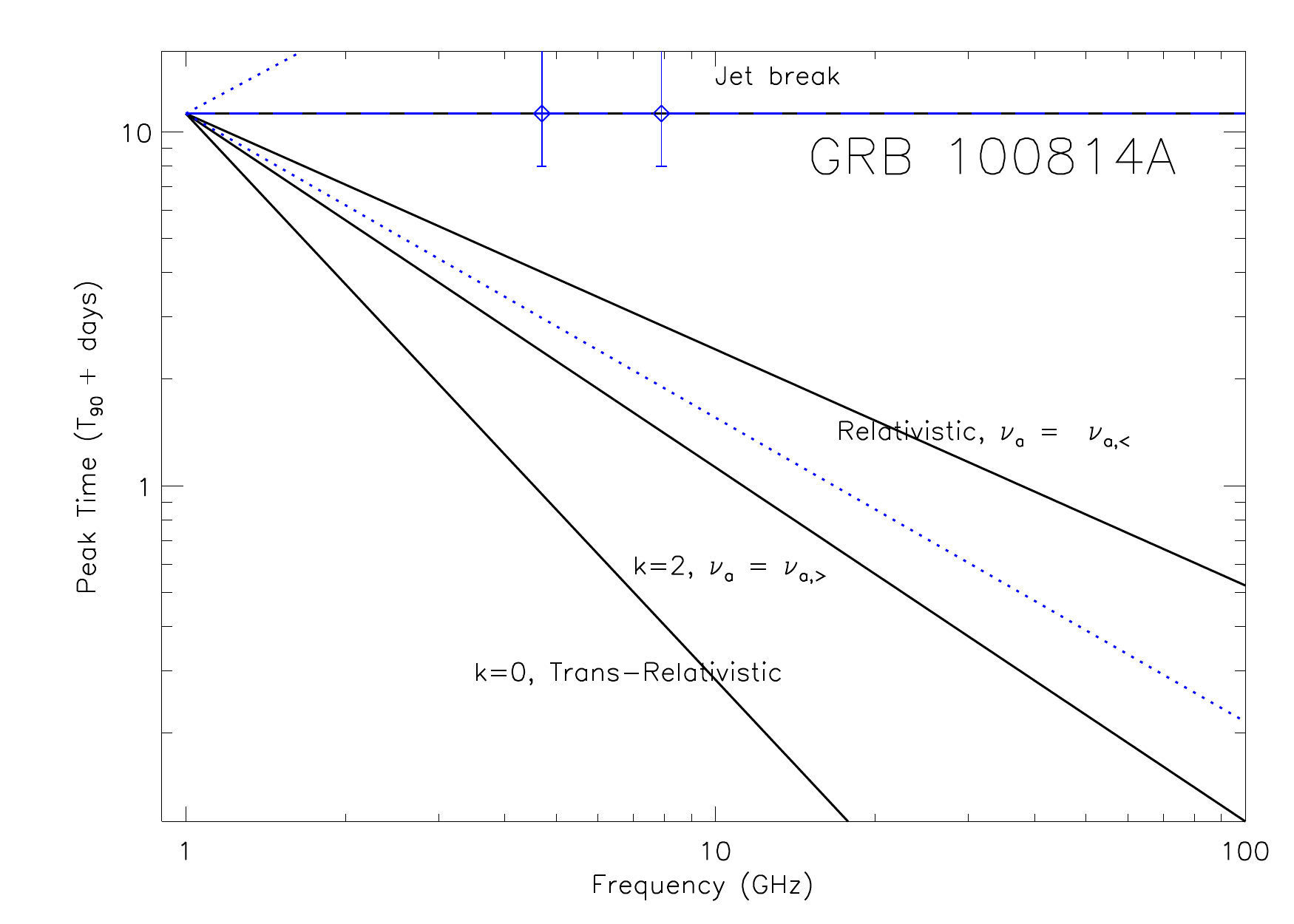}{0.31\textwidth}{}
          }
\small{Figure \ref{fig:tpeak_nu}. - Continued}
\clearpage
\gridline{\fig{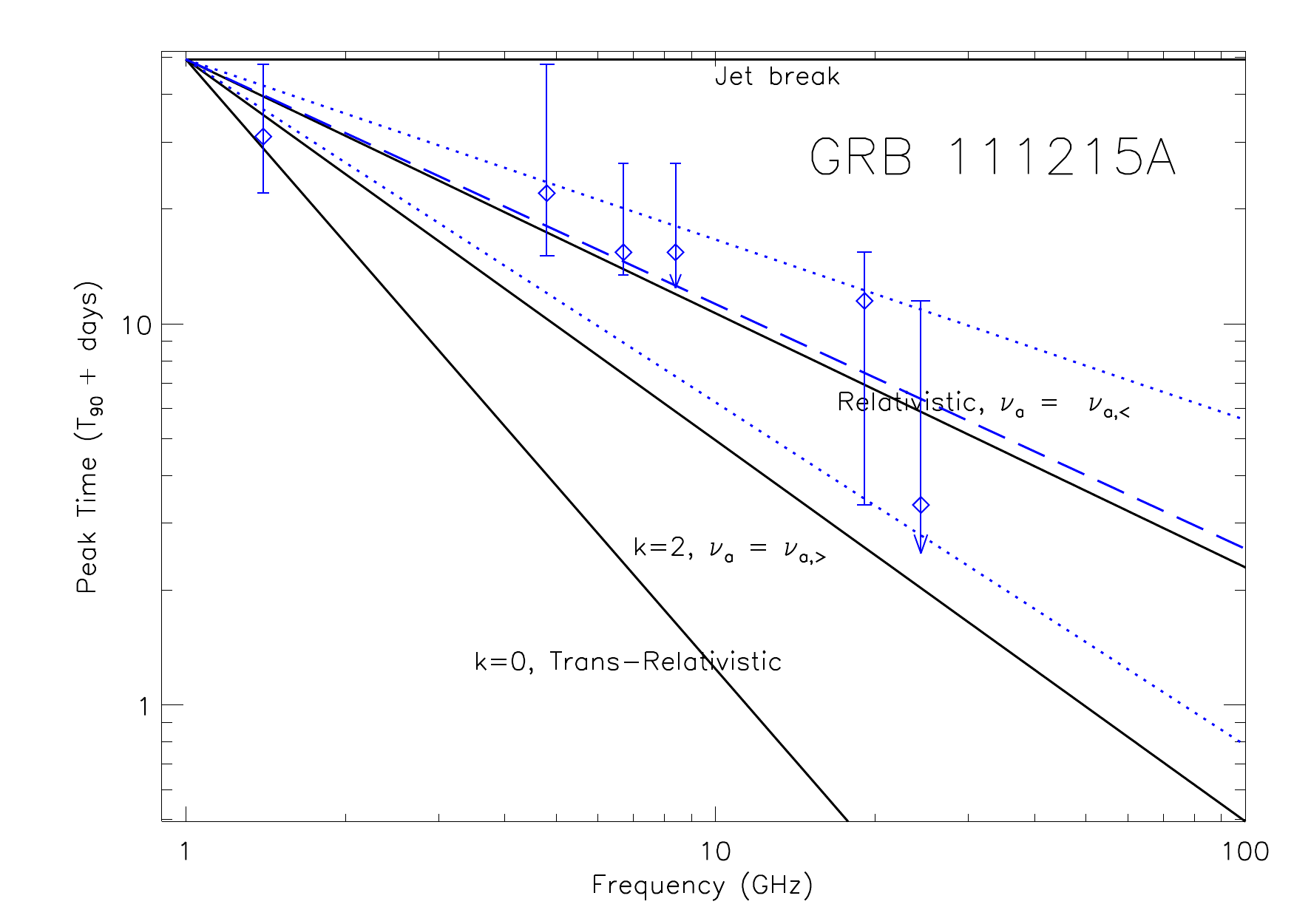}{0.31\textwidth}{}
          \fig{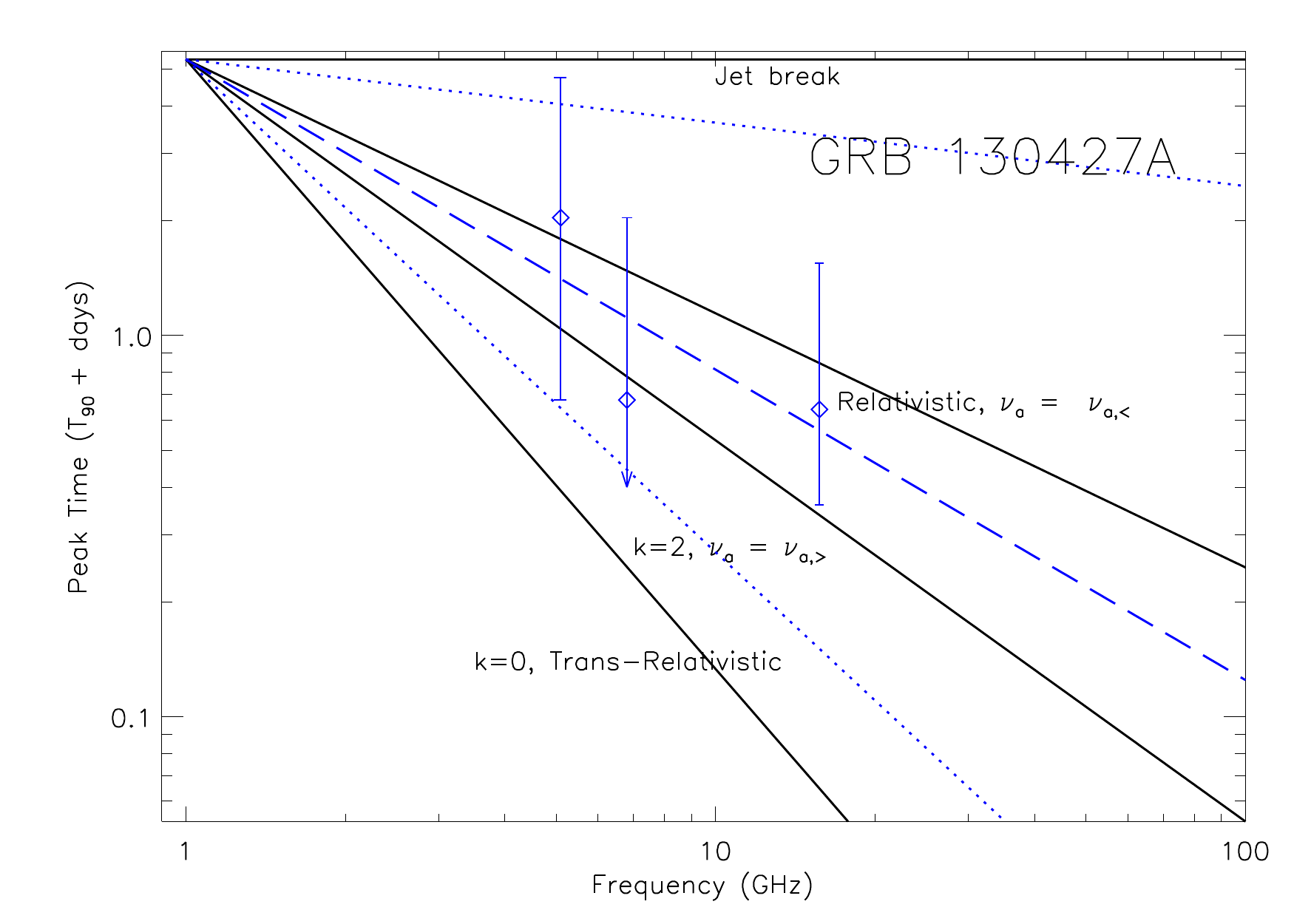}{0.31\textwidth}{}
          \fig{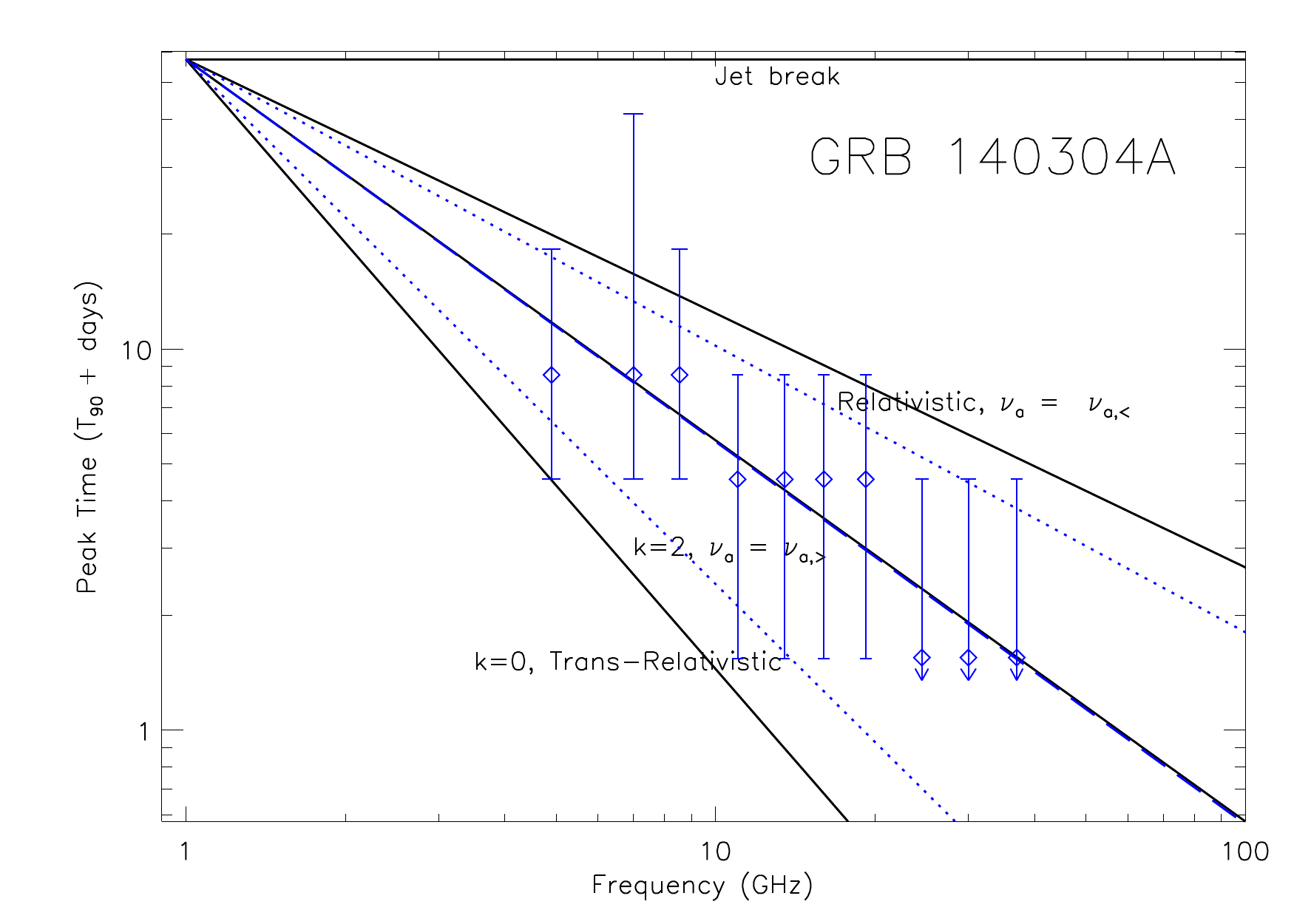}{0.31\textwidth}{}
          }
\gridline{\fig{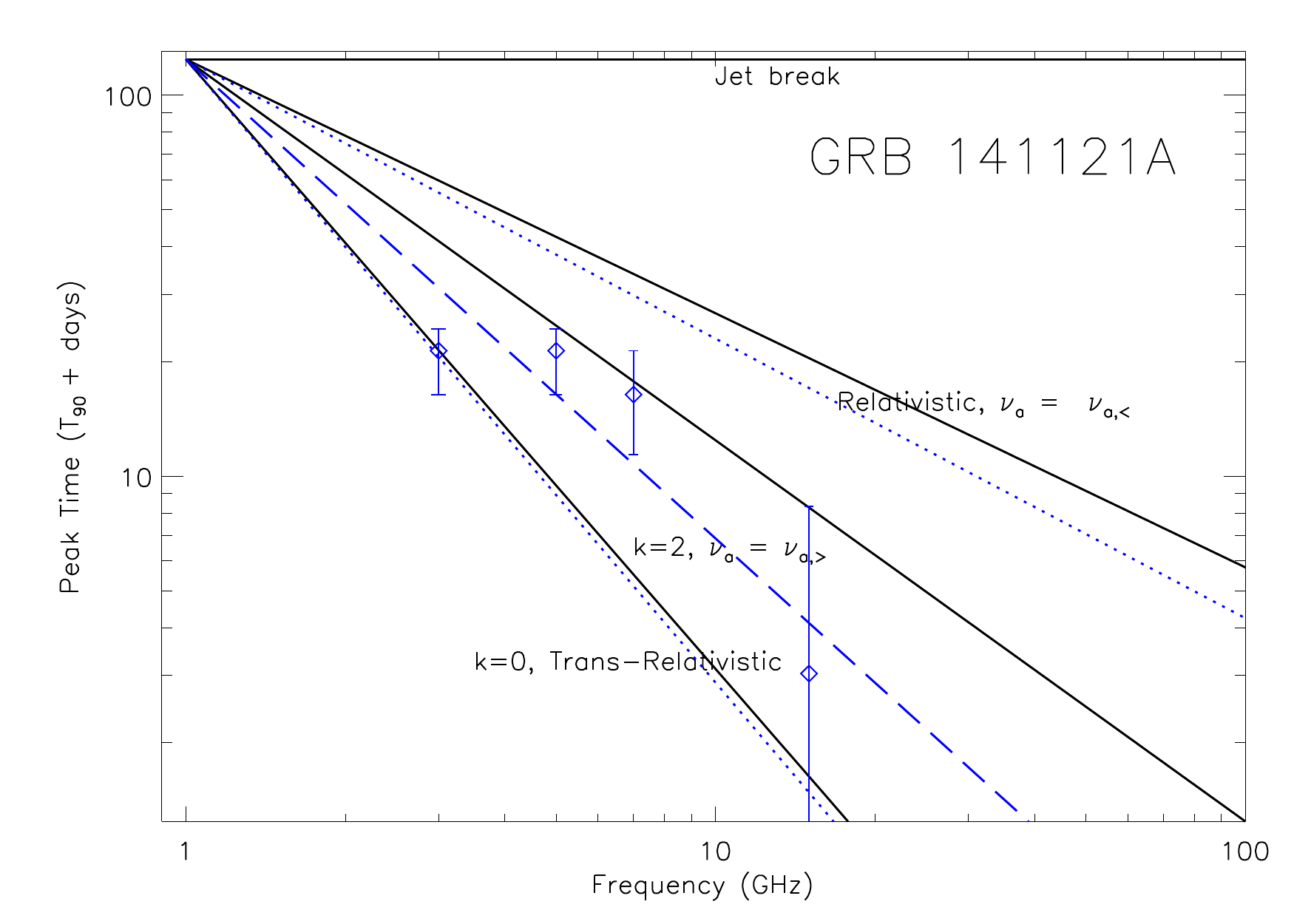}{0.31\textwidth}{}
          \fig{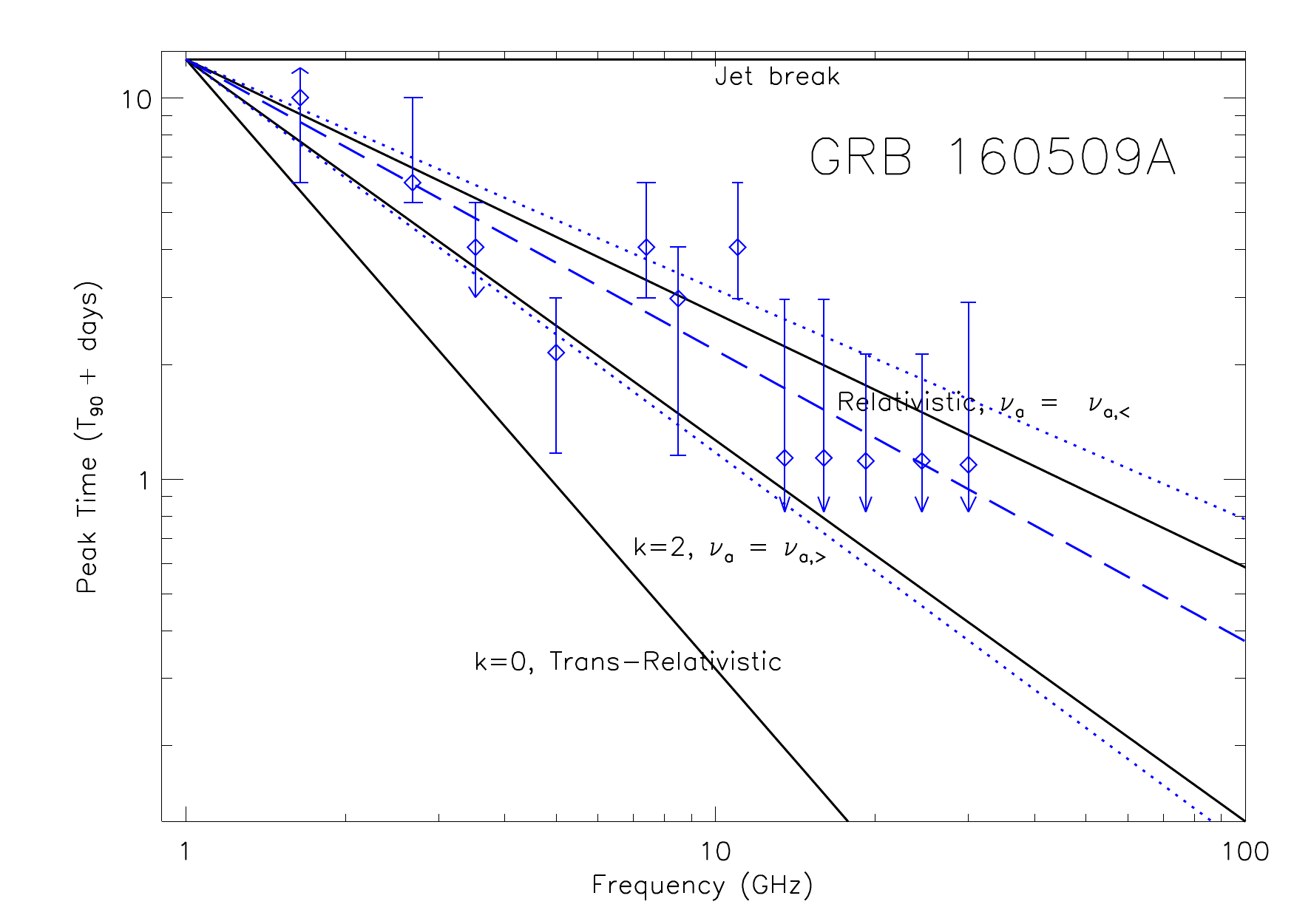}{0.31\textwidth}{}
          \fig{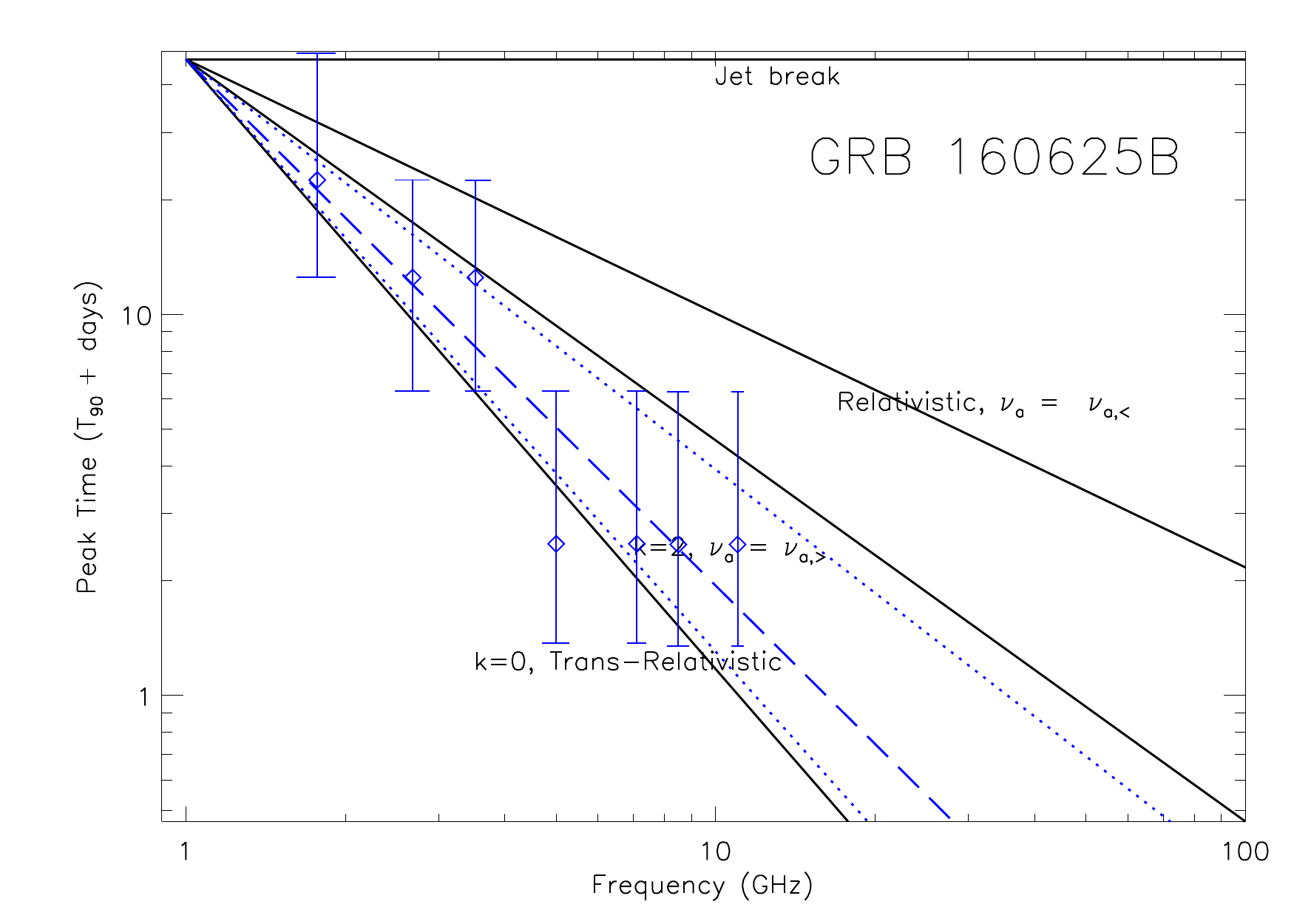}{0.31\textwidth}{}
          }
\gridline{\fig{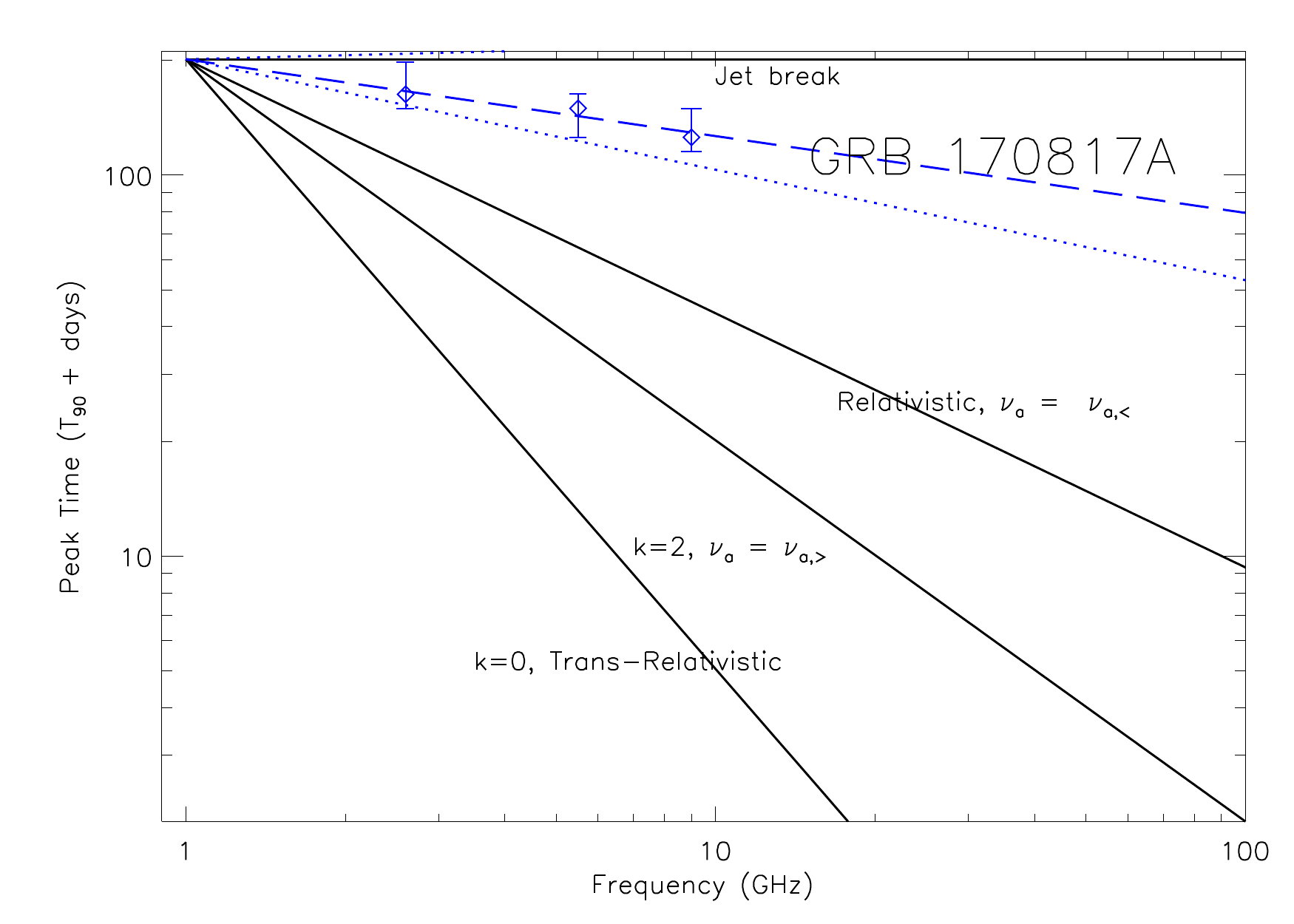}{0.31\textwidth}{}
          \fig{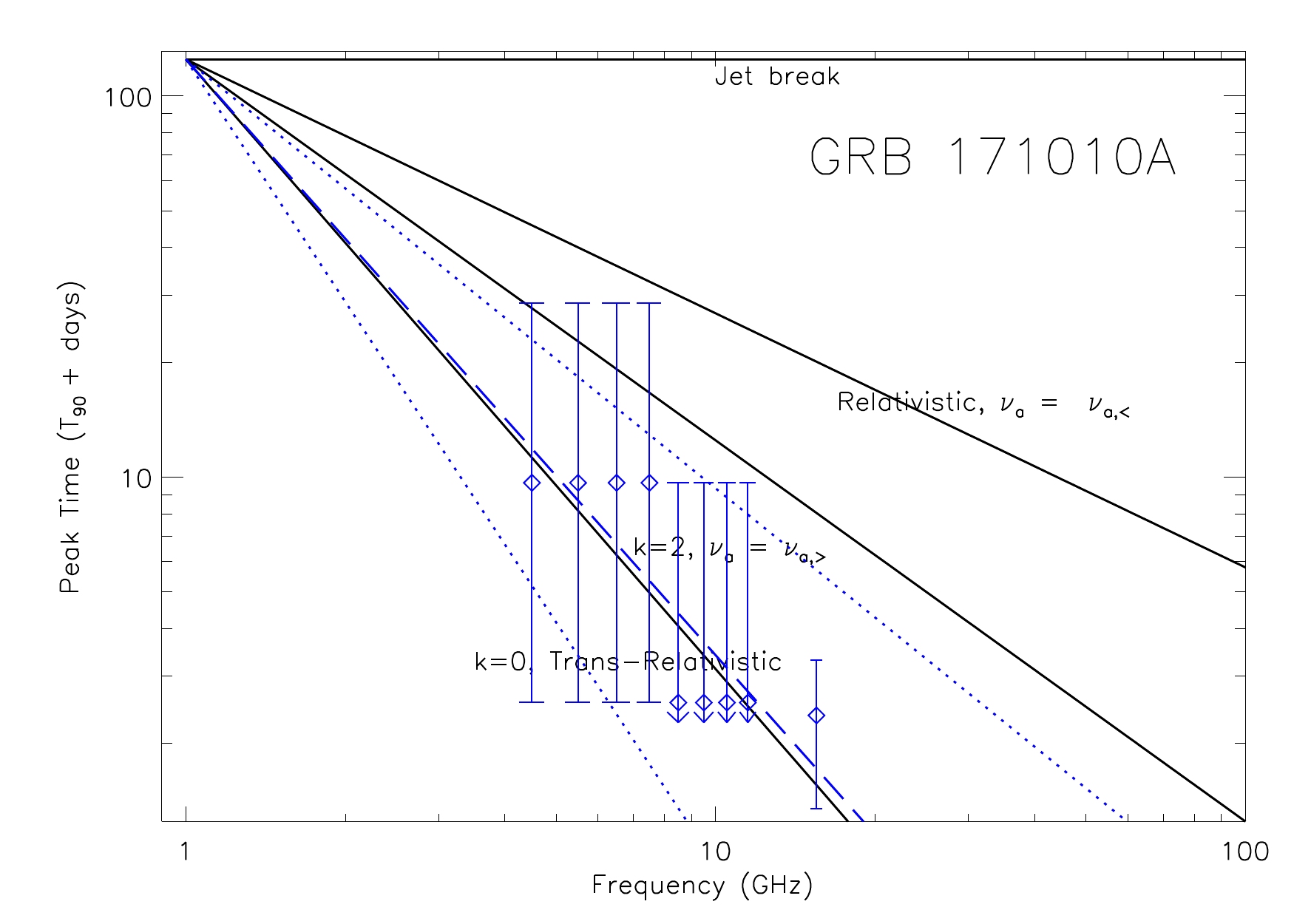}{0.31\textwidth}{}
          }
\small{Figure \ref{fig:tpeak_nu}. - Continued}

%\clearpage

\begin{figure*}
\gridline{\fig{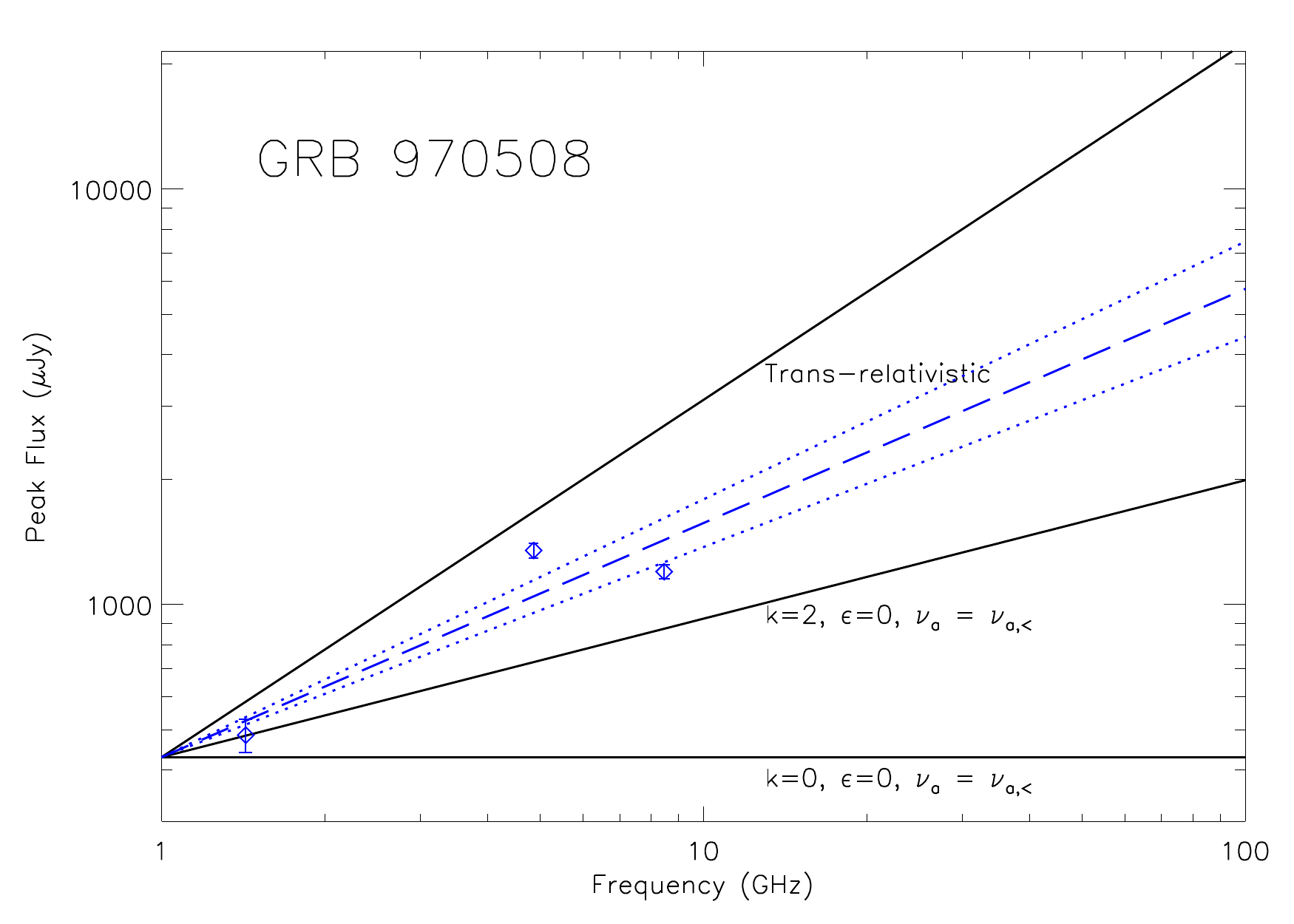}{0.31\textwidth}{}
          \fig{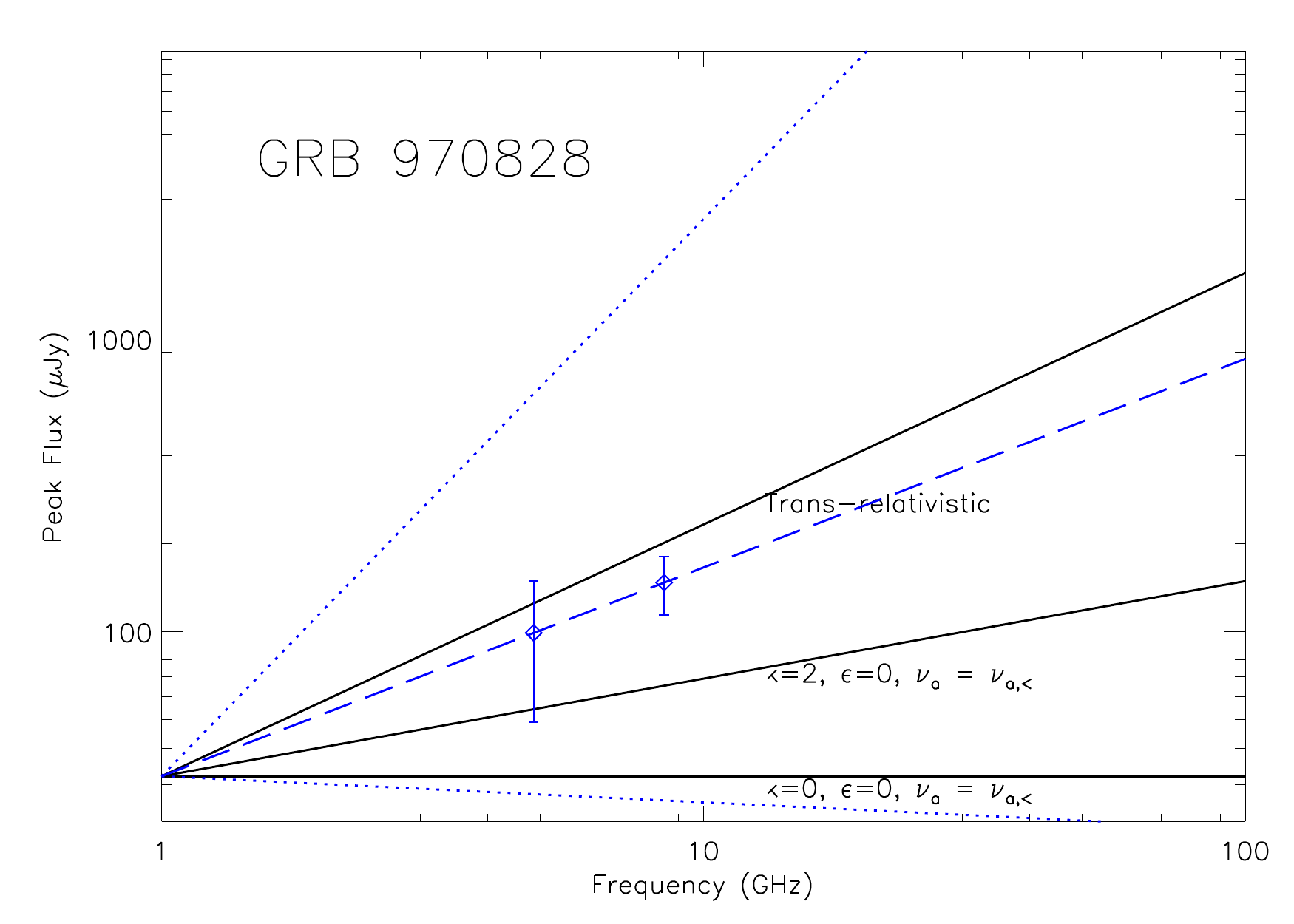}{0.31\textwidth}{}
          \fig{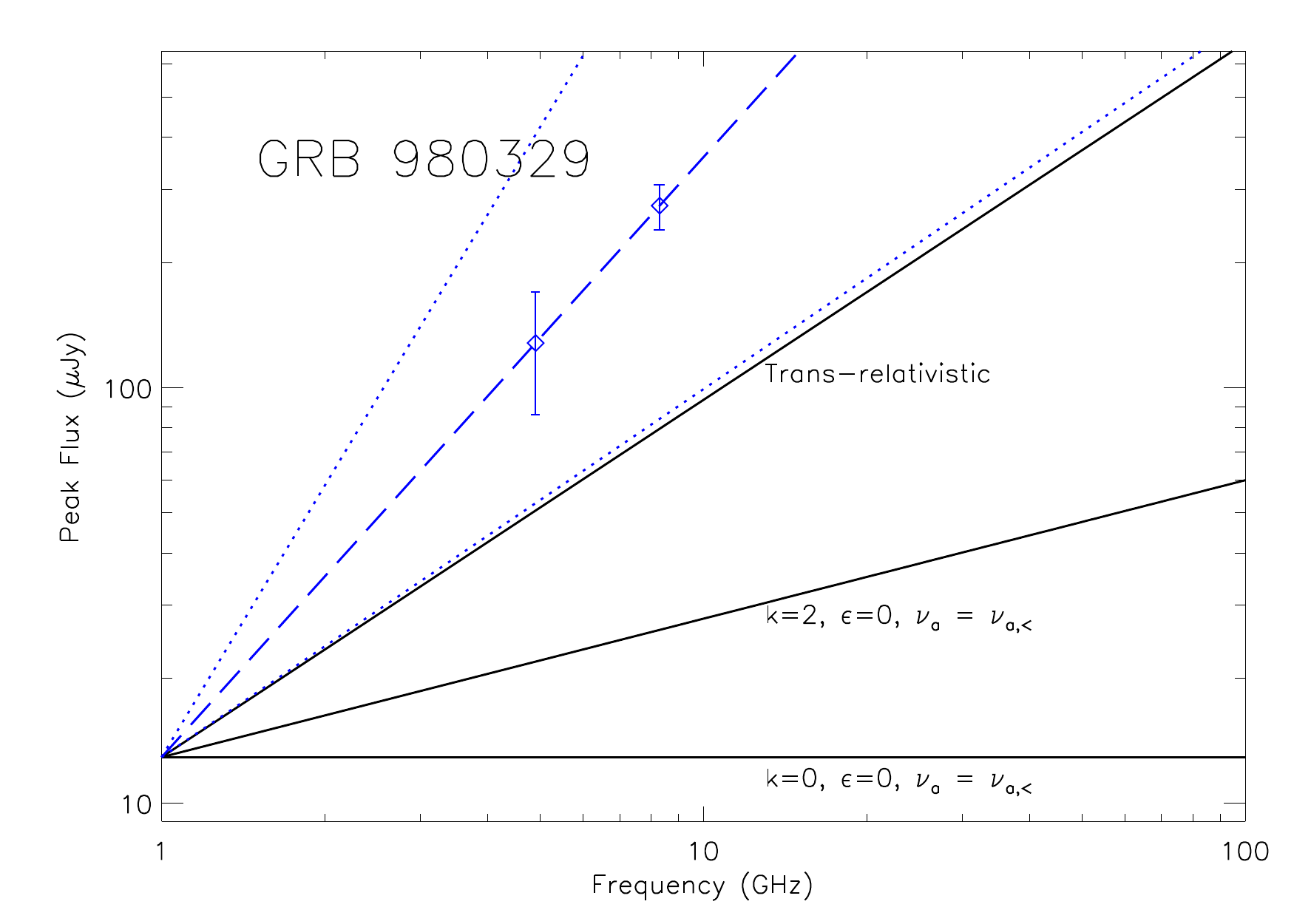}{0.31\textwidth}{}
         }
\gridline{\fig{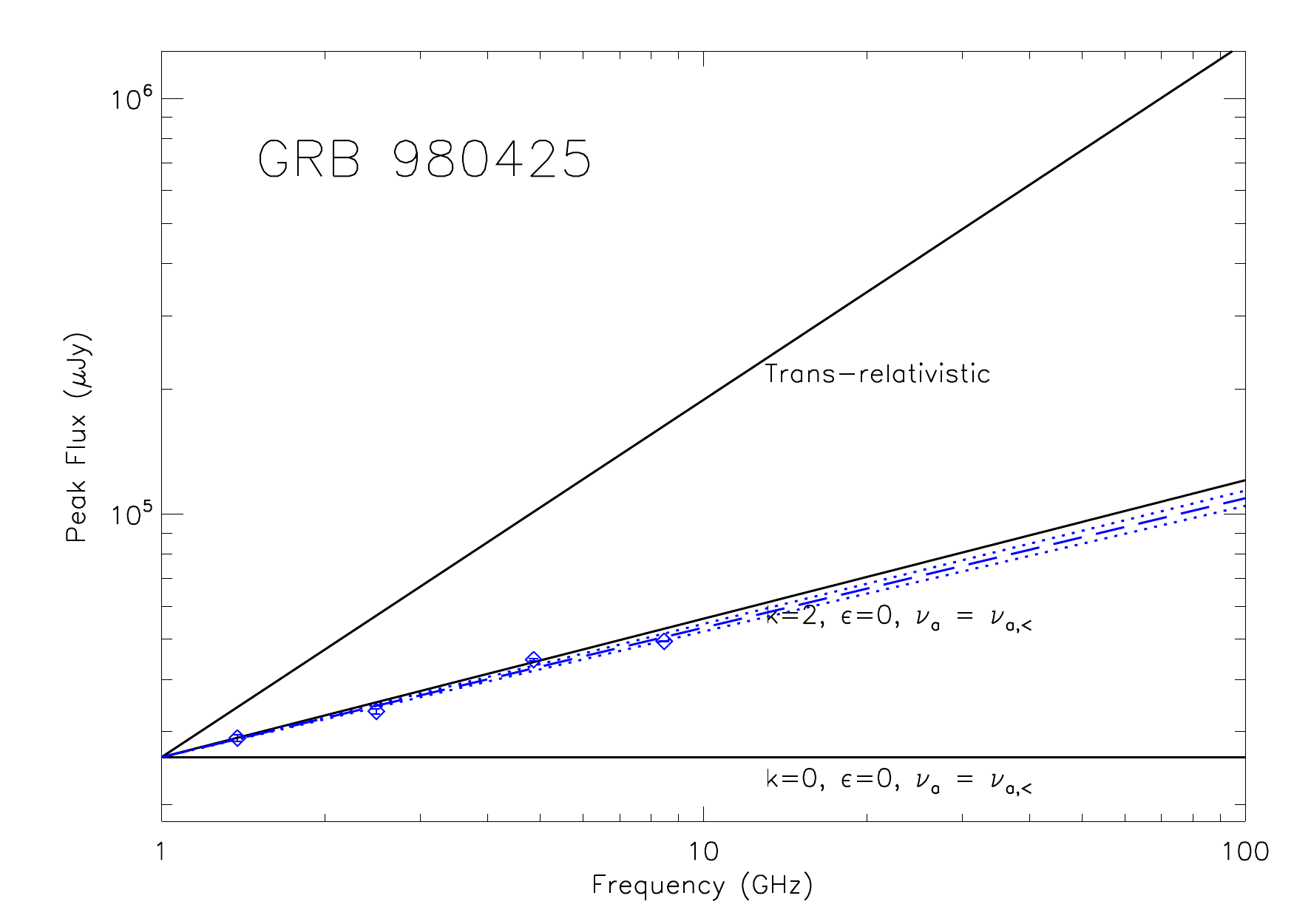}{0.31\textwidth}{}
          \fig{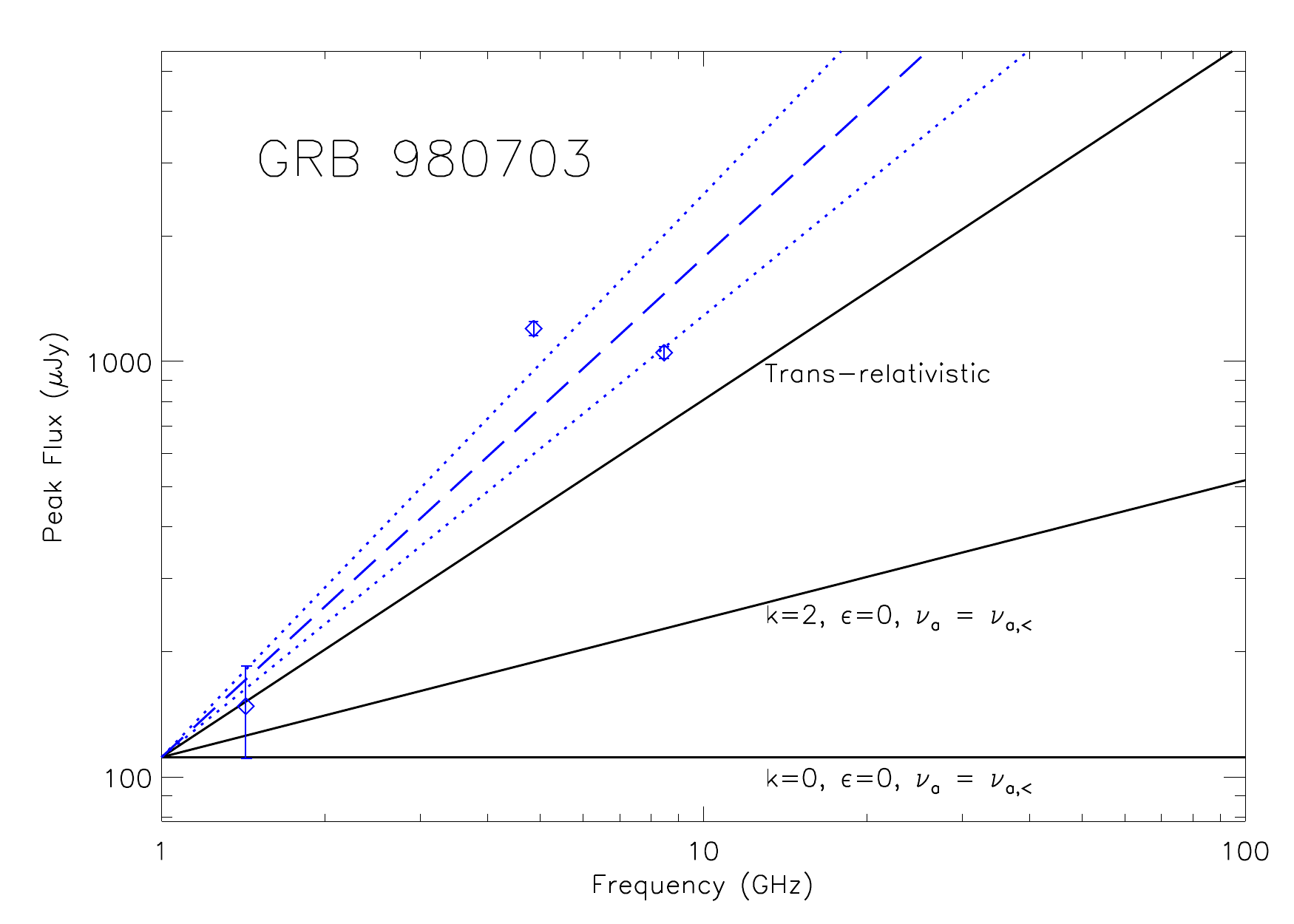}{0.31\textwidth}{}
          \fig{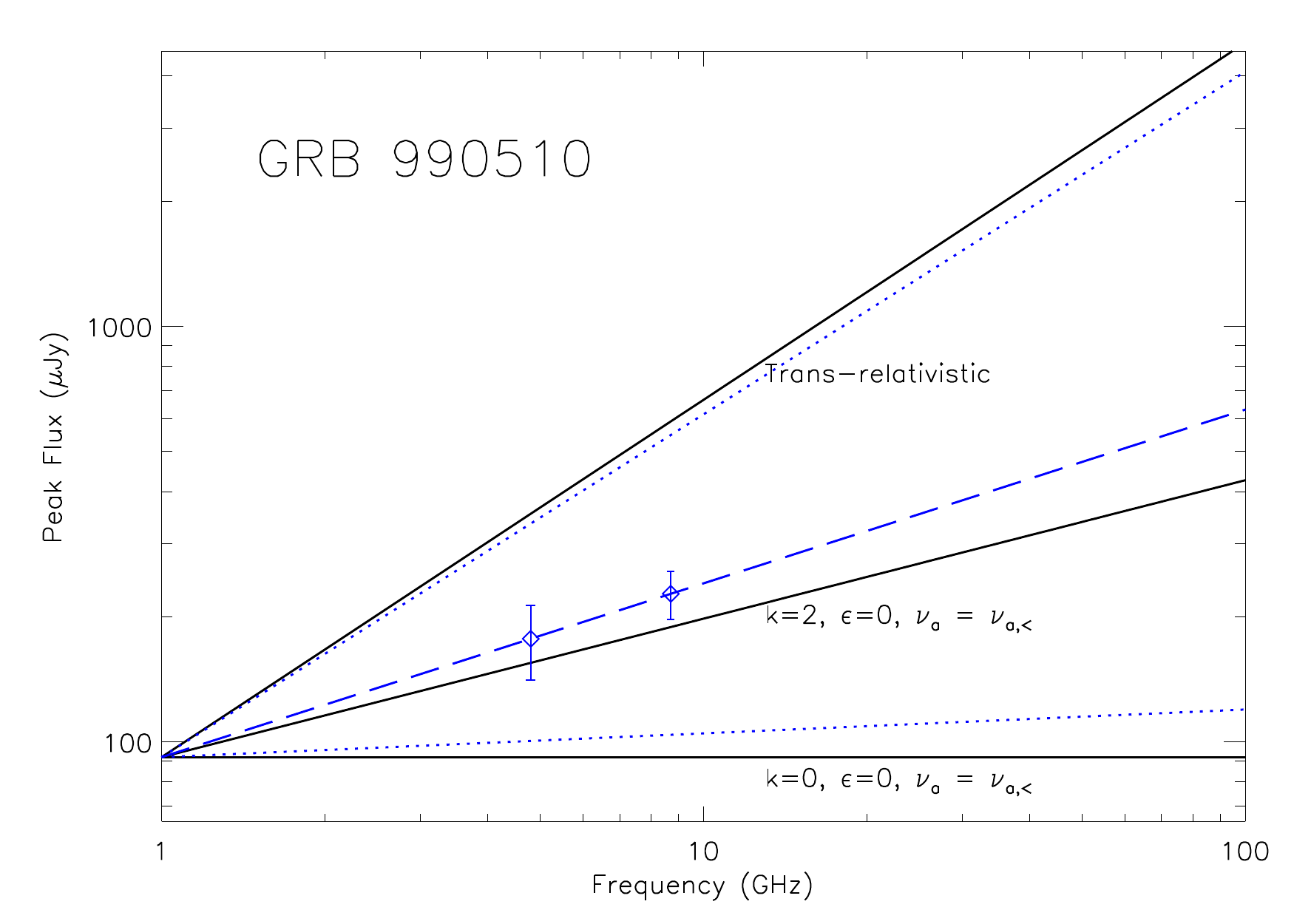}{0.31\textwidth}{}
          }
\gridline{\fig{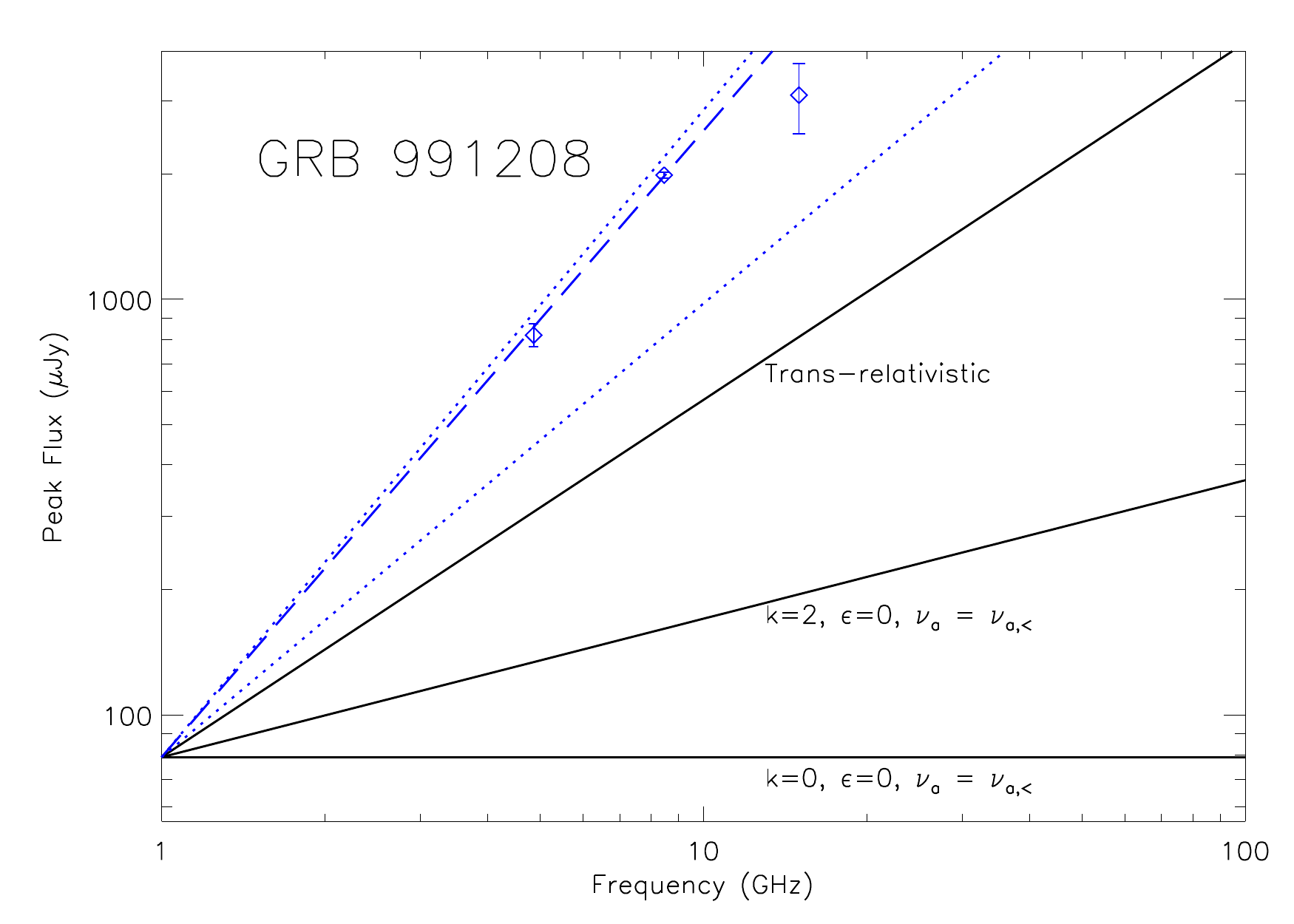}{0.31\textwidth}{}
          \fig{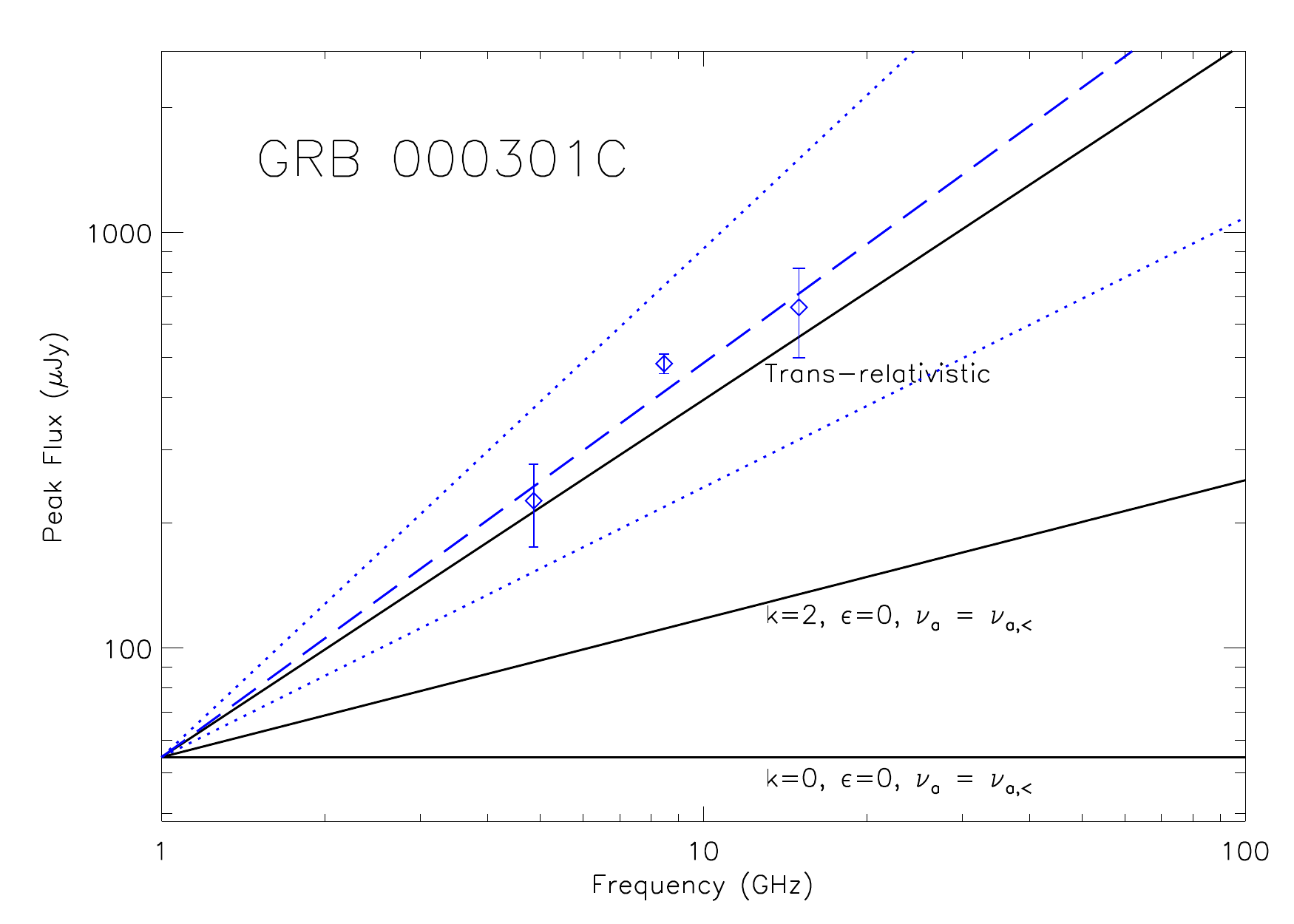}{0.31\textwidth}{}
          \fig{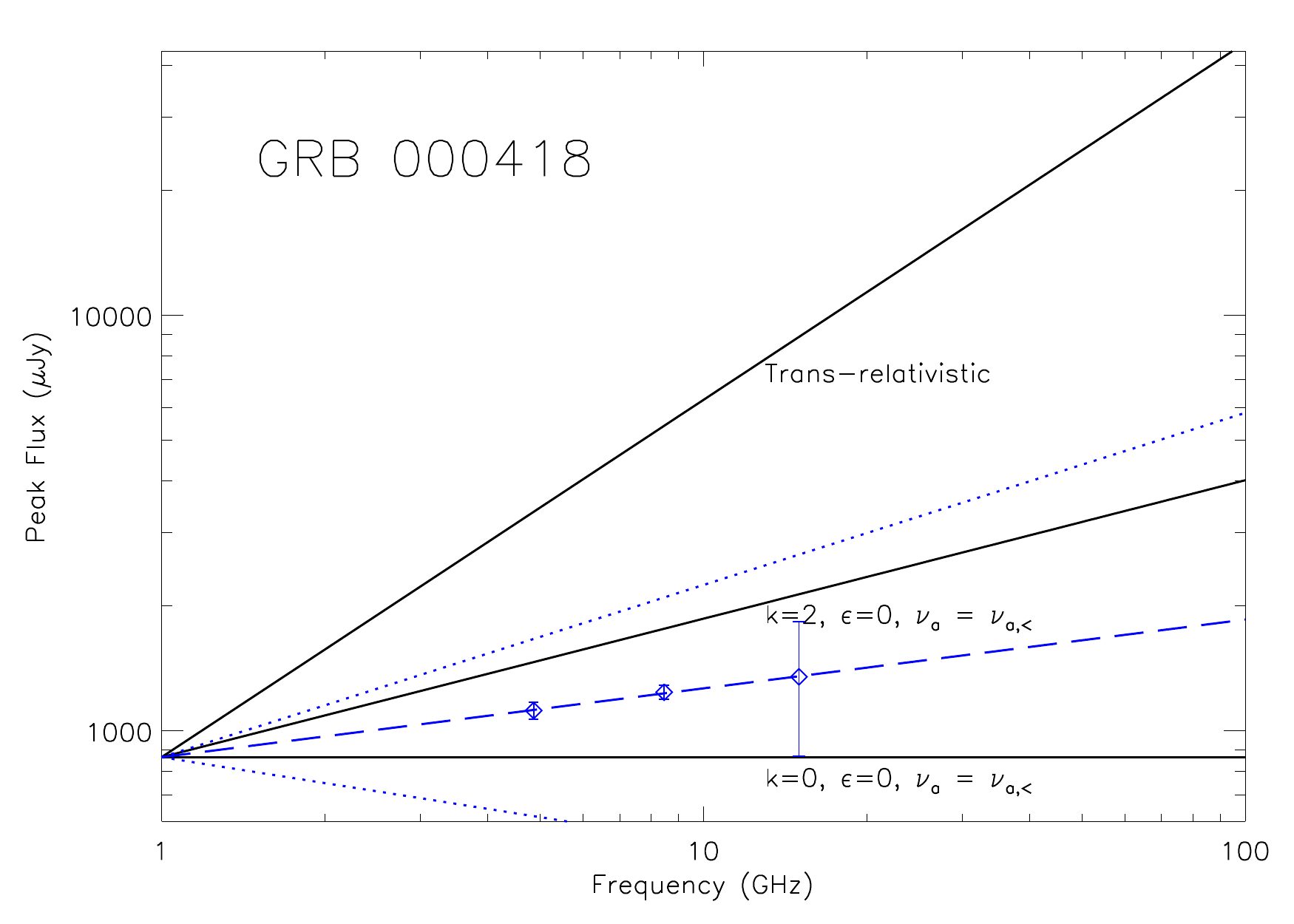}{0.31\textwidth}{}
          }
\gridline{\fig{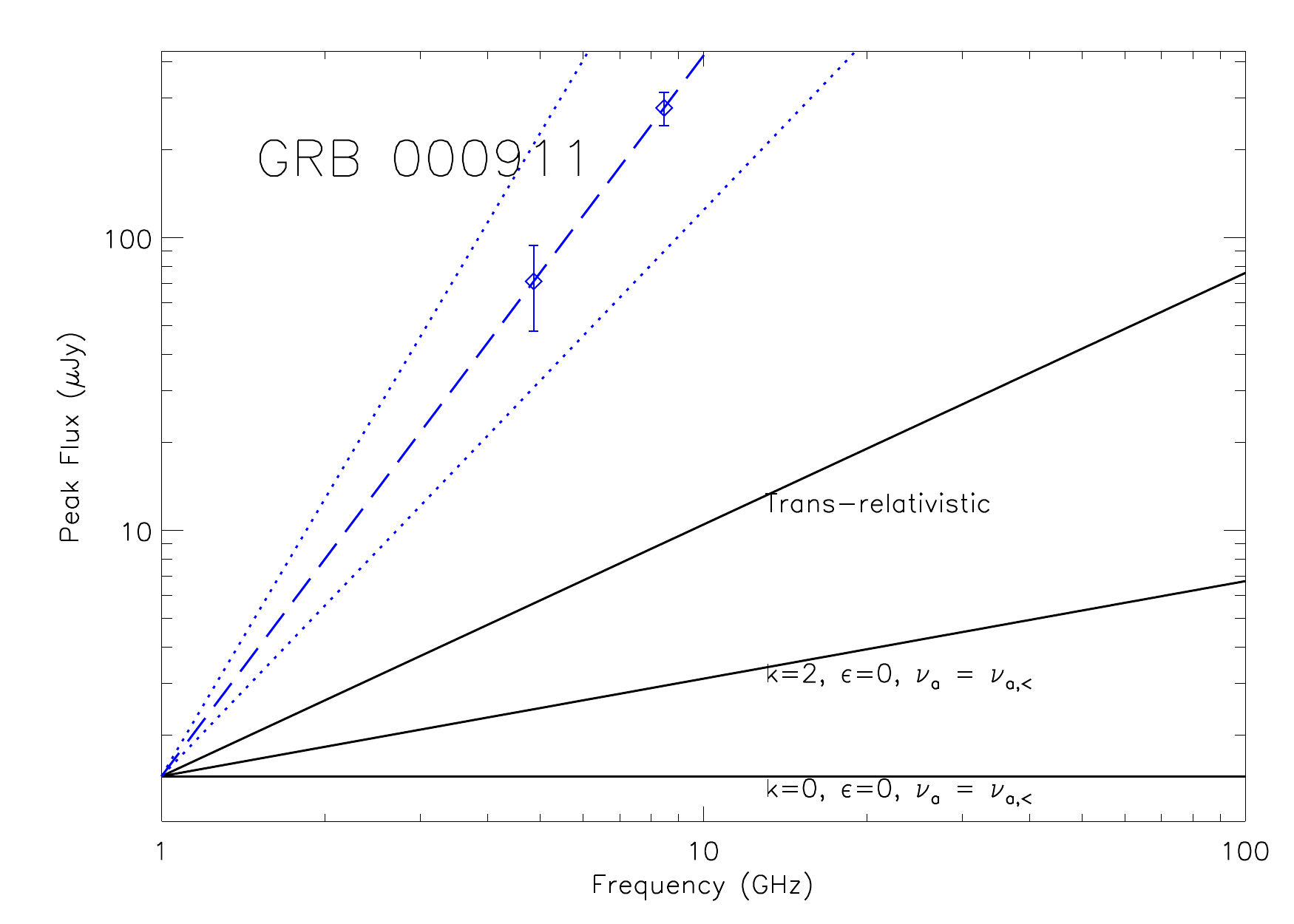}{0.31\textwidth}{}
          \fig{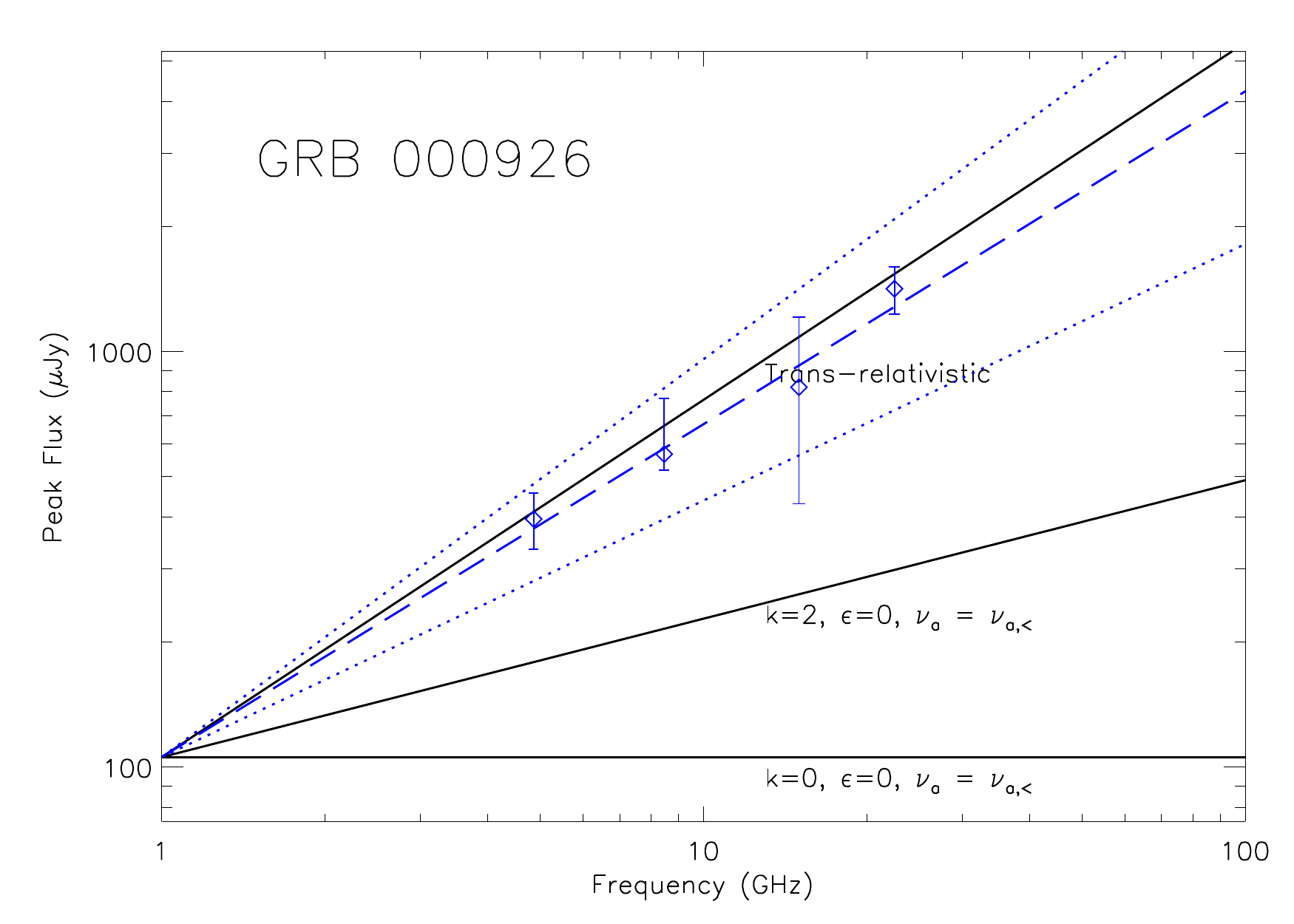}{0.31\textwidth}{}
          \fig{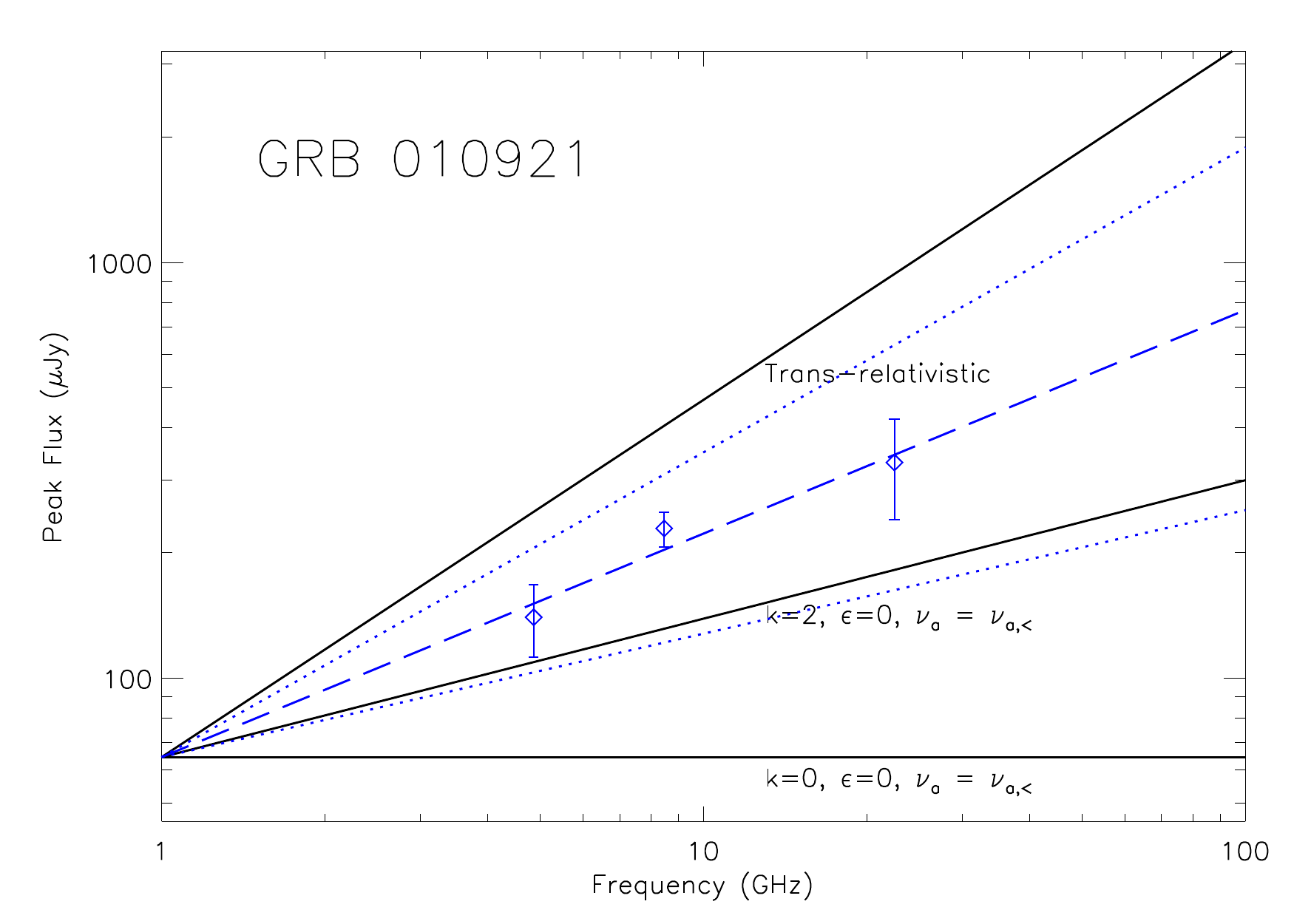}{0.31\textwidth}{}
          }
\caption{\small{The relations between radio afterglow peak fluxes $F_{{\rm peak},\nu}$ and observing frequencies $\nu/\nu_{{\rm min}}$ for our sample GRBs. The blue dashed line with data points in each figure shows the fitting results from observations, the blue dotted lines denote the possible range of fitting errors, while the black solid lines from top to bottom represent the theoretical predictions for transrelativistic, $k=2$, $\epsilon = 0$ relativistic, and $k=0$, $\epsilon=0$ relativistic cases, respectively. It should be noted that as seen in Section \ref{subsec:trans-relativistic}, the $b$ value for transrelativistic cases is insensitive to $k$. }}\label{fig:fpeak_nu}
\end{figure*}
\clearpage

\gridline{\fig{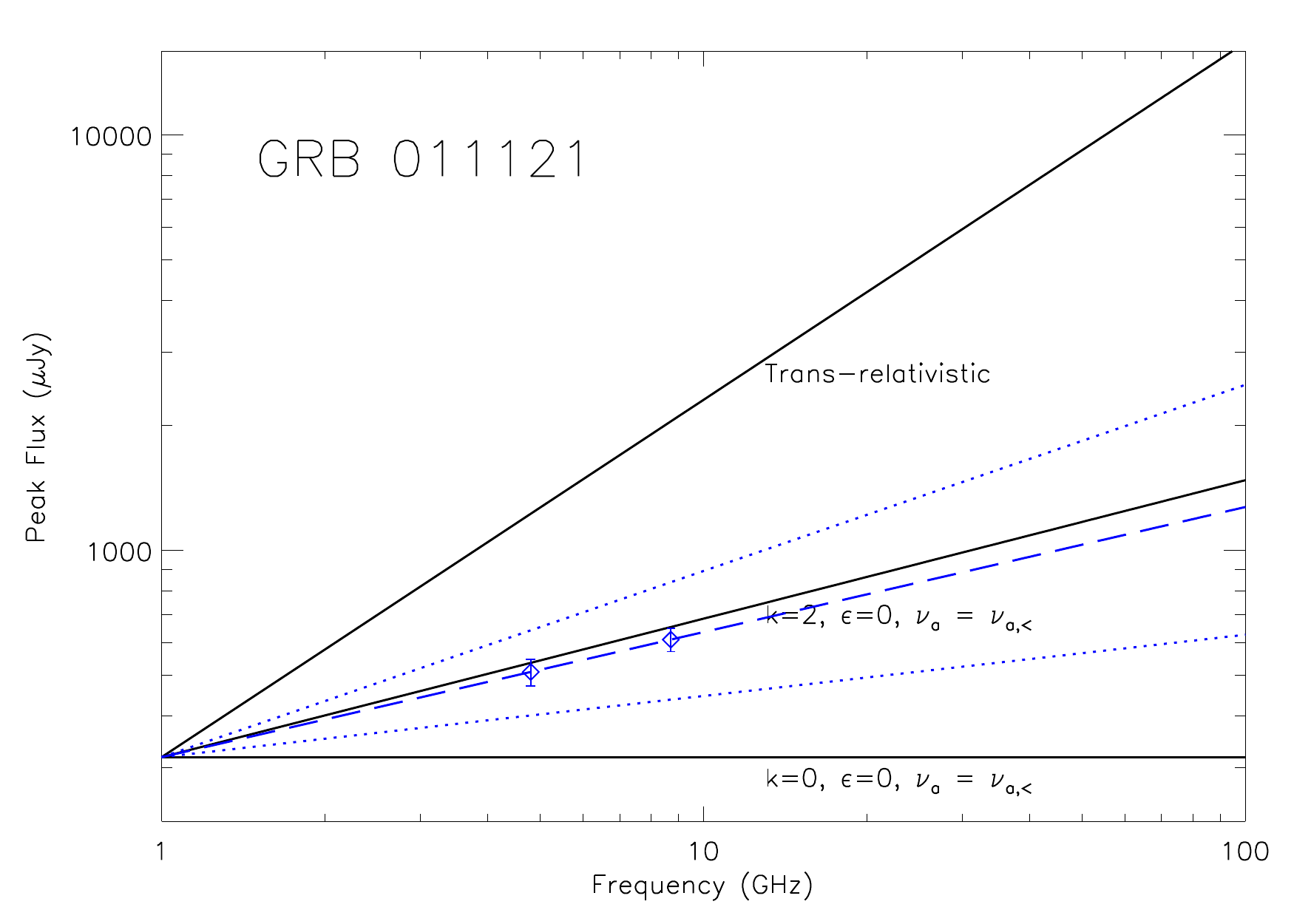}{0.31\textwidth}{}
          \fig{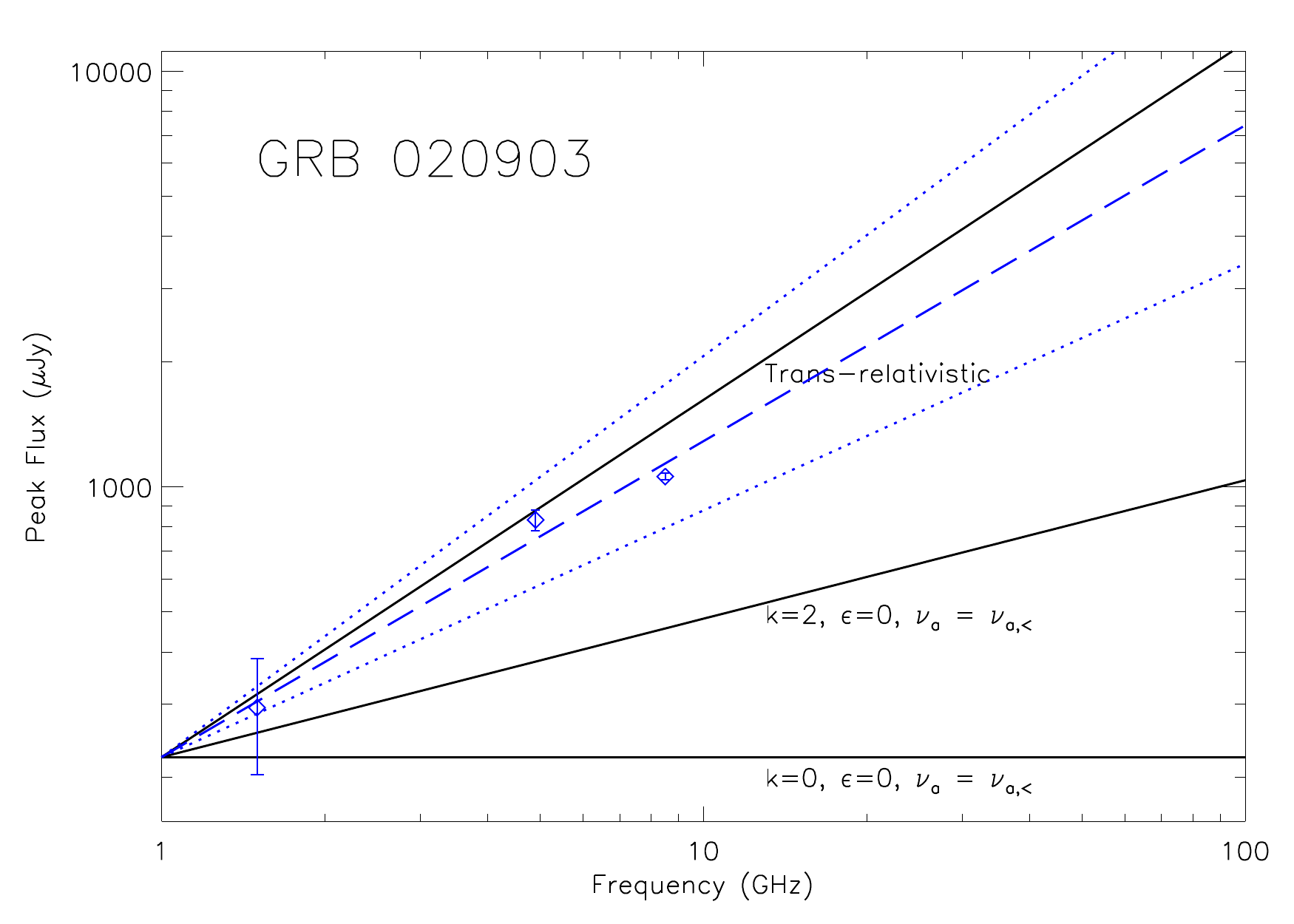}{0.31\textwidth}{}
          \fig{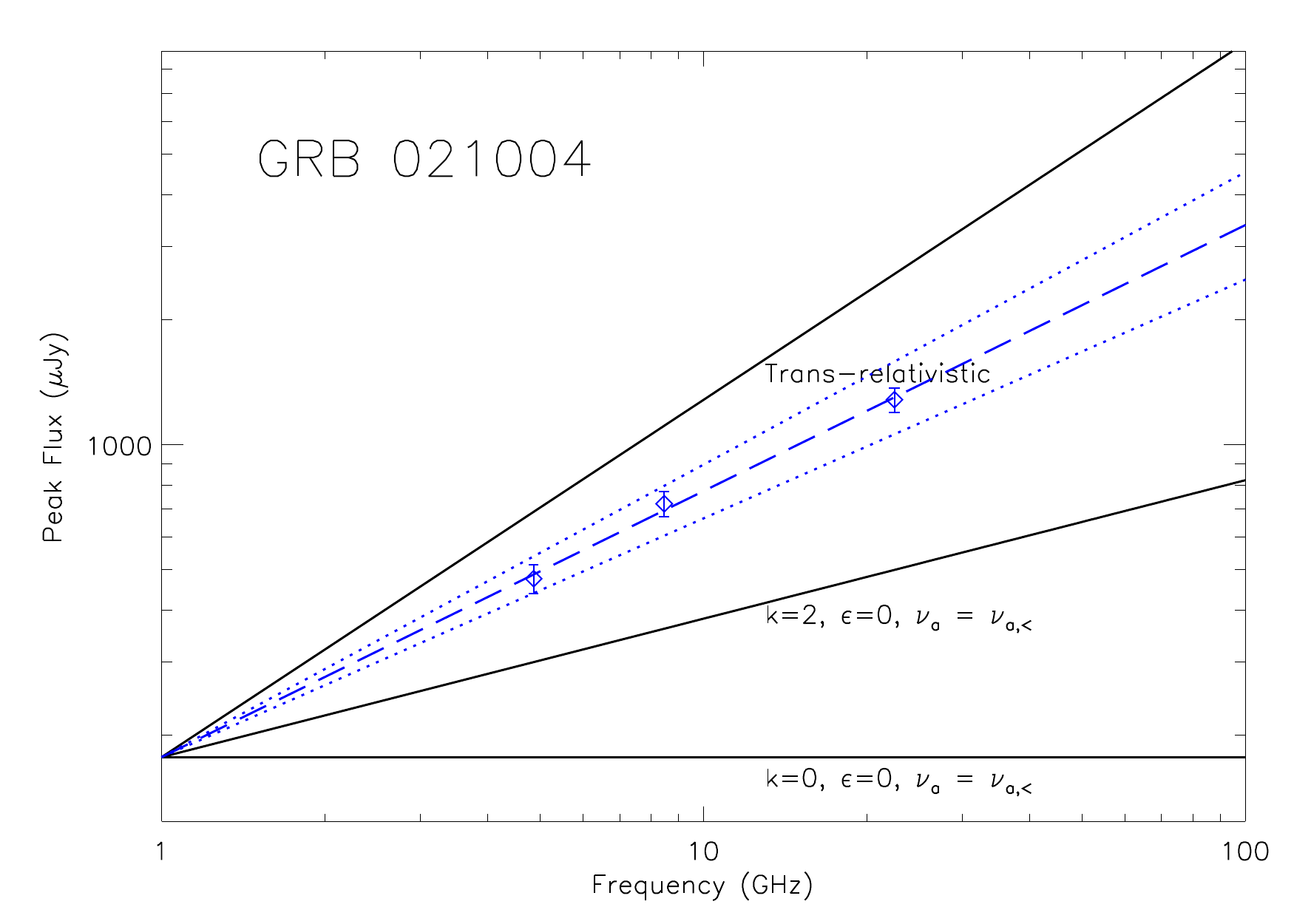}{0.31\textwidth}{}
          }
\gridline{\fig{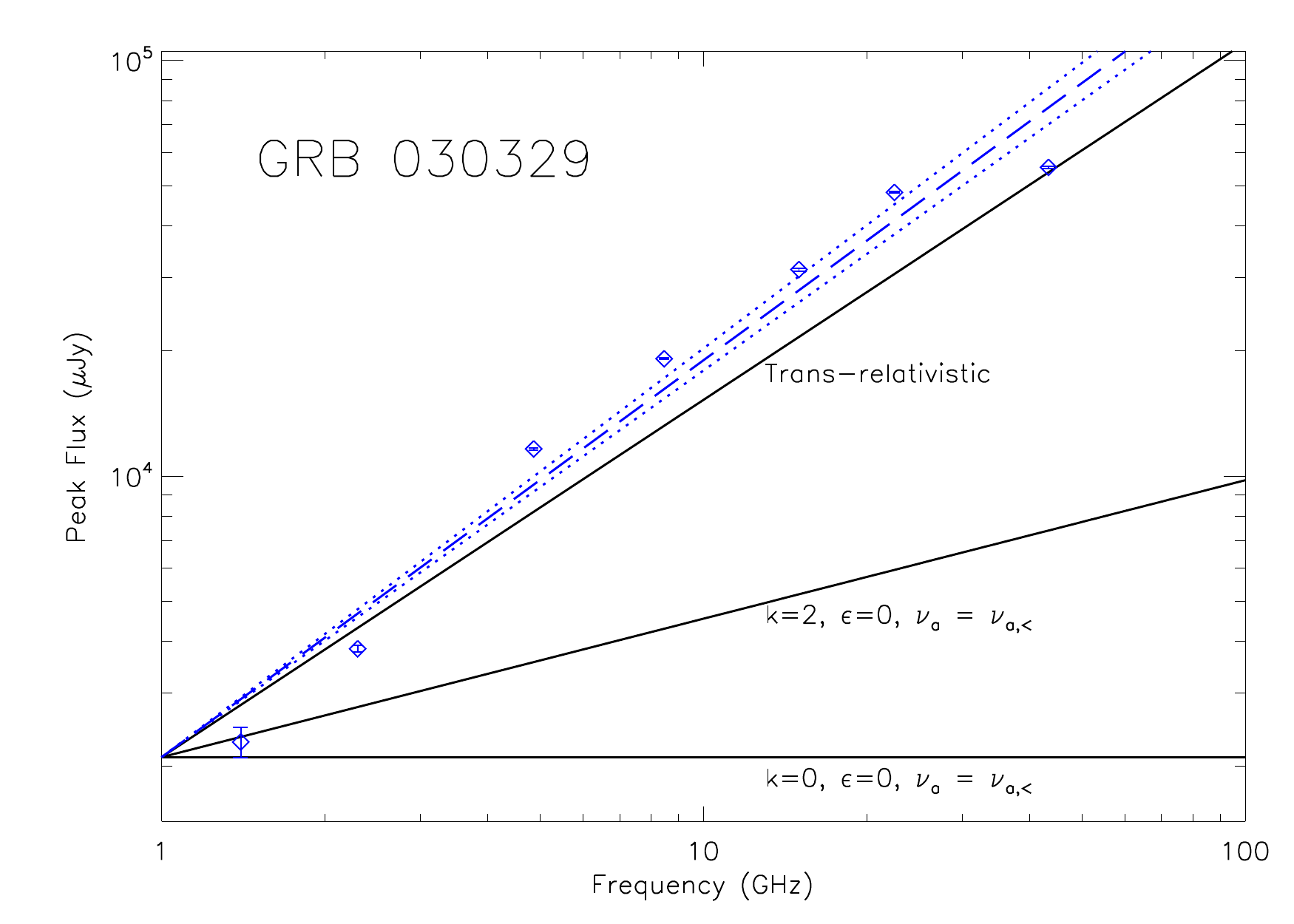}{0.31\textwidth}{}
          \fig{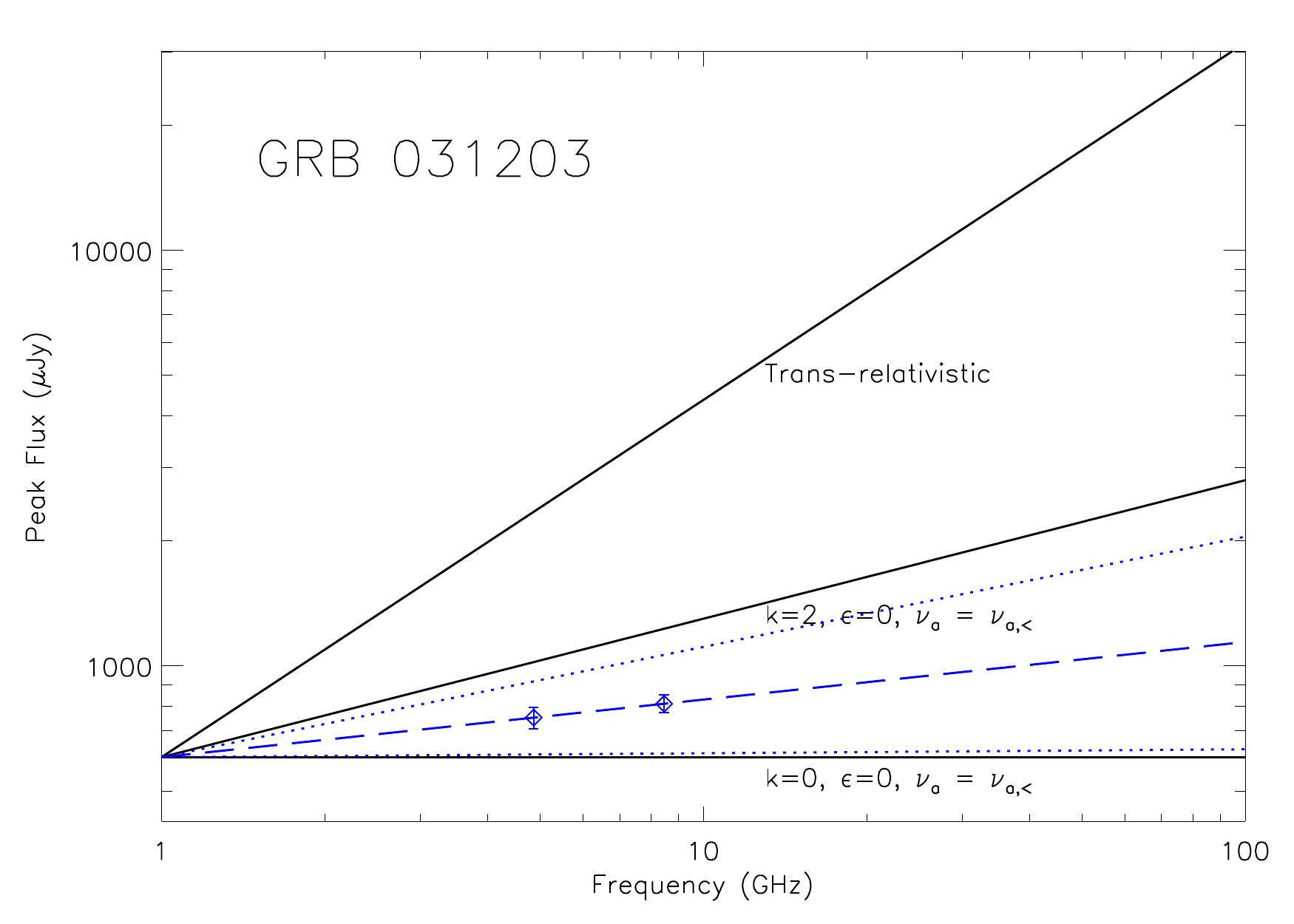}{0.31\textwidth}{}
          \fig{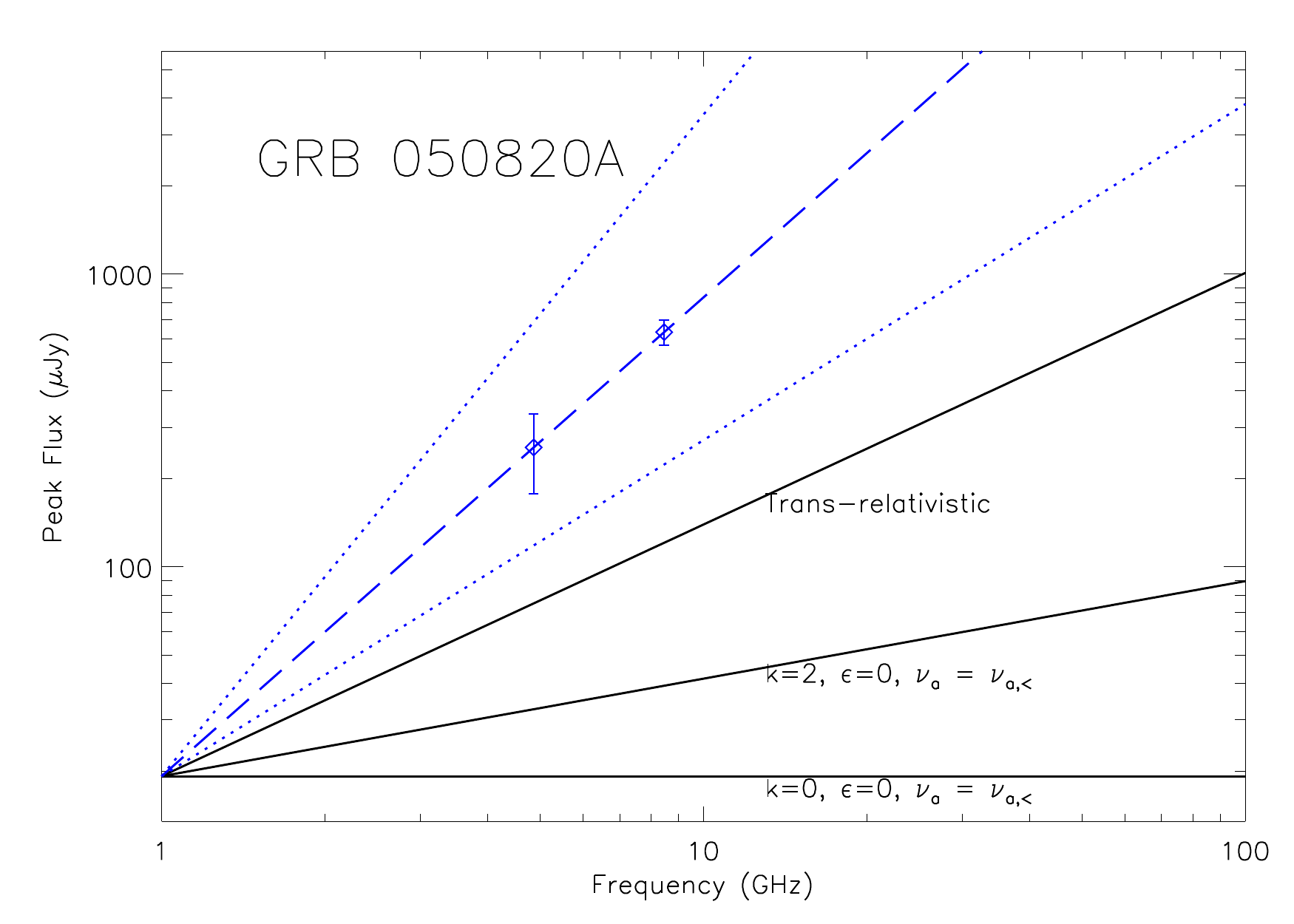}{0.31\textwidth}{}
          }
\gridline{\fig{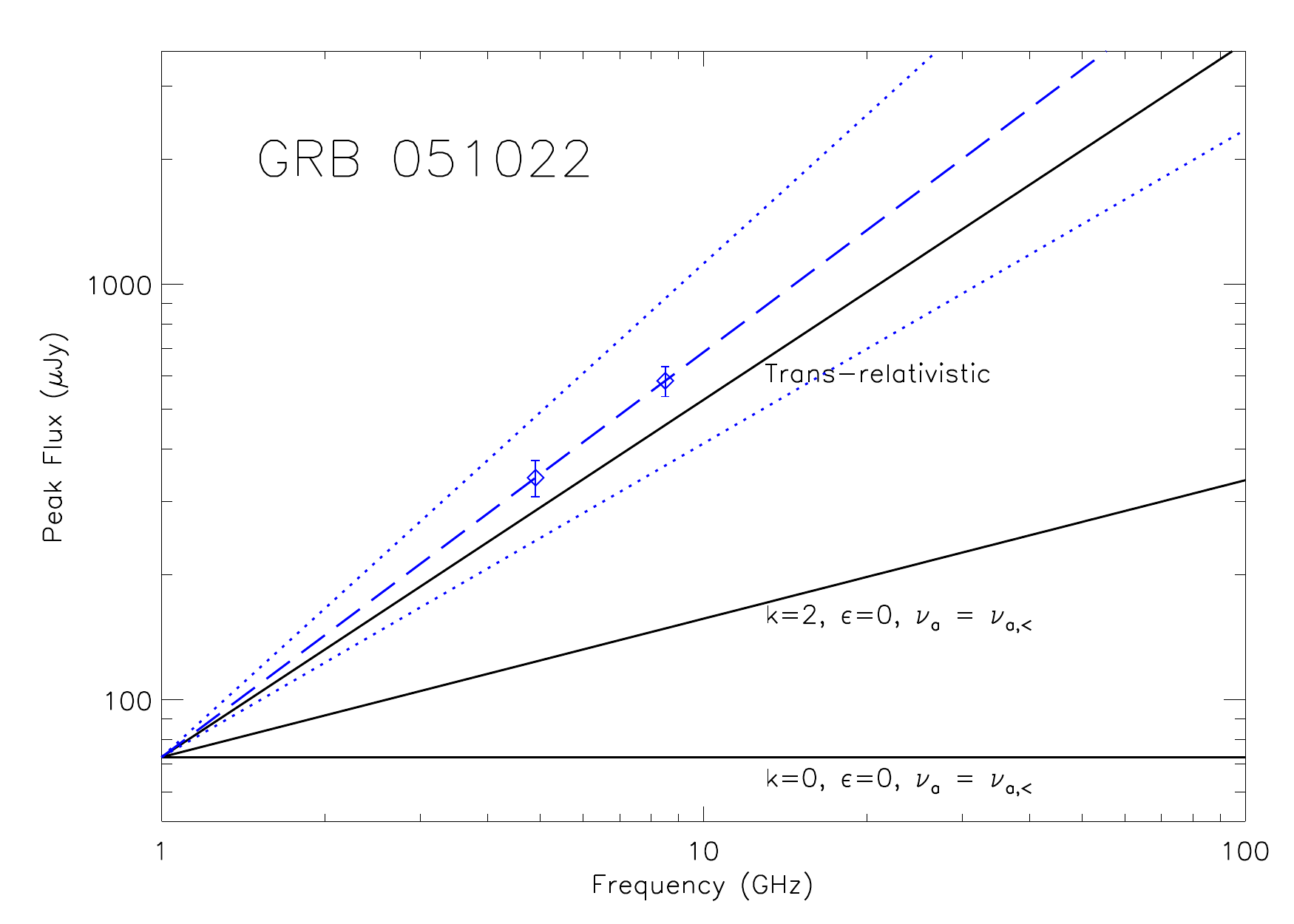}{0.31\textwidth}{}
          \fig{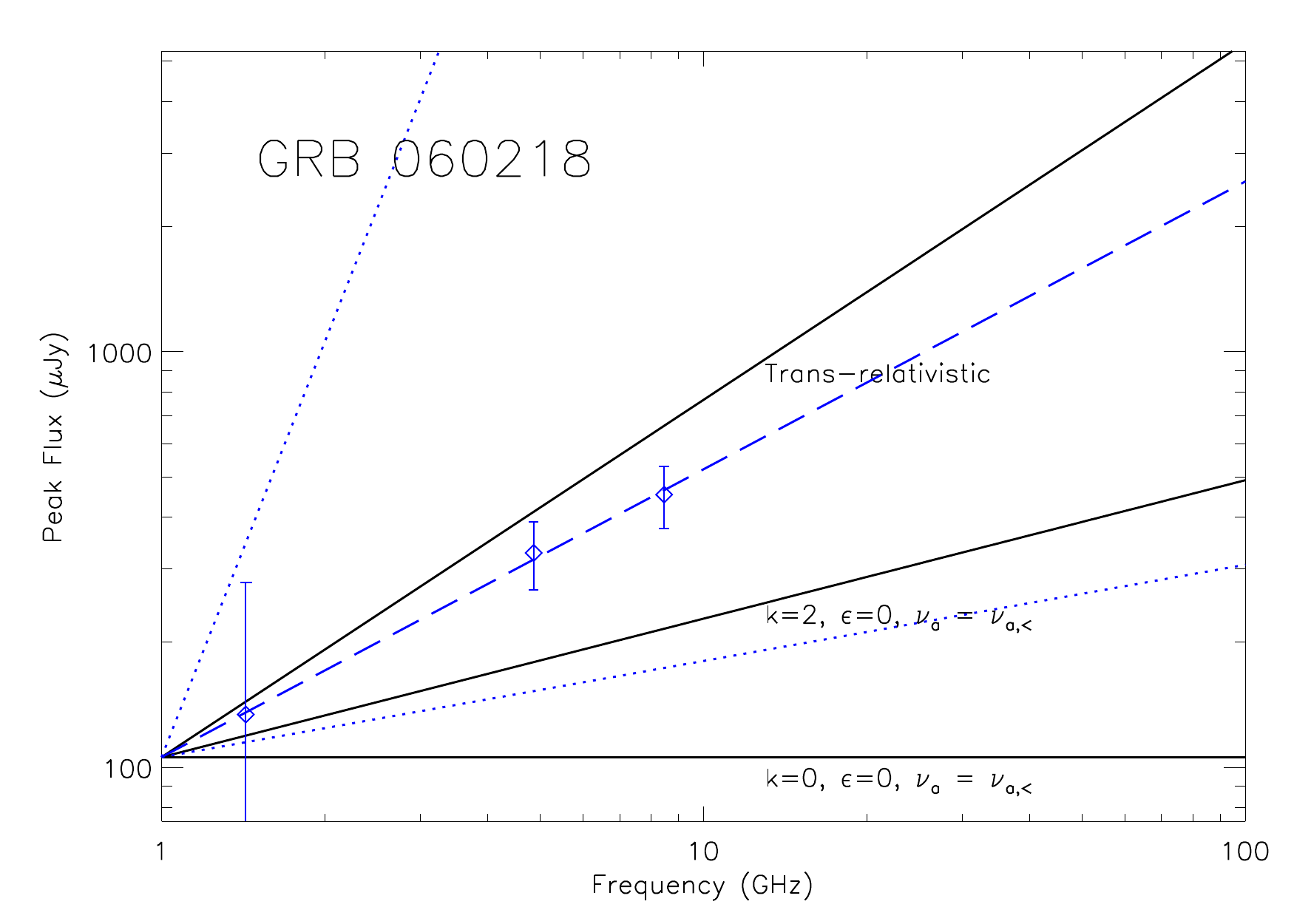}{0.31\textwidth}{}
          \fig{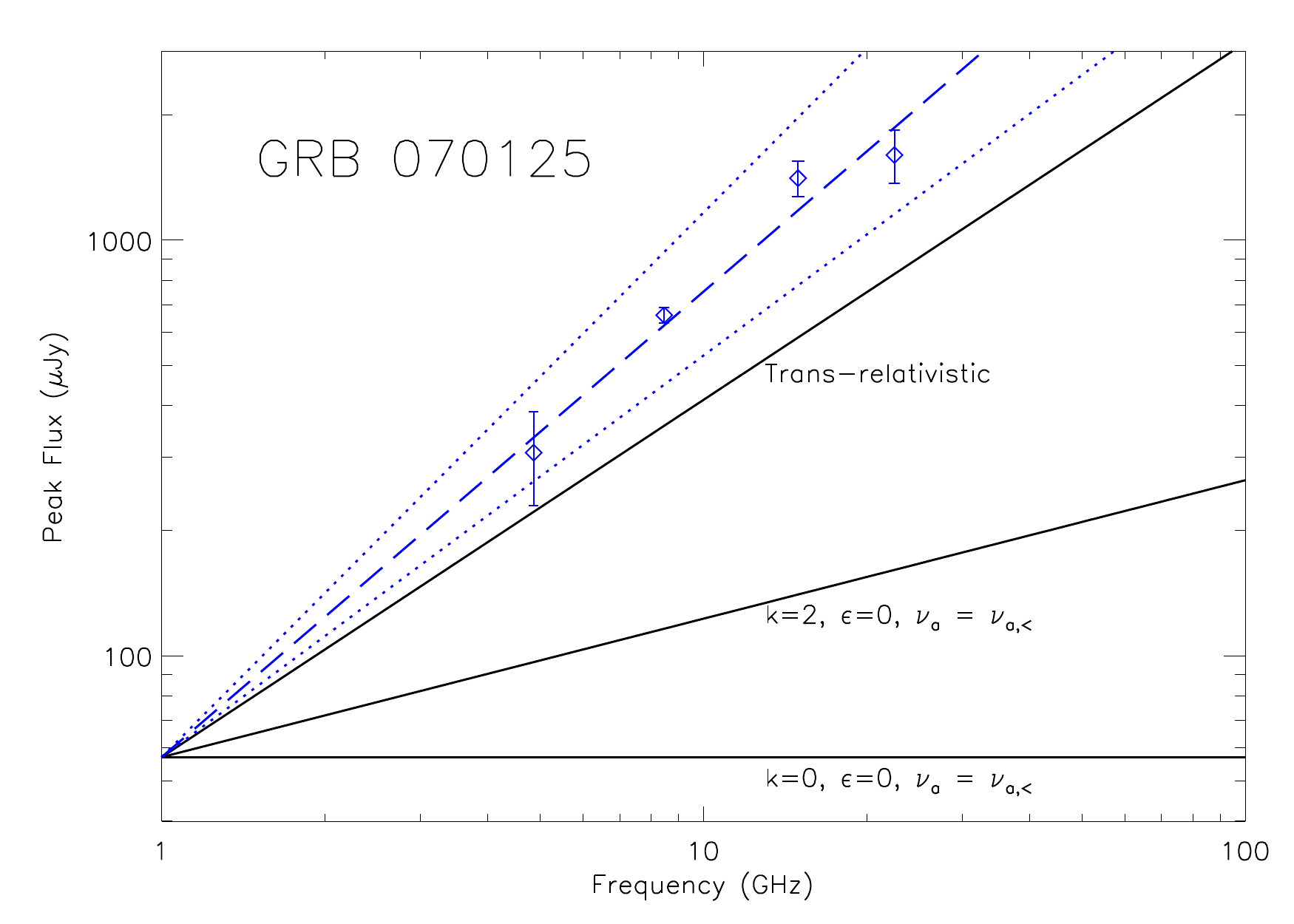}{0.31\textwidth}{}
          }
\gridline{\fig{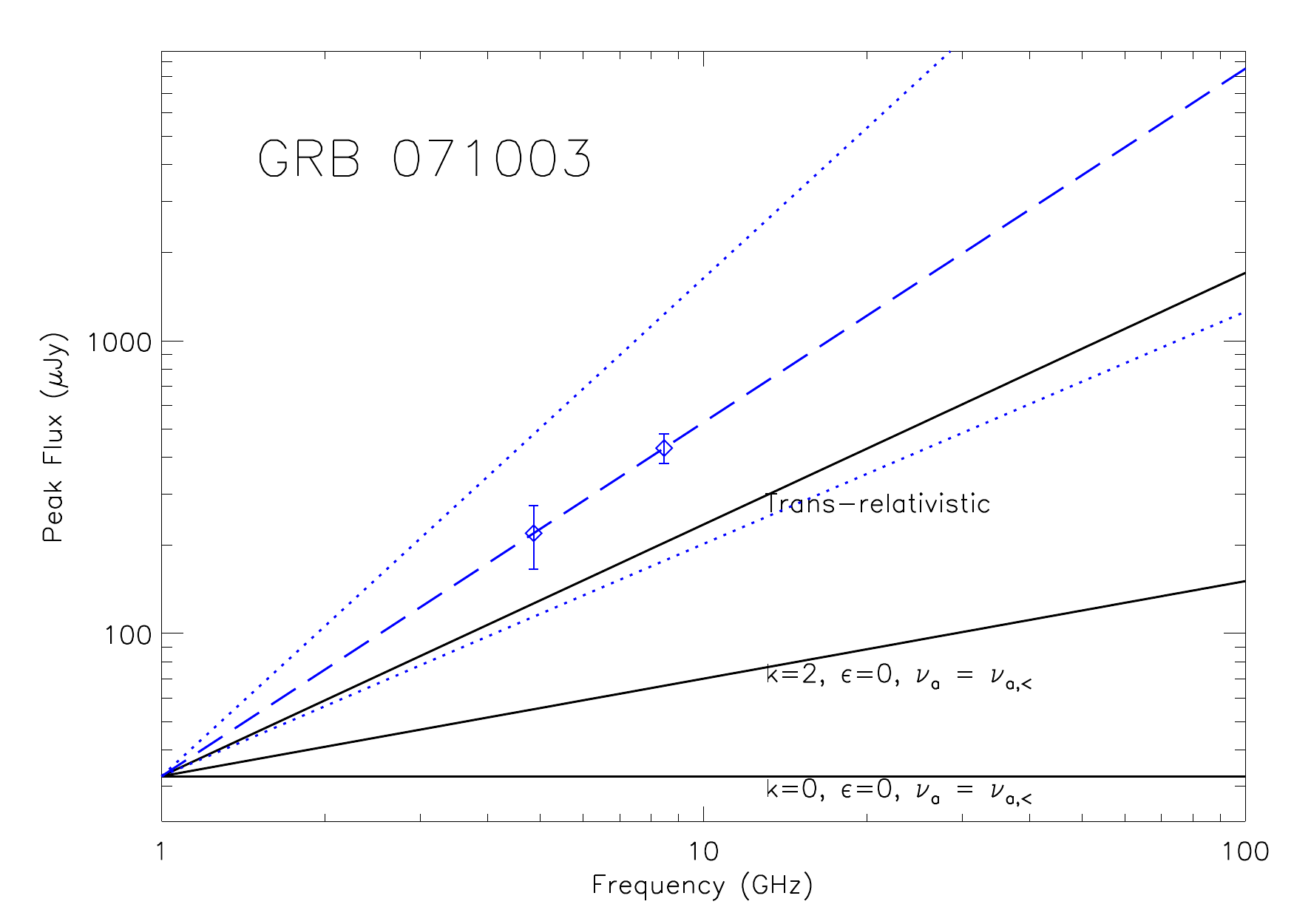}{0.31\textwidth}{}
          \fig{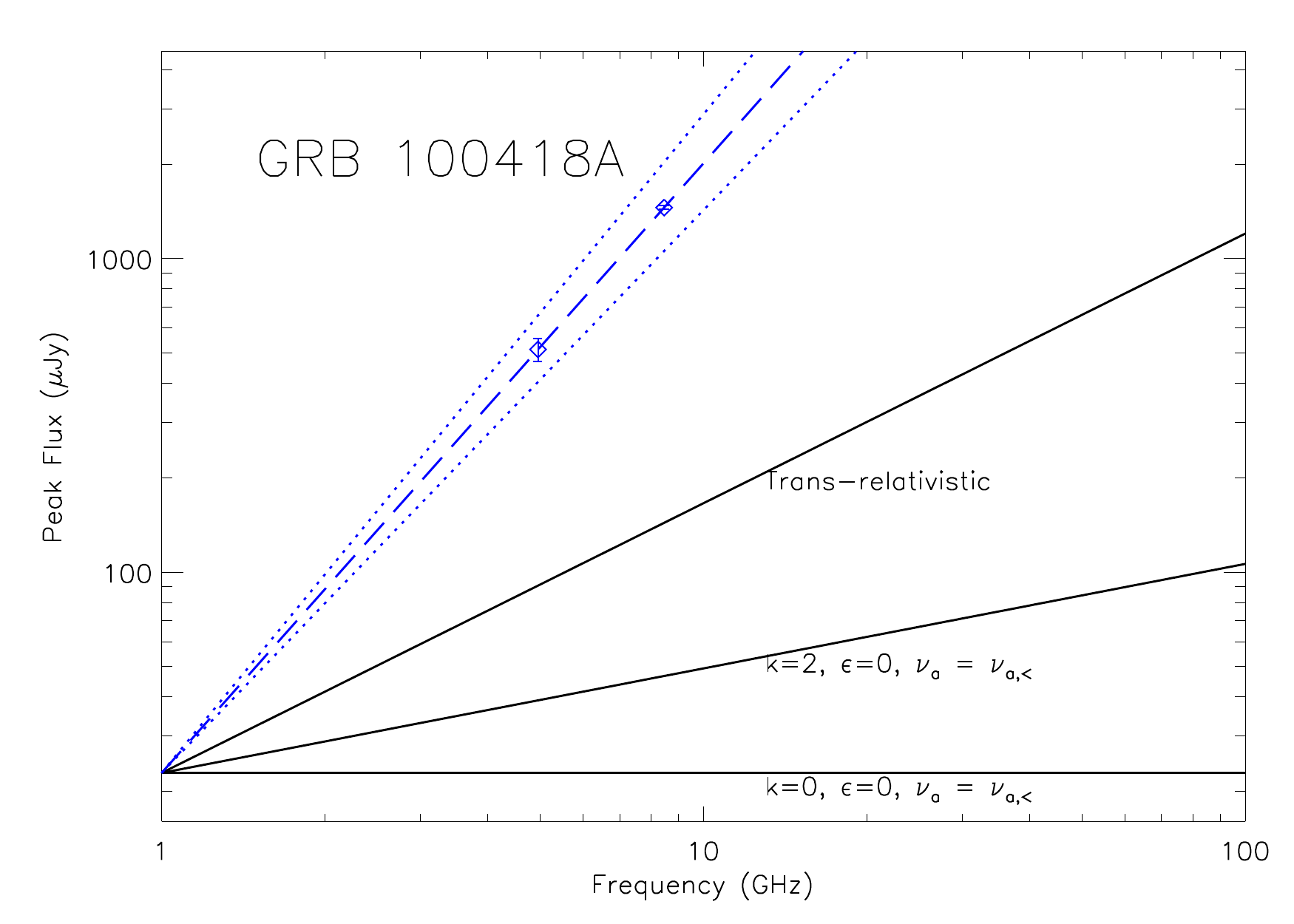}{0.31\textwidth}{}
          \fig{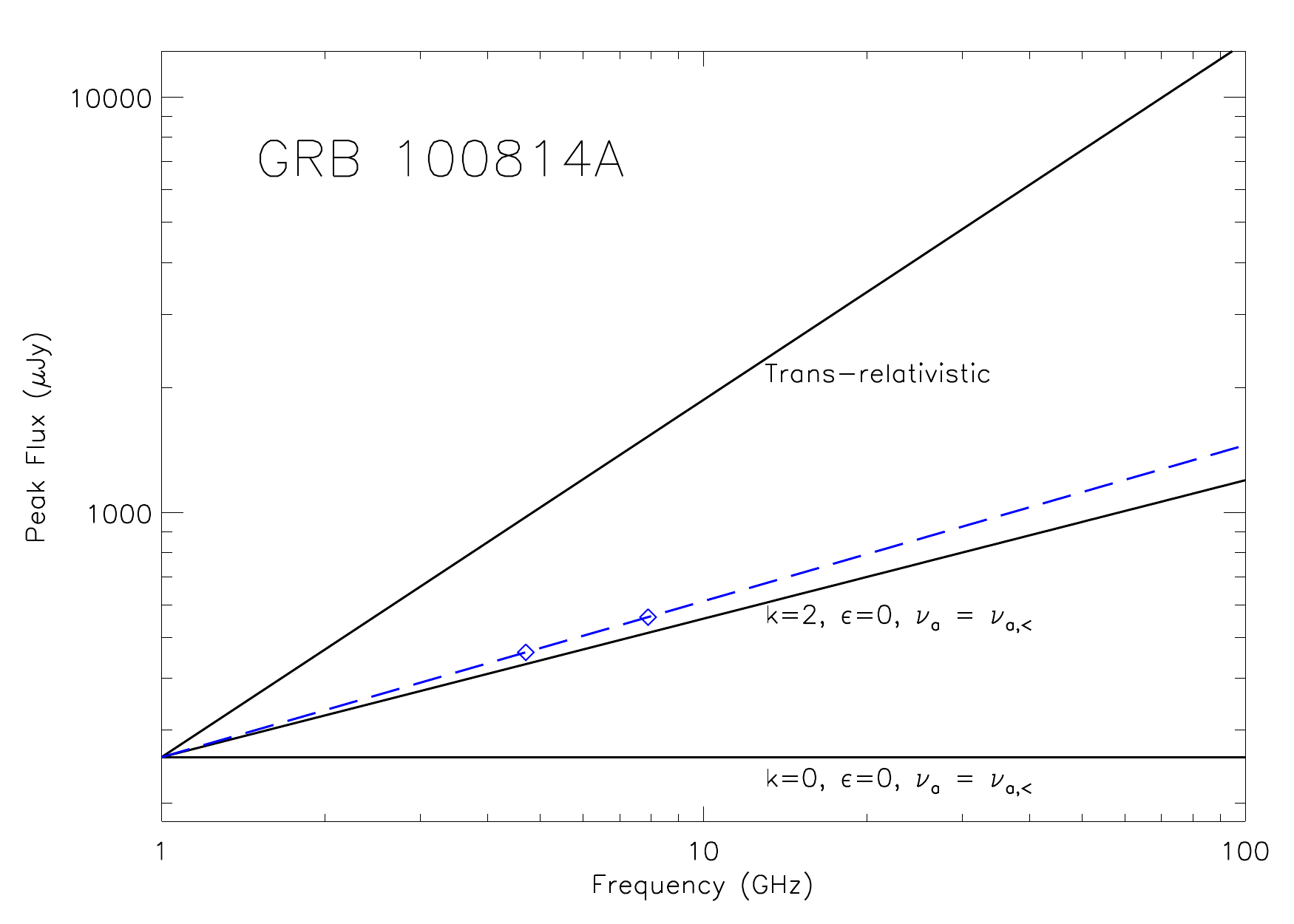}{0.31\textwidth}{}
          }
\small{Figure \ref{fig:fpeak_nu}. - Continued}
\clearpage

\gridline{\fig{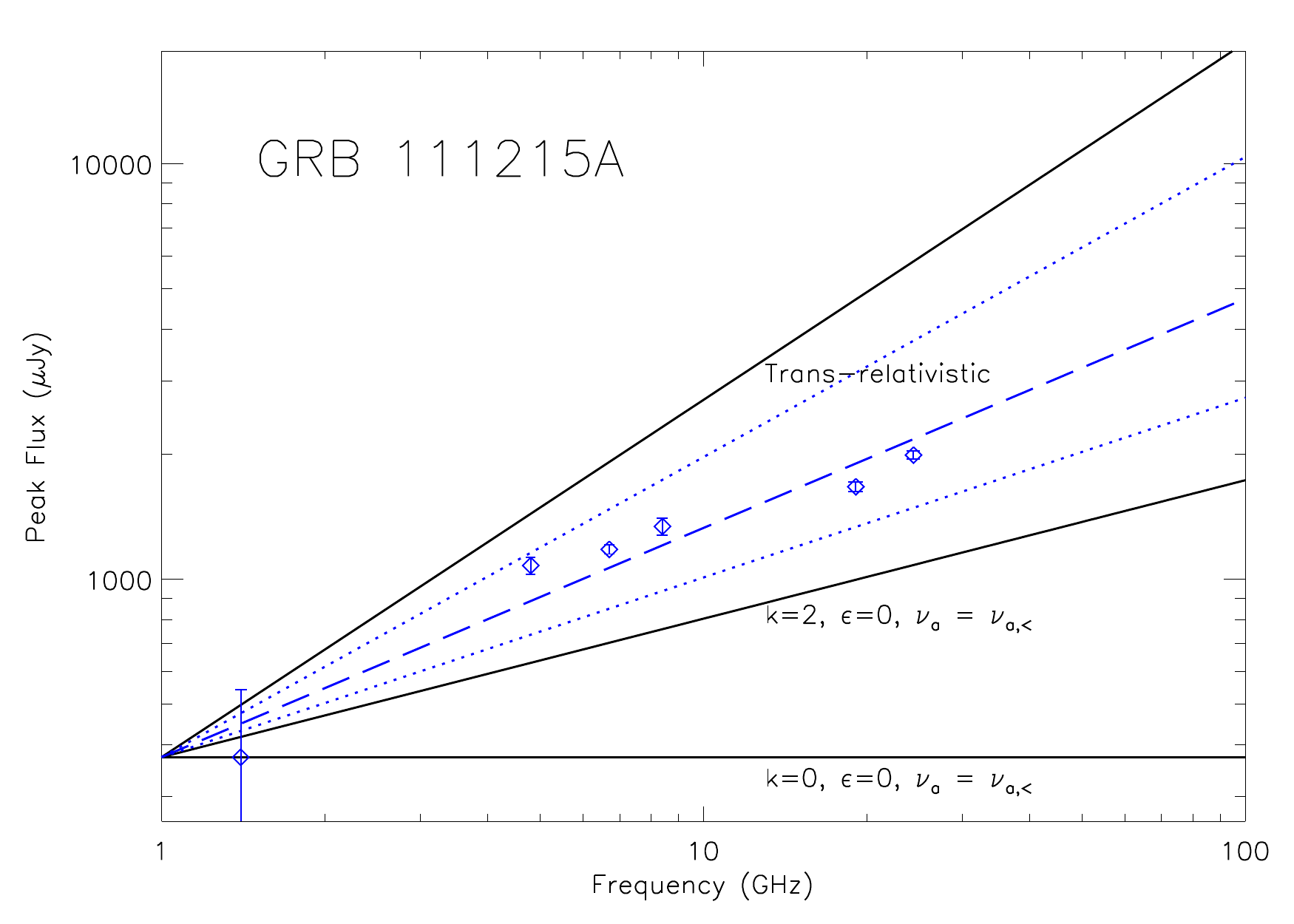}{0.31\textwidth}{}
          \fig{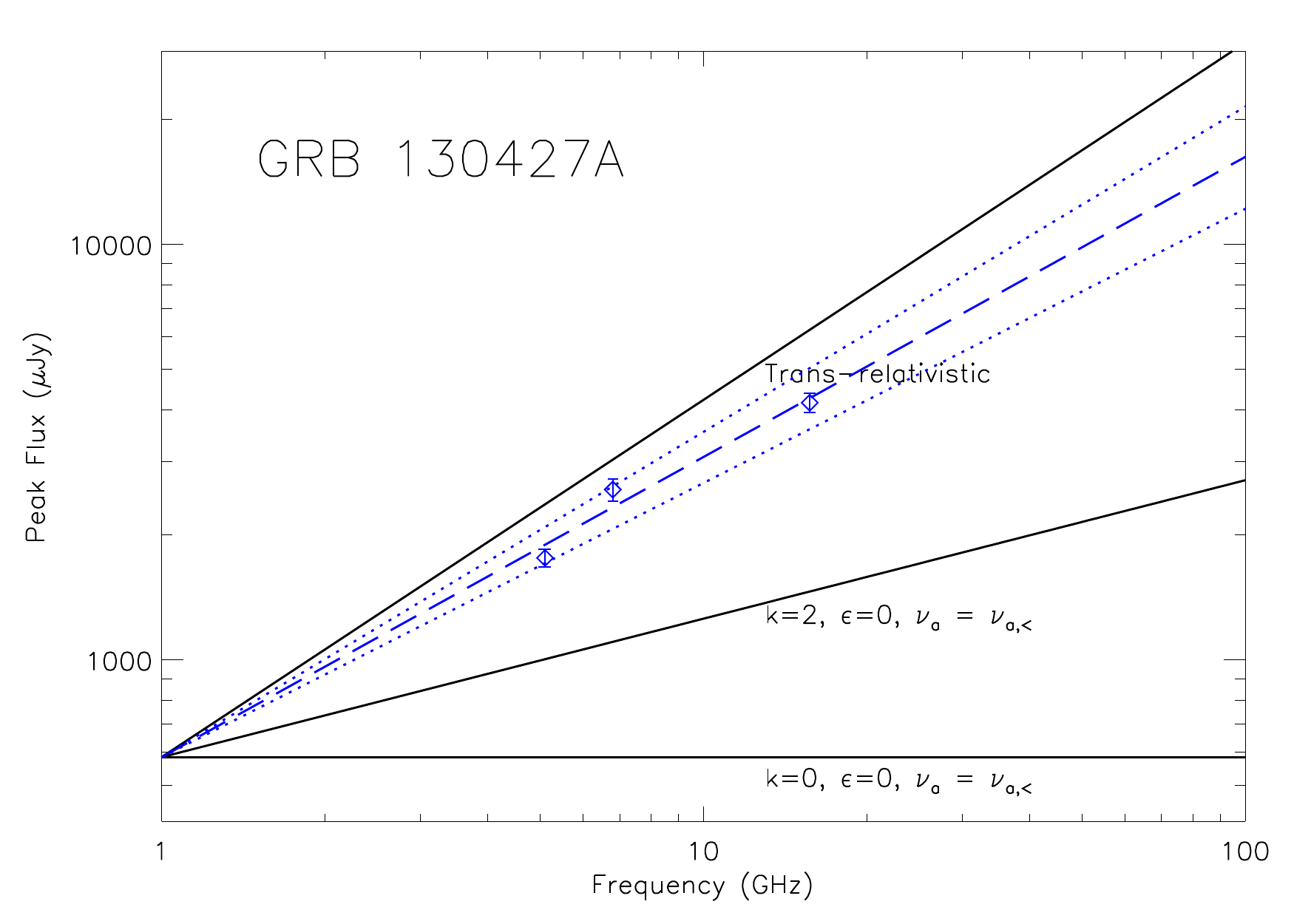}{0.31\textwidth}{}
          \fig{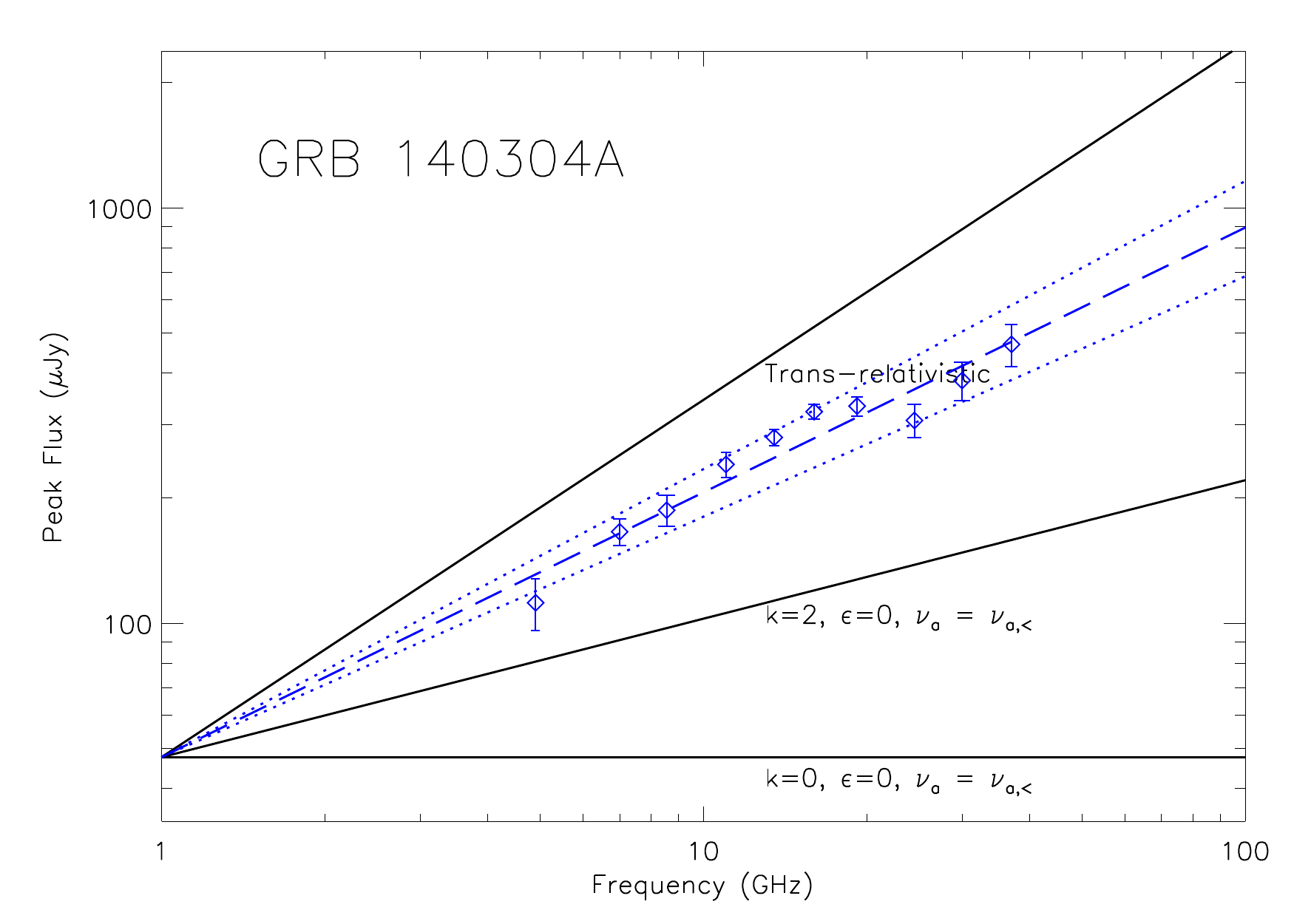}{0.31\textwidth}{}
          }
\gridline{\fig{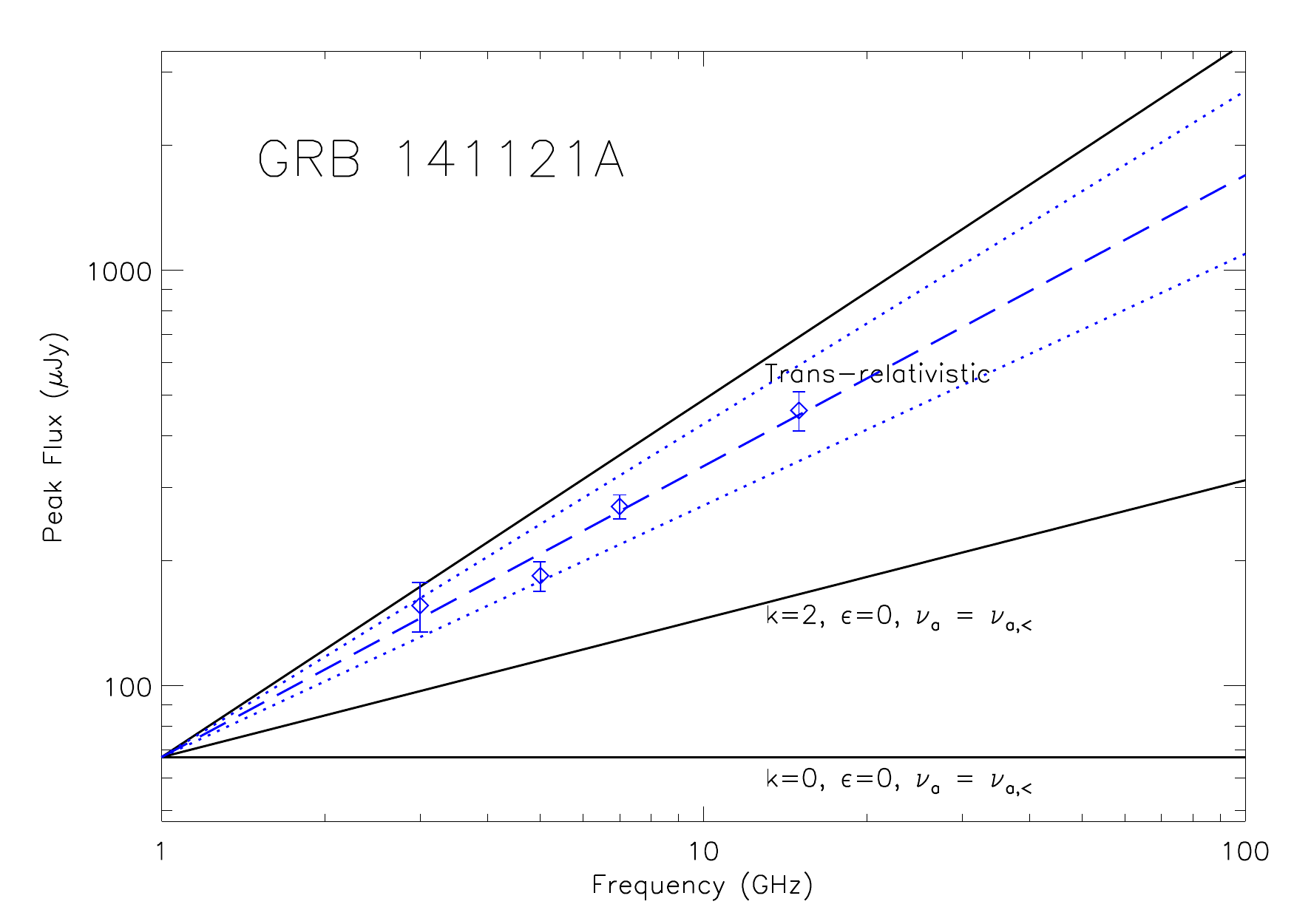}{0.31\textwidth}{}
          \fig{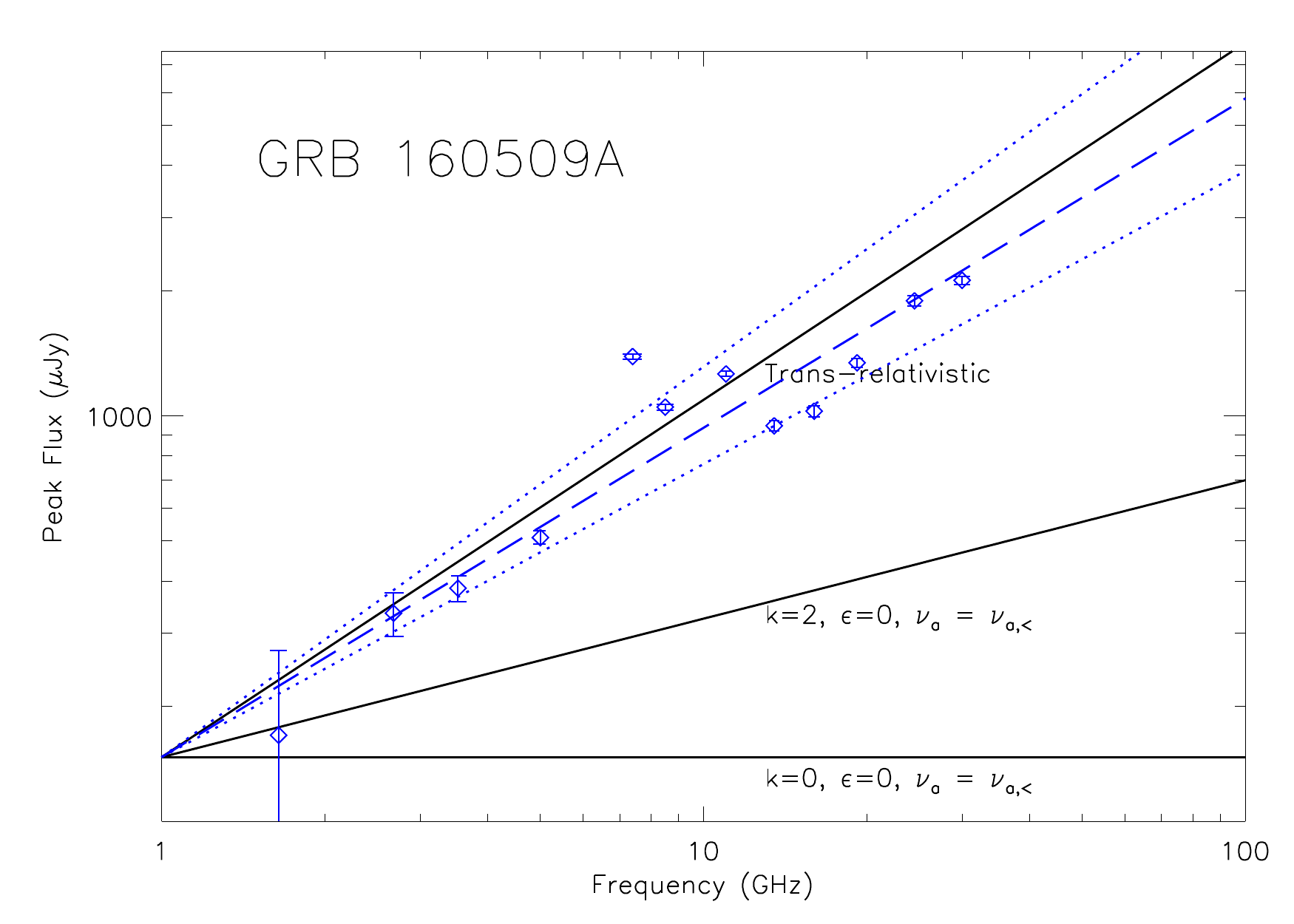}{0.31\textwidth}{}
          \fig{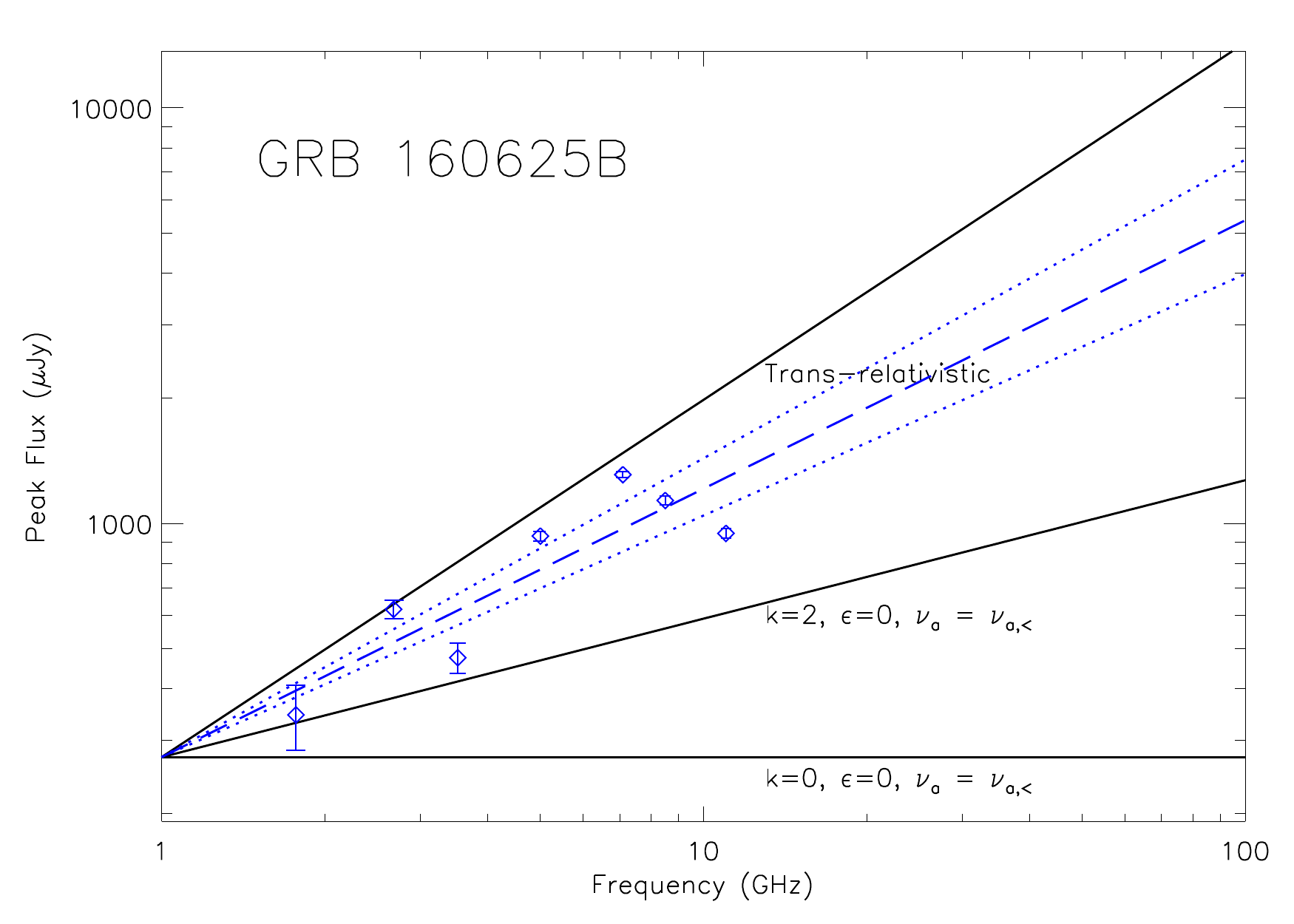}{0.31\textwidth}{}
          }
\gridline{\fig{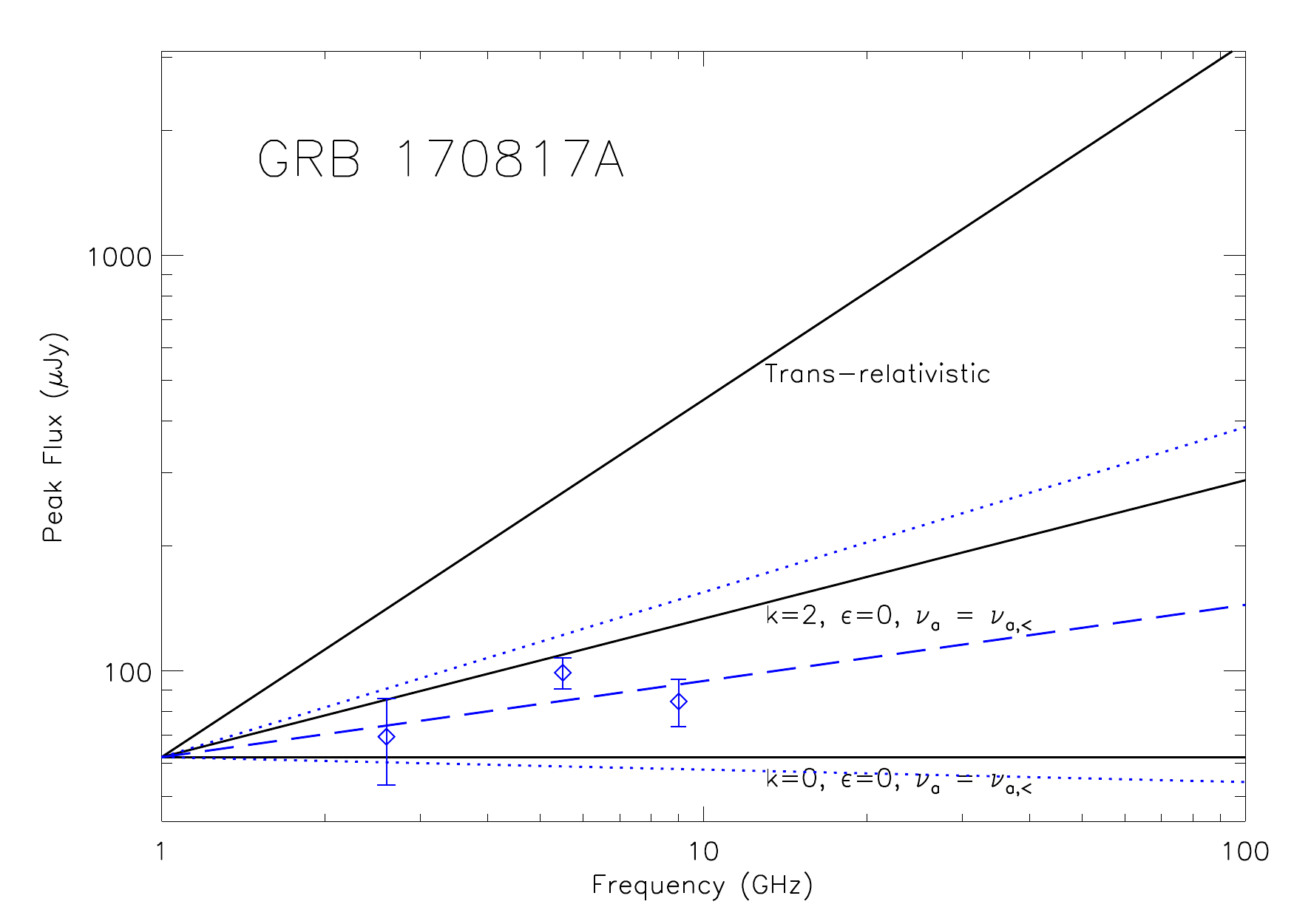}{0.31\textwidth}{}
          \fig{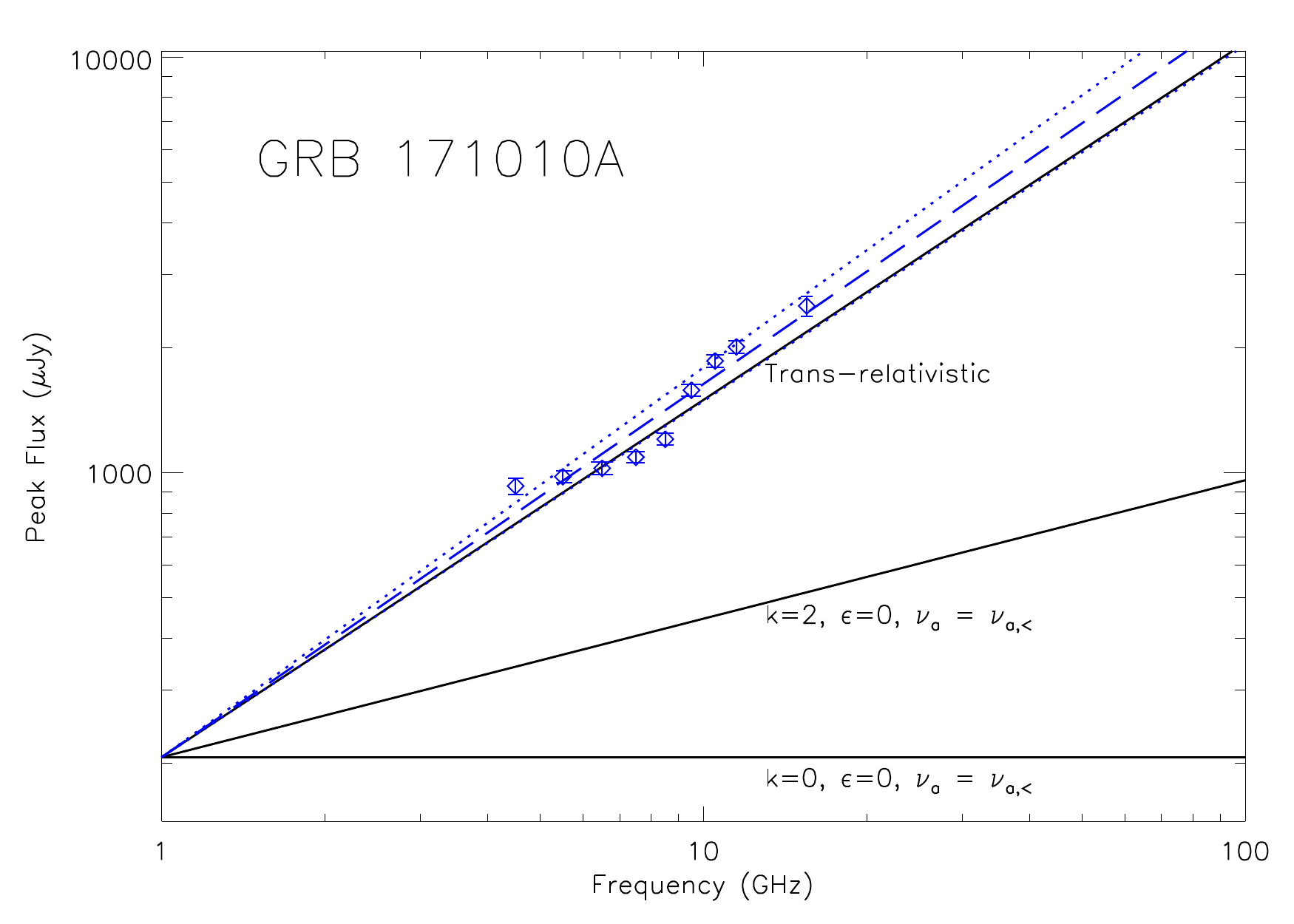}{0.31\textwidth}{}
          }
\small{Figure \ref{fig:fpeak_nu}. - Continued}
\clearpage

It can be seen that by comparing multiband radio-peak data, one can determine the value of the circumburst medium distribution index $k$. We carry out an analysis for 32 GRBs with multiband radio afterglow light-curve data, and find that half of them can be explained with transrelativistic afterglows under uniform ISM environments. Only $\sim20$\% of our sample GRBs can be certainly determined that they occur in wind-like environments with larger $k$, although it should also be noted that the radio peaks of nearly 30\% of our sample GRBs can be interpreted by a $k$ value between 0 and 2. Most of these results are compatible with existing analyses of individual bursts. However, \cite{Yi2013} found that nearly all their 19 sample GRBs show signs of $k \sim 0.4 -1.4$, with a typical value of $k=1$, by analyzing early forward-reverse-shock evolution. Obviously, our conclusion is different from that of \cite{Yi2013}. One of the possible explanations could be that the emission regions of GRB afterglows become larger after entering the self-similar phase. And circumburst medium distributions at these regions may be quite different from the early forward-reverse-shock-dominated areas, which are much closer to the GRB central engine. That is because the mass-loss processes of GRB progenitors induced by the stellar wind or other mechanisms have limited influence and may not change the density distribution farther away. Besides, the sample utilized by \cite{Yi2013} does not overlap with our radio sample. It is difficult to judge whether such inconsistency is due to diversity among individual bursts or changes in density distribution at different distances around a single GRB. If complete follow-up observations for one burst become available in the future, from early X-ray and optical detections to late-time radio observations, we can use these information to fully understand the GRB circumburst medium distributions at various distances and possible changes within.

As seen in Section \ref{sec:obs}, the main uncertainty in our analysis comes from errors in peak-time/flux estimation. As for each GRB we consider the maximum observed flux as its approximate peak flux and the corresponding observing time as an estimated peak time in our index fitting, while the real peak may be situated at any time between the two observing times adjacent to our estimation, with a flux no lower than our estimation, if the sampling of the light curve is not frequent enough during the follow-up observations, especially if the data points near the peak time are sparsely distributed, the fitted $a$ and $b$ values using our method do have larger errors, and the circumburst medium distribution cannot be pinned down without doubts. However, it is worth noting that although such uncertainty does exist, our constraints on $k$ are still better than those from single-band light-curve slope fitting. On one hand, in many cases, it is nearly impossible to fit the light-curve slopes reliably in order to calculate $k$ with only a handful of data points; while our approach described in Section \ref{sec:obs} can show some preliminary results at least. On the other hand, even if we can make enough observations for one GRB in a single band, those data far away from the emission peak often exhibit large errors due to their low fluxes and instrument limitations, and a precise fitting on $k$ cannot be guaranteed. On the contrary, with more data on hand, the uncertainties on peak estimation are greatly reduced this time. Hence, with more high-quality light-curve observations made available in the future, the parameters on multiband radio peaks can be better determined, and more stringent constraints on circumburst density distributions can be obtained.

Because multiband afterglow observations covering the low-frequency range may show signs of transrelativistic transitions, the values of $a$ and $b$ at various bands could be quite different. Because existing radio observations usually sample a few frequencies only, we can only see signs of such transitions in a handful of bursts. Thus, the uncertainties here cannot be ignored. If one can carry out radio follow-ups at more frequencies in the future, such transitions should be unveiled. Besides, the new generation of large radio telescopes, including the Five-hundred-meter Aperture Spherical radio Telescope (FAST) and the upcoming Square Kilometer Array (SKA), have much higher sensitivities at lower frequencies, and thus can be used to detect GRB late-time afterglows (e.g., see \citealt{Li2015}; \citealt{Zhang2015}; \citealt{Ruggeri2016}), and sample more frequency bands at longer wavelengths in order to get a more complete picture of transrelativistic shock behaviors.

\acknowledgments 

This work is supported by the National Natural Science Foundation of China (grant Nos. 11903056, 11725314, 12041306), as well as the Joint Research Fund in Astronomy (U1731125) under a cooperative agreement between the National Natural Science Foundation of China and Chinese Academy of Sciences. B.Z. is also supported by the Open Project Program of the Key Laboratory of FAST, National Astronomical Observatories, Chinese Academy of Sciences (NAOC), the Cultivation Project for FAST Scientific Payoff and Research Achievement of CAMS-CAS, as well as the Young Researcher Grant of NAOC. L.D.L. is also supported by the National Postdoctoral Program for Innovative Talents (grant No. BX20190044), China Postdoctoral Science Foundation (grant No. 2019M660515), and ``LiYun'' Postdoctoral Fellow of Beijing Normal University. FL is also supported by Shanghai Post-doctoral Excellence Program. The authors thank Ming Zhu, Cheng Cheng, Heng Yu, and Qing-Zheng Yu for helpful discussions.

\bibliography{sample63}{}
\bibliographystyle{aasjournal}



\end{document}